\documentclass[11pt,a4paper]{article}
\usepackage{amsmath,amsthm,amssymb,latexsym}
\usepackage[T1]{fontenc}    % Accents cods dans la fonte.
\usepackage{graphics}
\usepackage{graphicx}
\usepackage{pstricks,pst-coil,pst-fill,pst-plot}
\newcommand{\be}[1]{\begin{equation}\label{#1}}
\newcommand{\ba}[1]{\begin{multline}\label{#1}}
\newcommand{\ee}{\end{equation}}
\newcommand{\ea}{\end{eqnarray}}

\newcommand{\num}{\\\rule{0pt}{20pt}}
\newcommand{\numa}[1]{\\\rule{0pt}{#1pt}}
\newcommand{\dis}{\displaystyle}

\newcommand{\tr}{\mathop{\rm tr}}
\newcommand{\Res}{\mathop{\rm Res}}

\newcommand{\qp}{s^{\scriptscriptstyle +}}
\newcommand{\qm}{s^{\scriptscriptstyle -}}
\newcommand{\qpm}{s^{\scriptscriptstyle \pm}}
\newcommand{\Rp}{r^{\scriptscriptstyle +}}
\newcommand{\Rm}{r^{\scriptscriptstyle -}}
\newcommand{\Rpm}{r^{\scriptscriptstyle \pm}}
\newcommand{\Rmp}{r^{\scriptscriptstyle \mp}}
\newcommand{\hqp}{\hat s^{\scriptscriptstyle +}}
\newcommand{\hqm}{\hat s^{\scriptscriptstyle -}}
\newcommand{\hqpm}{\hat s^{\scriptscriptstyle \pm}}

\newcommand{\moy}[1]{\ensuremath{\langle #1 \rangle}_T}
\newcommand{\Mmoy}[1]{\ensuremath{\langle\hspace{-1mm}\langle #1 \rangle\hspace{-1mm}\rangle}_T}

\newtheorem{thm}{Theorem}[section]
\newtheorem{prop}{Proposition}[section]
\newtheorem{lemma}{Lemma}[section]

\def\qed{\hfill\nobreak\hbox{$\square$}\par\medbreak}
 \makeatletter
 \@addtoreset{equation}{section}
 \makeatother
 
\newcommand{\abs}[1]{\ensuremath{\left| #1 \right|}}
\newcommand{\e}[1]{\ensuremath{\text{#1}}}

\begin{document}
\begin{flushright}

LPENSL-TH-11/10

DESY-T- 11/10
\end{flushright}
\par \vskip .1in \noindent

\vspace{24pt}

\begin{center}
\begin{LARGE}
\vspace*{1cm}
  {Long-distance behavior of temperature correlation functions
  in the one-dimensional Bose gas }
\end{LARGE}

\vspace{30pt}

\begin{large}

{\bf K.~K.~Kozlowski}\footnote[1]{DESY, Hamburg, Deutschland,
 karol.kajetan.kozlowski@desy.de},~~
{\bf J.~M.~Maillet}\footnote[2]{ Laboratoire de Physique, UMR 5672
du CNRS, ENS Lyon,  France,
 maillet@ens-lyon.fr},~~
{\bf N.~A.~Slavnov}\footnote[3]{ Steklov Mathematical Institute,
Moscow, Russia, nslavnov@mi.ras.ru},~~
%
%{\bf V.~Terras}\footnote[5]{ Laboratoire de Physique, UMR 5672 du
%CNRS, ENS Lyon,  France, veronique.terras@ens-lyon.fr}
\par

\end{large}

\vspace{80pt}

\centerline{\bf Abstract} \vspace{1cm}
\parbox{12cm}{\small  We describe a Bethe ansatz based method to derive, starting from a multiple integral representation, the long-distance
asymptotic behavior at finite temperature of the density-density correlation function in the interacting one-dimensional Bose gas.
We compute the correlation lengths in terms of solutions of non-linear integral equations of the thermodynamic Bethe ansatz type.
Finally, we establish a connection between the results obtained in our  approach with
the correlation lengths stemming from the quantum transfer matrix method.}
\end{center}

\newpage%

\section{Introduction}

This article is devoted to the study the long-distance asymptotic
behavior of the correlation functions in the one-dimensional
Bose-gas at finite temperature. The method of analysis we apply
builds on the method developed in the work \cite{KitKMST09a}. There,
the long-distance asymptotic behavior of the zero temperature
correlation functions of massless quantum integrable models was
derived in the framework of the algebraic Bethe ansatz. This
setting allows one to present the correlation functions at zero
temperature as  series of multiple integrals \cite{KitMT00,KitMST02a,KitMST05me,KitMST05k,KitKMST07}. In the long-distance
regime, each of the multiple integrals appearing in one of these series  can
be evaluated asymptotically. The resulting asymptotic series can
then be summed up \cite{KitKMST09a}.

In \cite{KitKMST09a}, we have focused on the example of the
generating function for correlation functions of the third
components of local spin in the $XXZ$ spin-$1/2$ Heisenberg chain.
We have mentioned, however, that the method can be applied to other
integrable models as well. In particular we showed how one can use
the same approach for the model of one-dimensional bosons
described by the quantum nonlinear Schr\"odinger equation (QNLS
model) using  \cite{KitKMST07}.

We would like to stress that the algebraic Bethe ansatz approach of
\cite{KitKMST09a} was fit for the analysis of the asymptotic behavior of correlation functions
at zero temperature. However, in some models such as the QNLS one, it can also be applied
for the evaluation of the long-distance asymptotic behavior of
correlation functions at finite temperature. It is this problem that we consider in the
present article.

%\subsection{Formulation of the model}
%\label{Section Formulation of the model}

The Hamiltonian of the QNLS model is given by
 \begin{equation}\label{8-Ham}
 H=\int\limits_0^L\!\left(
 \partial_x\Psi^\dagger\partial_x\Psi
 +
 c\Psi^\dagger\Psi^\dagger\Psi\Psi-h\Psi^\dagger\Psi\right)\,d x\, .
 \end{equation}
Here $\Psi$ and $\Psi^\dagger$ are Bose-fields subject to canonical,
equal-time commutation relations, $c$ is a coupling constant and $h$
the chemical potential. We focus on the case of the repulsive regime (i.e. $c>0$)
 and
consider the model on a finite interval $[0,L]$ subject to
periodic boundary conditions. We will take the thermodynamic
limit $L\to\infty$ later on.

This model, for generic $c$, was first introduced and solved by Lieb
and Liniger in \cite{LieL63,LieM66}. It can be considered as a
generalization of the model of impenetrable bosons considered by
Girardeau \cite{Gir60}. The spectrum of this model can be obtained
by the algebraic Bethe ansatz \cite{FadST79,FadLH96,BogIK93L}. The
thermodynamics of the QNLS model were first studied in \cite{YanY69}
and this analysis was made rigorous in \cite{DorLP89}. It was shown
there that the state of thermal equilibrium is described by a
non-linear integral equation for the excitation energy
$\varepsilon(\lambda)$
 \begin{equation}\label{YY-eq}
 \varepsilon(\lambda)=\lambda^2-h-\frac
 T{2\pi}\int\limits_{\mathbb{R}}K(\lambda-\mu)\log\left(
 1+e^{-\frac{\varepsilon(\mu)}T}\right)\,d\mu,
 \end{equation}
where $T$ is the temperature and
 \begin{equation}\label{Lieb-kern}
 K(\lambda)=\frac{2c}{\lambda^2+c^2}.
 \end{equation}
Below we refer to equation \eqref{YY-eq} as the Yang--Yang equation. The state
of the thermal equilibrium is given by a Dirac sea filled with a certain
density of particles $\rho_p(\lambda)$ and an associated density of holes
$\rho_h(\lambda)$. The total density is denoted
$\rho_t(\lambda)=\rho_p(\lambda)+\rho_h(\lambda)$. The Fermi weight $\vartheta(\lambda)$, as usual, is defined as the ratio of the density of particles to the total density, namely,
 \begin{equation}\label{dens-FW}
 \frac{\rho_p(\lambda)}{
 \rho_t(\lambda)}=\vartheta(\lambda)=\left(1+e^{\frac{\varepsilon(\lambda)}T}\right)^{-1},
 \end{equation}
and the functions
$\rho_p(\lambda)$ and  $\rho_t(\lambda)$ are related to each other by the
integral equation
 \begin{equation}\label{inteq-rho}
 \rho_t(\lambda) -\frac1{2\pi}\int\limits_{\mathbb{R}}K(\lambda,\mu)
 \rho_p(\mu)\, d\mu=\frac1{2\pi}.
 \end{equation}

The temperature dependent correlation functions are defined in a
standard way
 \begin{equation}\label{def-Tcf}
 \langle{\cal O}\rangle_T=\frac{\tr\left({\cal O}e^{-H/T}\right)}{\tr\left(e^{-H/T}\right)}
 =\frac{\sum\langle\Omega|{\cal O}|\Omega\rangle
 e^{-E/T}}{\sum e^{-E/T}}.
 \end{equation}
Here $|\Omega\rangle$ are eigenstates of the Hamiltonian \eqref{8-Ham} and $E$
are their eigenvalues. The sums in \eqref{def-Tcf} are taken over the complete
set of the eigenstates $|\Omega\rangle$.

Exact representations for the temperature dependent correlation
functions of the QNLS model where obtained for the impenetrable Bose
gas ($c=\infty$) in \cite{Len66,Kor87,KorS90}  by using the free fermion structure of the model.
The long-distance asymptotic behavior of the two-point functions in
the impenetrable Bose gas has been derived in
\cite{ItsIK90,ItsIK90a,ItsIKV91} with the use of Riemann--Hilbert
problem method. The case of general coupling constant $0<c<+\infty$ was
considered in \cite{BogK84} by the algebraic Bethe ansatz. There a
series of multiple integrals was obtained for thermal
density-density correlation function of the QNLS model. The
method of dual fields was applied in \cite{Kor87,Kor89,KorS91,KojKS97}
for the derivation of determinant-type representations for various
correlation functions. Those representations were used in
\cite{ItsS99,Sla99} for the asymptotic analysis. One should also mention the method of the asymptotic analysis based on
the functional integral approach \cite{PopL83}. Yet another approach
to estimate these long-distance asymptotic behavior of correlation
functions at finite temperature is provided by the conformal field
theory \cite{BelPZ84,Car84,Car86,Car96} and references therein. These last two methods
are, however, restricted to the low-temperature regime only.

Recently, in the work \cite{SeeBGK07}, the quantum transfer matrix
(QTM) approach was applied to the QNLS model. We recall that,
originally, this method was developed for quantum spin chains
(see the nice review \cite{Klu04} and references therein), where one can
construct the QTM $T_q$ explicitly. The diagonalization of the QTM
gives access to its leading eigenvalue (the logarithm of which
corresponds to the free energy) as well as to the subdominant ones
(which in their turn give access to the correlation lengths at
$T>0$). In the infinite size limit, these eigenvalues are expressed
as weighted integrals involving a counting function %$\mathfrak{a}$
which satisfies a thermodynamic Bethe ansatz (TBA) non-linear
integral equation (see also \cite{TakSK2001}). To the best of our knowledge, the analog of $T_q$ in the QNLS model is
not known nowadays. Therefore,  in \cite{SeeBGK07}, the last model
was treated as some special, continuous limit of the XXZ spin chain.
It was shown that, in this limit, the non-linear integral equation
describing the maximal eigenvalue of the QTM goes to the Yang--Yang
equation. This observation allowed the authors of \cite{SeeBGK07} to
obtain a multiple integral representations for the temperature
correlation functions in the QNLS model.

The first steps of our method are close to the ones adopted in
\cite{BogK84}. Namely, we build on the arguments given in
\cite{BogIK85} (see also \cite{BogK84,BogIK93L}):  in the
thermodynamic limit, the representation \eqref{def-Tcf} can be
replaced, at least for local or quasi-local operators $\cal O$, by a single expectation value
 \begin{equation}\label{Tcf-OO}
 \langle{\cal O}\rangle_T=\frac{\langle\Omega_T|{\cal O}|\Omega_T\rangle}
 {\langle\Omega_T|\Omega_T\rangle},
 \end{equation}
where $|\Omega_T\rangle$ is any one of the eigenstates of the
Hamiltonian corresponding to the thermal equilibrium. This
representation constitutes the starting point for our calculations.
The matter is that the algebraic Bethe ansatz provides multiple
integral representations for the expectation values of a wide class
of operators with respect to an {\sl arbitrary} eigenstate of the
Hamiltonian. In particular, one can obtain such representations for
the expectation values with respect to the state of thermal
equilibrium. In this way, we recast $ \langle{\cal O}\rangle_T$ as a
series of multiple integrals. Once this is done, it remains to
evaluate the asymptotic behavior (in the distance) of the obtained
integrals. It is remarkable that our results are given in terms of
solutions to TBA non-linear integral equations. These can be
understood as the limiting (in the sense of \cite{SeeBGK07})
equations describing the sub-leading eigenvalues of the QTM. Thus,
our method provides a link between the method based on the
representation \eqref{Tcf-OO} and the QTM approach. In particular,
we obtain an extension  of the QTM-based results
\cite{Klu93,KluWZ93} for the correlation lengths in the XXZ spin
chain here for the QNLS model. By applying the scaling proposed in \cite{SeeBGK07} one can
actually map the correlation lengths obtained for the XXZ spin chain
to the ones obtained by our asymptotic analysis.

We have already mentioned that the representation for temperature
correlation functions of QNLS model as a series of multiple
integrals were obtained also in \cite{BogK84}. Later it was analyzed
in \cite{BogK85,BogK85a,KorS86}. However the authors of those works
restricted themselves to the analysis of the first few terms of the
series. This led them to the wrong conjecture for the long-distance
asymptotic behavior of the correlation functions. We would like to
stress that the series of multiple integrals obtained in
\cite{BogK84},  as well as ours are not \textit{well ordered} in respect to the
$x\to\infty$ limit. That is to say all the summands in both series contribute
to the leading asymptotic behavior as well as to the
corrections. Our method allows us to compute all such
contributions and re-sum them.

This article is organized as follows. In section~\ref{S-PtSMR}, we state the problem to solve and the main results of this paper. In section~\ref{Section
MultIntRep} we derive a series of multiple integral representation
for the generating function of the density-density correlation
function at finite temperature. In section~\ref{Sec-aAMI} we perform
the asymptotic analysis of the individual multiple integrals
appearing in the series. We re-sum these asymptotic expressions in
section~\ref{section Lagrange series} by the use of Lagrange series.
Then, in section~\ref{ABoCF} we obtain the long-distance asymptotic
behavior of the density-density correlation function. There, we
discuss the leading term  and connect  the various correlation lengths
we obtain with the quantum transfer matrix method. In
appendix~\ref{Appendix Numerics} we present a numerical analysis of
the expressions for the correlation lengths that we have obtained. Technical
aspects relative to the formulae for the amplitudes are gathered in
appendix~\ref{AEebQ}. We discuss certain integral identities in
appendix~\ref{appendix Deformation of integrals}. Finally, in
appendix~\ref{appendix Lagrange series} we briefly review the
results we need on continuous Lagrange series.

%%%%%%%%%%%%%%%%%%%%%%%%%%%%%%%%%%%%%%%%%%%%%%%%%%%%%%%%%%%%%%%%%%%%%%%%%%%%%%%%
\section{The problem to solve and the main results\label{S-PtSMR}}

To calculate the long-distance asymptotic behavior of the density-density correlation functions we consider its
 generating function
 \begin{equation}\label{def-Gf}
 \langle e^{2\pi i\alpha {\cal Q}_x}\rangle_T=\frac{\langle\Omega_T|e^{2\pi i\alpha {\cal Q}_x}|\Omega_T\rangle}
 {\langle\Omega_T|\Omega_T\rangle},
 \end{equation}
were $\alpha$ is a complex number\footnote[1]{%
We draw the reader's attention  to the fact that the combination $2\pi i\alpha$
was denoted by $\beta$ in \cite{KitKMST09a}. We did not use this notation  here
so as to avoid a confusion with the inverse temperature, that is traditionally
denoted by $\beta$. }. This generating function is defined in terms of the
number of particles on $[0;x]$ operator ${\cal Q}_x$
 \begin{equation}\label{8-Q1}
 {\cal  Q}_x=\int\limits_0^xj(z)\,d z, \qquad \text{with} \qquad
 j(x)=\Psi^\dagger(x)\,\Psi(x),
 \end{equation}

 Then, the density-density correlation function
$\langle j(x)j(0)\rangle_{T}$ is obtained from \eqref{def-Gf} by
 \begin{equation}\label{8-cor-fun}
 \langle j(x)j(0)\rangle_T =\left.\frac{-1}{8\pi^2}\frac{\partial^2}{\partial x^2}
 \frac{\partial^2}{\partial \alpha^2}\langle e^{2\pi i\alpha {\cal Q}_x}\rangle_T \right|_{\alpha=0}.
 \end{equation}

To describe the large-$x$ asymptotic expansion of $\langle e^{2\pi
i\alpha {\cal Q}_x}\rangle_T$, we first consider a set of functions
$u_i(\lambda)$ satisfying the non-linear integral equations
 \begin{equation}\label{inteq-u-main} %u_{\boldsymbol{j};\boldsymbol{k}}
 u_{i}(\lambda)=\lambda^2-h_\alpha-\frac{T}{2\pi}\int\limits_{\mathbb{R}}
 K(\lambda-\mu)\log\left(1+e^{-\frac{u_{i}(\mu)}T}
 \right)\,d\lambda%\\
 -iT\sum_{\ell=1}^n [ \theta(\hqp_{\ell}-\lambda)-\theta(\hqm_{\ell}-\lambda) ],
 \end{equation}
where $n=0,1,\dots$, the kernel $K(\lambda-\mu)$ is given by \eqref{Lieb-kern},
$h_\alpha =h+2\pi i\alpha T$, and
 \begin{equation}\label{def-theta}
 \theta(\lambda)=i\log\left(\frac{ic+\lambda}{ic-\lambda}\right) \; ,
 \qquad \theta'(\lambda)=K(\lambda).
 \end{equation}
The parameters $\{\hqpm_{\ell}\}$ belong to the upper (resp. lower)
half-plane and are the roots of the equation
 \begin{equation}\label{cond}
 %1+\exp\Bigl(-{u_{i}(\hqpm_{\ell})}/T\Bigr)=0.
 1+e^{-{u_{i}(\hqpm_{\ell})}/T}=0.
 \end{equation}
The subscript $i$ in $u_i$  serves as a way to label all the
possible choices of $n$ roots $\{\hqpm_{\ell}\}_i$. In the framework
of our analysis, the roots $\hqpm_{\ell}$ arise as
deformations  of the poles $\Rpm_\ell$ of the Fermi weight
\eqref{dens-FW}. Namely, there exist functions
$\hqpm_{\ell}(\gamma)$ of some parameter $\gamma$ such that $\hqpm_{\ell}(0)=\Rpm_{\ell}$ and
$\hqpm_{\ell}(1)=\hqpm_{\ell}$. Thus, one can associate every set
$\{\hqpm_{\ell}\}$ with its pre-image $\{\Rpm_\ell\}$.  Every choice of the $\{\Rpm_\ell\}_i$ uniquely yields the image
$\{\hqpm_{\ell}\}_i$ and the corresponding function $u_i(\lambda)$.
Thus, the subscript $i$ in the equation \eqref{inteq-u-main}, in
fact, enumerates different subsets of the Fermi weight poles
$\{\Rpm_\ell\}_i$. Note that, for a given function $u_i(\lambda)$,
there may exist other points $w_p\ne \hqpm_\ell$ such that
$1+e^{-{u_{i}(w_p)}/T}=0$. Following the terminology of the
QTM-based approach, we call such roots hole-type solutions.

\begin{figure}[!h]
\begin{center}
\includegraphics[width=15cm]{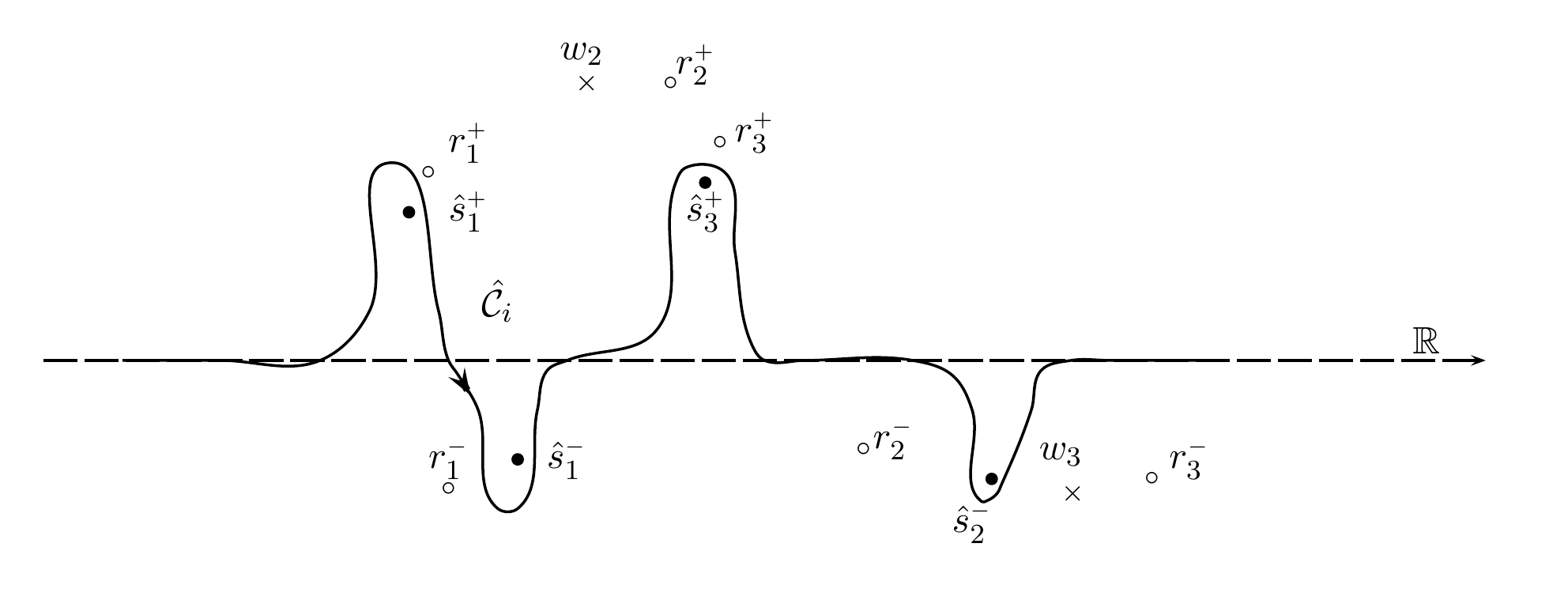}
\caption{\label{hG1212} The poles of the Fermi weight $\Rpm_\ell$
are denoted $\scriptstyle\circ$, the zeros $\hqpm_\ell$
 by $\scriptstyle\bullet$, the hole-type solutions $w_p$
  by $\scriptstyle \times$. The integration contour $\hat{\cal
C}_{i}$  bypasses the points $\hqp_1$, $\hqp_3$ in the upper
half-plane from above and bypasses the points $\hqm_1$, $\hqm_2$ in
the lower half-plane from below avoiding the points $\Rp_i$ and $w_i$.
%The contour $\Gamma^{(1)}_{2;1}$ (dotted line) skirts the points $\qp_2$ in the upper %half-plane and $\qm_1$ in the lower half-plane.
}
\end{center}
\end{figure}

Equation \eqref{inteq-u-main} differs from the Yang--Yang equation
\eqref{YY-eq} by a shift of the chemical potential and the term depending on the functions
$\theta(\hqpm_\ell-\lambda)$.
Due to \eqref{cond}, one can get rid of this term by changing  the
integration contour. Namely, let the contour $\hat{\cal C}_{i}$ be a
deformation of the real axis such that moving from $\mathbb{R}$ to
$\hat{\cal C}_{i}$ one only crosses the roots $\{\hqpm_{\ell}\}_i$
while all other solutions $w_p$ as well as all the poles of the
Fermi weight $\Rpm_\ell$ are not crossed (see Fig.~\ref{hG1212}).
Then, equation \eqref{inteq-u-main} turns into
 \begin{equation}\label{inteq-u-main1}
 u_{i}(\lambda)=\lambda^2-h_\alpha-\frac{T}{2\pi}\int\limits_{\hat{\cal C}_{i}}
 K(\lambda-\mu)\log\left(1+e^{-\frac{u_{i}(\mu)}T}
 \right)\,d\lambda,
 \end{equation}
and has the form of the Yang--Yang equation up to the shift of
the chemical potential and the change of the integration contour.
Hence, different functions $u_i(\lambda)$ are
enumerated by different  contours $\hat{\cal C}_{i}$:
$u_i(\lambda)=u(\lambda,\hat{\cal C}_{i})$. It is interesting to notice that this method of recasting equation \eqref{inteq-u-main} into \eqref{inteq-u-main1} is very similar to the one used in \cite{DorT96} to generate higher excited states within the TBA equations by analytic continuation.

%%%%%%%%%%%%%%%%%%%%%%%%%%%%%%%%%%%%%%%%%%%%%%%%%%%%%%%%%%%%%%%
% HERE IS RISUNOK-2
%%%%%%%%%%%%%%%%%%%%%%%%%%%%%%%%%%%%%%%%%%%%%%%%%%%%%%%%%%%%%%

%
%

The  large $x$ asymptotic expansion of $\langle e^{2\pi i\alpha
{\cal Q}_x}\rangle_T$ has the following form
 \begin{equation}\label{fin-answ}
 \moy{e^{2\pi i\alpha {{\cal Q}_x}}}\to    \sum_{i}e^{-xp_i}B[u_i],
 \qquad x\to\infty,
%
%
%     \; + \; \e{o}\left(e^{-Ax}\right) , \qquad \e{with} \quad  A=- \inf_{\{\boldsymbol{j};\boldsymbol{k}\}}
%    \Re ( p[\hat{\cal C}_{\boldsymbol{j},\boldsymbol{k}}] ) \; .
 \end{equation}
 where
 \begin{equation}\label{p-decay}
 p_i=-\frac1{2\pi }\int\limits_{\hat{\cal C}_i}\log\left(\frac{1+
 e^{-\frac{u_i(\lambda)}T} }{
 1+ e^{-\frac{\varepsilon(\lambda)}T}} \right)\,d\lambda \; ,
 \end{equation}
and the constant in $x$ amplitudes $B[u_i]$ are functionals of
$u_i(\lambda)$. Their explicit but rather cumbersome form is given
in section~\ref{ABoCF}. It follows from \eqref{8-cor-fun} that the correlation lengths
$p_i$ at $\alpha=0$ drive the long-distance exponential behavior of the two-point function.
Our numerical computations (\textit{cf} appendix
\ref{Appendix Numerics}) together with the low-temperature
expansions \cite{KozMS10c} confirm that
given $\alpha=0$ and any $i$, i.e. for any solution $u_i$ to
\eqref{inteq-u-main1}, $\Re(p_i)\geq 0$ and even $p_i=0$ if
$\hat{\cal C}_i=\mathbb{R}$ and $\Re(p_i) > 0$ for all other
contours $\hat{\cal C}_i \not= \mathbb{R}$.

The sum in \eqref{fin-answ} is taken with respect to all possible solutions
$u_i(\lambda)$, or what is equivalent, with respect to all different contours
$\hat{\cal C}_i$ including $\hat{\cal C}_i=\mathbb{R}$. In the last case, the
function $u_i(\lambda)$ is equal to the excitation energy
$\varepsilon(\lambda)$ with the chemical potential shifted by $2\pi i\alpha T$.
It is precisely this term that gives the leading contribution to the asymptotic behavior of the
density-density correlation function (see section~\ref{ABoCF}).

In \eqref{fin-answ}, we did not indicate the upper limit of the summation as
well as the possible corrections. We postpone the discussion of these topics
to section~\ref{ABoCF}. Here, we would like to draw the reader's attention to
the obvious analogy of our results for the correlation lengths with the ones obtained by the QTM approach
for the temperature correlation functions of spin chains \cite{Klu04}.
 We stress however that our method is completely different and that it is applied directly to
the QNLS model without any limiting procedure from the $XXZ$ Heisenberg chain like the one in \cite{SeeBGK07}.

%%%%%%%%%%%%%%%%%%%%%%%%%%%%%%%%%%%%%%%%%%%%%%%%%%%%%%%%%%%%%%%%%%%%%%%%%%%%%%%%%%%%%%%%%%%%%%%%%%%%%%%%%%%%%%%%%%%%%%%%%%%%%%%%%%%%%%%%%%%%%%%%%%%%
%%%%%%%%%%%%%%%%%%%%%%%%%%%%%%%%%%%%%%%%%%%%%%%%%%%%%%%%%%%%%%%%%%%%%%%%%%%%%%%%%%%%%%%%%%%%%%%%%%%%%%%%%%%%%%%%%%%%%%%%%%%%%%%%%%%%%%%%%%%%%%%%%%%%

\section{Multiple integral representation}
\label{Section MultIntRep}

The master equation \cite{KitMST05me,KitKMST07} is a single multidimensional
contour integral representation valid for a wide class of integrable
models. It allows one to obtain various types of series of multiple integral
representations for certain correlation functions. In the case of
the one-dimensional Bose gas at finite temperature, a series of multiple
integral representation was obtained in \cite{SeeBGK07} through a scaling limit from the one of the XXZ chain. Here, we
work with another one that we obtain from \cite{KitKMST07} by literarily repeating
the steps given in \cite{KitKMST09a} for the case of the $XXZ$
Heisenberg spin chain.

Prior to taking the thermodynamic limit, one can describe the eigenstates
$|\Omega\rangle$ of the Hamiltonian \eqref{8-Ham} by sets of real parameters
$\Lambda=\{\lambda_1,\dots,\lambda_N\}$, $N=0,1,\dots$, i.e.
$|\Omega\rangle=|\Omega(\Lambda)\rangle$. These parameters $\lambda_j$ are
solutions to the Bethe equations \cite{FadLH96,BogIK93L}, and different
sets $\Lambda$ yield different eigenstates $|\Omega(\Lambda)\rangle$.

We remind that the master equation refers to a multidimensional contour integral representation for the
normalized expectation value of the $e^{2\pi i\alpha {\cal Q}_x}$ operator with respect to {\sl any} eigenstate $|\Omega(\Lambda)\rangle$
of the Hamiltonian. Moreover, the expansion of this multidimensional contour integral
representation into the aforementioned series of multiple integrals does not depend on the specific
choice of the state $|\Omega(\Lambda)\rangle$. Hence, even when computing the thermal averages, one can still apply
the results of \cite{KitKMST09a}. Therefore,
 \begin{multline}\label{3-prefin-answ}
 \frac{  \langle\Omega(\Lambda)|e^{2\pi i\alpha {\cal Q}_x}|\Omega(\Lambda)\rangle } { {\langle\Omega(\Lambda)|\Omega(\Lambda)\rangle} }=
 \frac{1}
       {\det_N \Theta}\sum_{n=0}^N\frac{1}{n!}
  \oint\limits_{\Gamma(\Lambda)}\prod_{j=1}^n \frac{d z_j}{2\pi i}
 \sum_{\mu_1,\dots,\mu_n\in\Lambda}
 {\cal F}_n\biggl(\begin{array}{c} \{\mu\}\\ \{z\}\end{array}\biggr)
    \\
 \times \det_n \left(\frac1{z_k-\mu_j}\right)
 \prod_{j=1}^n \left(\frac1{2\pi iL\tilde\rho(\mu_j)}\frac{e^{ix(\mu_j-z_j)}}{\mu_j-z_j}\right),
 \end{multline}
where
 \begin{equation}\label{3-trho}
 2\pi
 L\tilde\rho(\lambda_j)=L+\sum_{a=1}^NK(\lambda_{j}-\lambda_a),\qquad
 \Theta_{jk}=\delta_{jk} -\frac{K(\lambda_j-\lambda_k)}{2\pi
 L\tilde\rho(\lambda_k)}.
 \end{equation}
The functions ${\cal F}_n$ appearing in \eqref{3-prefin-answ} are
symmetric functions in the $n$ variables $\{z\}$ and in the $n$
variables $\{\lambda\}$. Their detailed description will be given
later on. What only matters for the moment, is that these functions are
holomorphic with respect to every variable in some open strip of
fixed width $a$ around the real axis. As for the discrete
summations, each parameter $\mu_k$ runs independently through the
entire set $\Lambda$. The integrals over $z_j$ are taken with
respect to a counterclockwise oriented, bounded and closed contour
$\Gamma(\Lambda)$ surrounding the set $\Lambda$ in such a way that
the only singularities of the integrand within the contour are the
poles at $z_j \in \Lambda$. In particular, the contour
$\Gamma(\Lambda)$ lies inside of the strip of width $a$.

In order to obtain  the thermodynamic limit $N,L\to\infty$,
$N/L=D=const$ of the expectation value of the $e^{2\pi i\alpha {\cal
Q}_x}$ operator, one starts by choosing a particular set $\Lambda$
and then introduces the  density of particles defined by
$\rho_p^{-1}(\lambda_j)=\lim_{N,L\to\infty}L(\lambda_{j+1}-\lambda_j)$.
In the QNLS model, it can be shown that for sets $\Lambda$ having a
physical interpretation (the ground state, excited states of finite
energy above the ground state, states of thermal equilibrium)
this limit always exists. The density of particles allows one to
replace the discrete sums over $\mu_k \in \Lambda$ by integrals,
 \begin{equation}\label{rule}
 \lim_{N,L\to\infty}\frac1L\sum_{\mu\in\Lambda}f(\mu)=\int\limits_{\mathbb{R}}
 f(\lambda)\rho_p(\lambda)\,d\lambda  \; .
 \end{equation}
Above, we have made the assumption that $f(\lambda) \rho_p(\lambda)$
is integrable on $\mathbb{R}$.

We now consider the case where the state $|\Omega(\Lambda)\rangle$ corresponds
to any finite $N,L$ representative of the state of thermal equilibrium
$|\Omega\rangle_T$. Then, due to \eqref{inteq-rho}, it follows that the
thermodynamic limit of the function $\tilde\rho(\lambda)$ coincides with the
total density $\rho_t(\lambda)$. Respectively, the determinant of the matrix
$\Theta_{jk}$ goes to the Fredholm determinant
 \begin{equation}\label{lim-detTh}
 \lim_{N,L\to\infty}\det_N\Theta_{jk}=\det\bigl[I-{\textstyle\frac1{2\pi}}K^{(\varepsilon)}\bigr],
 \end{equation}
where we used \eqref{dens-FW} and have introduced the kernel
$K^{(\varepsilon)}(\lambda,\mu)=\vartheta(\mu)K(\lambda-\mu)$. Finally, using
\eqref{rule}  we can replace the discrete sums over the set $\Lambda$ by
integrals in every $n^{\text{th}}$ term of \eqref{3-prefin-answ}. Then
agreeing upon
 \begin{equation}\label{5-DoubleangleT}
 \Mmoy{e^{2\pi i\alpha {\cal Q}_x}}
 =\moy{e^{2\pi i\alpha {\cal Q}_x}}\cdot
       \det\left[I-{\textstyle\frac1{2\pi}}K^{(\varepsilon)}\right]\,,
 \end{equation}
we obtain
 \begin{equation}\label{3-fin-answ-TD}
 \Mmoy{e^{2\pi i\alpha {\cal Q}_x}}=\sum_{n=0}^\infty\frac{1}{n!}
 \int\limits_{\mathbb{R}}\frac{d^n \lambda}{ (2\pi i)^n }
  \oint\limits_{\Gamma(\{\lambda\})} \frac{d^n z}{ (2\pi i)^n}
 \cdot
 %\prod_{j=1}^n
% \left\{
 \prod_{j=1}^n \left( \frac{\vartheta(\lambda_j)e^{ix(\lambda_j-z_j)}}{\lambda_j-z_j}\right)\cdot
 {\cal F}_n\biggl(\begin{array}{c} \{\lambda\}\\ \{z\}\end{array}\biggr)
  \cdot\det_n\left(\frac1{z_k-\lambda_j}\right),
 \end{equation}
where the contour $\Gamma(\{\lambda\})$ surrounds counterclockwise the
 variables $\lambda_1,\dots, \lambda_n $  avoiding any other
singularity of the integrand. Observe that apart from the replacement of the
discrete sums by integrals we also have replaced the finite sum over $n$ in
\eqref{3-prefin-answ} by the infinite series in \eqref{3-fin-answ-TD}. The
question of convergence of this series will be discussed later, after the
description of the functions ${\cal F}_n$ entering the representation
\eqref{3-fin-answ-TD} for $\Mmoy{e^{2\pi i\alpha {\cal Q}_x}}$ will be given.

These functions depend on $2n$ variables $\lambda_1,\dots,\lambda_n$ and
$z_1,\dots,z_n$ and read,
 \begin{equation}\label{6-F-VW}
 {\cal F}_n\biggl(\begin{array}{c} \{\lambda\}\\ \{z\}\end{array}\biggr)
 =W_n\biggl(\begin{array}{c} \{\lambda\}\\ \{z\}\end{array}\biggr) \cdot
 \prod_{j=1}^n {\cal V}_n\biggl(\lambda_j \mid\begin{matrix} \{\lambda\} \\  \{z\} \end{matrix} \biggr).
 \end{equation}
The explicit formulae for the functions $W_n$ and ${\cal V}_n$
involve the set  of  auxiliary functions:
 \begin{equation}\label{1-Vpm}
 V_{\sigma;n}(\mu)=\prod_{a=1}^n\frac{\mu-\lambda_a+ic\sigma }{\mu-z_a+ ic\sigma },
 \qquad \sigma=0,\pm,\qquad\mbox{and}\qquad
 K_\alpha(\lambda)=\frac1{\lambda+ic}-\frac{e^{2\pi i\alpha}}{\lambda-ic} \; .
 \end{equation}
Namely,
\begin{equation}\label{5-mcV}
 {\cal V}_n\biggl(\mu \mid\begin{matrix} \{\lambda\} \\  \{z\} \end{matrix} \biggr)
 ={e^{2\pi i\alpha}}\,\frac{V_{+;n}(\mu)}{V_{-;n}(\mu)}-1 \; .
\end{equation}
The expression for $W_n$ is more involved. It can be represented as
 \begin{equation}\label{2-W}
  W_n\biggl(\begin{array}{c} \{\lambda\}\\ \{z\}\end{array}\biggr)
 =\widetilde W[V_{\sigma;n}]\cdot \prod_{k=1}^{n}\frac{V_{-;n}( z_k)}{V_{-;n}( \lambda_k)}
  \; ,
 \end{equation}
where
 \begin{equation}\label{2-W1}
   \widetilde W[V_{\sigma;n}]=\frac{({e^{2\pi i\alpha}}-1)^2
 \det[I+\frac1{2\pi i}U^{(1)}]
 \det[I+\frac1{2\pi i}U^{(2)}] }
 {\bigl( V_{+;n}^{-1}(\theta_1)-{e^{2\pi i\alpha}}  V_{-;n}^{-1}(\theta_1) \bigr)\bigl( V_{-;n}(\theta_2)-{e^{2\pi i\alpha}}
 V_{+;n}(\theta_2) \bigr) }.
 \end{equation}
Above appear two Fredholm determinants of integral operators whose kernels are
 \begin{equation}\label{2-U1}
 U^{(1)}(w,w')=- V_{0;n}^{-1}(w)\cdot
 \frac{K_\alpha(w-w')-K_\alpha(\theta_1-w')}
                            {V_{+;n}^{-1}(w)-{e^{2\pi i\alpha}} V_{-;n}^{-1}(w)},
 \end{equation}
and
 \begin{equation}\label{2-U2}
 U^{(2)}(w,w')=V_{0;n}(w')\cdot
 \frac{K_\alpha(w-w')-K_\alpha(w-\theta_2)}{V_{-;n}(w')-{e^{2\pi i\alpha}} V_{+;n}(w')}.
 \end{equation}
These operators act on a counterclockwise oriented contour $\Gamma(\mathbb{R})$
surrounding the real axis. By definition, the only singularities of $U^{(k)}(w,w^{\prime})$ inside of
$\Gamma(\mathbb{R})$ are the zeros of $V_{0;n}(w)$ (resp. the poles of
$V_{0;n}(w')$). In the following, when we will slightly deform the integration
contour in \eqref{3-fin-answ-TD}, we will always keep this prescription for
$\Gamma(\mathbb{R})$, making it larger, if necessary. The parameters $\theta_1$
and $\theta_2$ in \eqref{2-W1}--\eqref{2-U2} are arbitrary complex numbers
lying inside of the contour $\Gamma(\mathbb{R})$. It was proved in \cite{KitKMST09a}
that the overall combination appearing in the r.h.s. of \eqref{2-W1} does not
depend on a specific choice of these parameters.

Due to the rather complicated form of the functions ${\cal F}_n$, we
are unable to provide a proof of the convergence of the series
\eqref{3-fin-answ-TD} in the case of general $c$. In fact, the
situation is here quite analogous to the one occurring in the zero
temperature case \cite{KitKMST09a}. One can easily prove the
convergence in the special case corresponding to the free fermion
point $c=\infty$. Indeed,  then ${\cal F}_n=\left(e^{2\pi
i\alpha}-1\right)^n$ and the integrals over $z_j$ can be taken
explicitly. We obtain
 \begin{equation}\label{4-FD-exp}
 \left.\moy{e^{2\pi i\alpha{\cal Q}_x}}\right|_{c=\infty}
 =\sum_{n=0}^\infty\frac{(e^{2\pi i\alpha}-1)^n}{n!}
 \int\limits_{\mathbb{R}}\det_n\left[\frac{\sin\frac{x}2(\lambda_j-\lambda_k)}
 {\pi(\lambda_j-\lambda_k)} \vartheta(\lambda_k)\right]\prod_{j=1}^n \,d\lambda_j .
 \end{equation}% \end{multline}
The series \eqref{4-FD-exp} is an expansion of the Fredholm determinant of the
integral operator $I+V_{0}$ with
 \begin{equation}\label{4-kernel}
 V_{0}(\lambda,\mu)
 =(e^{2\pi i\alpha}-1)
 \frac{\sin\frac{x}2(\lambda-\mu)}
 {\pi(\lambda-\mu)}\vartheta(\mu)\, .
 \end{equation}
The general theory of Fredholm determinants ensures that the series
\eqref{4-FD-exp} is absolutely convergent and  defines an entire function of
$e^{2\pi i\alpha}$. It particular, the expansion coefficients decay faster
than exponentially.

In the case  $c<\infty$, the series \eqref{3-fin-answ-TD} cannot be reduced to
a simple form as in \eqref{4-FD-exp}. Nevertheless, taking into account that
the QNLS model with a general coupling constant can be considered as a smooth
deformation of the free fermions case, we shall assume in the following that
the series \eqref{3-fin-answ-TD} is absolutely convergent.

In order to study the series \eqref{3-fin-answ-TD} in the next sections, let us make two important remarks concerning the functions $W_n$, ${\cal V}_n$ and  ${\cal F}_n$ present in \eqref{3-fin-answ-TD}. Although their precise expressions are rather cumbersome, it is easy to show that $W_n$ and ${\cal V}_n$ (and hence ${\cal F}_n$) are
holomorphic functions within some multi-dimensional  strip  $\frak{S}$:
$|\Im(z_j)|\leq a$, $|\Im(\lambda_j)| \leq a$, $j=1,\dots,n$. Though
we can not determine explicitly the width $a$ of this strip, it is
clear that the latter is temperature independent, because the Fermi
weight $\vartheta(\lambda)$ \eqref{dens-FW} is the only function
present in \eqref{3-fin-answ-TD} that depends on $T$. In the case of
positive chemical potential, the poles of the Fermi weight
accumulate, in the $T\to 0$  limit, on certain points of
$\mathbb{R}$. Hence, there exists a crossover  temperature $T_0$ such
that, for $T<T_0$, there are always poles of the Fermi weight that
are located inside of the strip $|\Im(\lambda)|<a$. The precise
value of $T_0$ depends on the value of $a$ which, in its turn, is
fixed by the analytic properties of ${\cal F}_n$. For instance, in
the free fermion point, one has $T_0=+\infty$ as ${\cal
F}_n=\left(e^{2\pi i\alpha}-1\right)^n$. We will not study this question further.
Simply, when $c<\infty$ we shall limit ourselves to the regime
$T<T_0$. In such a case, the nearest (to the real axis) singularity of the integrand in
\eqref{3-fin-answ-TD}  always correspond to the
poles of the Fermi weight.

The other important features of the functions  $W_n$ and ${\cal V}_n$ concerns their
reduction properties,  namely,
\begin{equation}\label{8-recurs}
\left.{\cal V}_n\biggl(\omega\mid\!\begin{array}{c} \{\lambda\} \\
\{z\}
\end{array}\! \biggr) \right|_{\lambda_j=z_k } \hspace{-4mm} =
 {\cal V}_{n-1}\biggl(\omega\mid\!\begin{array}{c} \{\lambda\}\setminus \lambda_j
              \\ \{z\}\setminus z_k   \end{array}\! \biggr),
\quad \left. W_n\biggl(\!\begin{array}{c} \{\lambda\} \\ \{z\}
\end{array} \!\biggr) \right|_{\lambda_j=z_k } \hspace{-4mm} =
 W_{n-1}\biggl(\!\begin{array}{c} \{\lambda\}\setminus \lambda_j
             \\ \{z\}\setminus z_k   \end{array} \!\biggr).
\end{equation}
In the following, we will see that these reduction properties
\eqref{8-recurs} are crucial for the re-summation the series \eqref{3-fin-answ-TD} in
the asymptotic regime  $x\to\infty$.

%%%%%%%%%%%%%%%%%%%%%%%%%%%%%%%%%%%%%%%%%%%%%%%%%%%%%%%%%%%%%%%%%%%%%%%%%%%%%%%%%%%%%%%%%%%%%%%%%%%%%%%%%%%%%%%%%%%%%%%%%%%%%%%%%%%%%%%%%%%%%%%%%%%%
%%%%%%%%%%%%%%%%%%%%%%%%%%%%%%%%%%%%%%%%%%%%%%%%%%%%%%%%%%%%%%%%%%%%%%%%%%%%%%%%%%%%%%%%%%%%%%%%%%%%%%%%%%%%%%%%%%%%%%%%%%%%%%%%%%%%%%%%%%%%%%%%%%%%

\section{Asymptotic behavior of multiple integrals\label{AppCI}}
\label{Sec-aAMI}

The asymptotic analysis of the series \eqref{3-fin-answ-TD} can be done
along the lines of the zero-temperature case \cite{KitKMST09a}.
The first step of that method consists in extracting the large-$x$
asymptotic behavior of each $2n$-fold integral appearing in the series
\eqref{3-fin-answ-TD}. The main difference between the $T=0$ and $T>0$ situations is
that, in the first case, the corrections to the leading terms have a power-law
behavior in $x$, whereas, at $T>0$, they are exponentially small in $x$.
Apart from this difference, the general strategy of the asymptotic analysis is the same with however several interesting technical simplifications in the case $T>0$. Below, we
recall the general framework of our method.

The $2n$-fold multiple integrals of interest have the form
 \begin{equation}\label{def-CI}
 {\cal I}_n[{\cal F}_n]=   \int\limits_{\mathbb{R}} \frac{d^n \lambda}{(2\pi i)^{n}}\; \oint\limits_{\Gamma(\{\lambda\})}
\hspace{-2mm} \frac{d^n z}{(2\pi i)^{n}} \;
 \prod_{j=1}^n \left( \frac{\vartheta(\lambda_j)e^{ix(\lambda_j-z_j)}}{\lambda_j-z_j} \right)
 \cdot\det_n \left(\frac1{z_j-\lambda_{k}} \right)  \cdot
  {\cal F}_n\biggl(\begin{array}{c} \{\lambda\} \\ \{z\} \end{array} \biggr),
 \end{equation}
where ${\cal F}_n$ are holomorphic functions in some
 multi-dimensional strip $\frak{S}$: $|\Im(z_j)|\le
a$, $|\Im(\lambda_j)| \le a$, $j=1,\dots,n$, $a$ being $n$-independent.
%${\cal F}_n$ is also
%symmetric in respect to the integration
%variables $\lambda_1,\ldots,\lambda_n$ and $z_1,\ldots,z_n$ taken separately
%and satisfies the reduction properties \eqref{8-recurs}.
The nearest to the real axis singularities of the integrand in
\eqref{def-CI} are the poles of the Fermi weight
$\vartheta(\lambda)$.

\subsection{The Fredholm determinant as a multiple integrals generating function}

In this subsection, we briefly recall the connection between multiple integrals of
the type \eqref{def-CI} and  Fredholm determinants of integral operators
\cite{KitKMST09a}.

Let us assume for a moment that the functions ${\cal F}_n$ take the form of a pure product as
 \begin{equation}\label{CI-factor-0}
 {\cal F}_{n}\biggl(\begin{array}{c} \{\lambda\} \\ \{z\} \end{array} \biggr)
 =\prod_{p=1}^n\varphi(\lambda_p) e^{-g(z_p)} \; ,
 \end{equation}
where $\varphi(\lambda)$ and $e^{-g(\lambda) }$ are  holomorphic in
the strip $|\Im(\lambda)|\le a$. Then in \eqref{def-CI}, the integrals over the $z_j$
separate and can be computed by taking the residues at
$z_j=\lambda_j$ and $z_j=\lambda_{k}$. A simple calculation leads to
 \begin{equation}\label{CI-fact}
 {\cal I}_n[{\cal F}_n]=\int\limits_{\mathbb{R}}\det_n
 [V(\lambda_j,\lambda_k)]
 \,d^n\lambda = \left. \partial_{\gamma}^n \det_{\mathbb{R}}[I+ \gamma V]\right|_{\gamma=0} \; ,
 %=\left. \partial_{\gamma}^s \det(I+ \gamma V)\right|_{\gamma=0}  ,
 \end{equation}
where
 \begin{equation}\label{Vjk-kern}
 V(\lambda,\mu)=\frac{ \vartheta(\lambda) F(\lambda)}
 {2\pi i(\lambda-\mu)}\left(e^{\frac{ix}2(\lambda-\mu)+\frac12(g(\lambda)-g(\mu))}
 - e^{-\frac{ix}2(\lambda-\mu)-\frac12(g(\lambda)-g(\mu))}\right),
 \end{equation}
and
 \begin{equation}\label{Fg}
 F(\lambda)=\varphi(\lambda)e^{-g(\lambda)} \; . %\qquad e^{-g(\lambda)}=\phi(\lambda).
 \end{equation}
Thus, in \eqref{CI-fact}, we have identified the multiple integral \eqref{def-CI} as the  $n^{\text{th}}$ $\gamma$-derivative of the Fredholm
determinant of the operator $I+\gamma V$ acting on $\mathbb{R}$ with the
integral kernel \eqref{Vjk-kern}.

The large-$x$ asymptotic behavior  of the Fredholm determinant of
the operator \eqref{Vjk-kern}, up to $O(e^{-ax})$ corrections, has
been established in \cite{Sla10a}. This asymptotic expansion as well as any of its finite order $\gamma$-derivatives is uniform in  $\gamma$
provided  $\gamma$ belongs to a sufficiently small neighborhood
of the origin. By taking the $n$-th $\gamma$-derivative of this asymptotic
expansion  at $\gamma=0$, we obtain the asymptotic
behavior of \eqref{CI-fact}.

In the case of interest, ${\cal F}_n$ cannot be represented as
pure product functions.  However, one can proceed as in \cite{KitKMST09a}
and apply the {\it density procedure} (see  \cite{KitKMST09b} for
more details), that is, to decompose ${\cal F}_n$ on the class of
pure product functions :
 \begin{equation}\label{CI-factor}
 {\cal F}_{n}\biggl(\begin{array}{c} \{\lambda\} \\ \{z\} \end{array} \biggr)
 =\sum_{k=1}^\infty \; \prod_{p=1}^n\varphi_k(\lambda_p)e^{-g_k(z_p)} ,
 \end{equation}
where the functions $\varphi_k(\lambda)$ and $e^{-g_k(\lambda)}$ are
holomorphic in the strip $|\Im(\lambda)|\le a$.  We thus define
 \begin{equation}\label{pp-gF-1}
 F_k(\lambda)=\varphi_k(\lambda)e^{-g_k(\lambda)}
 \end{equation}
and denote $V_k(\lambda,\mu)$ the kernel obtained from
$V(\lambda,\mu)$ by the replacement $F \hookrightarrow F_k$ and $g
\hookrightarrow g_k$. Then
 \begin{equation}\label{5-Is-Fredholm}
 {\cal I}_{n}[{\cal F}_n]=\sum_{k=1}^{\infty}
\left.\frac{\partial^n }{\partial{\gamma}^n}\det[I+
 \gamma V_k]\right|_{\gamma=0}  .
 \end{equation}
We stress that in \eqref{5-Is-Fredholm} there are no problems with
permuting the symbols of integration along $\mathbb{R}$ (contained in the Fredholm determinant) and
summation over $k$ induced by the density procedure as the
convergence in \eqref{CI-factor} holds in the supremum norm on the strip
$\{|\Im(z)|\leq a\}$ and the weights $\vartheta(\lambda)$ ensure the
convergence of the integrals\footnote[1]{%
We would like to point out that  one could avoid the manipulation of
infinite sums related to the density procedure as follows. Namely,
one first considers the case of functions $\mathcal{F}_n$ that are
represented as \textit{finite} linear combinations of pure product
functions and proceeds through all the steps in the re-summation
below. At the end, once that the function one started with has been
reconstructed, it is enough to observe that the answer is expressed
in terms of a linear continuous functional defined on a dense
subspace of a Banach space. Therefore, the obtained asymptotic
expansion can be extended, by continuity, to more general classes of
functions $\mathcal{F}_n$. As this does not alter the final
conclusions, we chose not to insist on that point later on.}.

\subsection{Large-$x$ asymptotic behavior of the Fredholm
determinant\label{S-LxABFD}}

We have seen that the problem of the calculation of the asymptotic
behavior of the multiple integrals \eqref{def-CI} boils down to the
asymptotic analysis of the Fredholm determinant of the integral
operator with the kernel \eqref{Vjk-kern}. In this section, we
present the results of this analysis \cite{Sla10a}. Recall that the
Fermi weight $\vartheta(\lambda)$  \eqref{dens-FW} appearing in
\eqref{Vjk-kern} is a meromorphic function in the strip $|\Im(\lambda)| \leq
a$ with simple poles at $\lambda=\Rpm_j$, where $j=1,\dots,M$. The
superscripts $\pm$ indicate that the corresponding pole lies in the
upper (resp. lower) half-plane. The total number of poles $2M$ is
not essential for our analysis, however we have assumed that $M>0$.
It is also important that the Fermi weight decays as a Gaussian when
$\lambda\to\pm\infty$.

Concerning the functions $g(\lambda)$ and $F(\lambda)$ entering the kernel
\eqref{Vjk-kern}, in addition to their analyticity in the strip
$|\Im(\lambda)|\le a$ we assume that $g^{\prime}(\lambda)\vartheta(\lambda)F(\lambda)$ and
$\vartheta(\lambda)F(\lambda)$ are integrable along any curve avoiding the poles of $\vartheta$ in this strip.  This, in particular,
means that $|\tr V|<\infty$.

Now we give several definitions and introduce  new notations
necessary for the formulation of the theorem about the large-$x$
asymptotic behavior of the Fredholm determinant $\det[I+\gamma V]$.

\begin{figure}[h]
\begin{center}
\includegraphics[width=15cm]{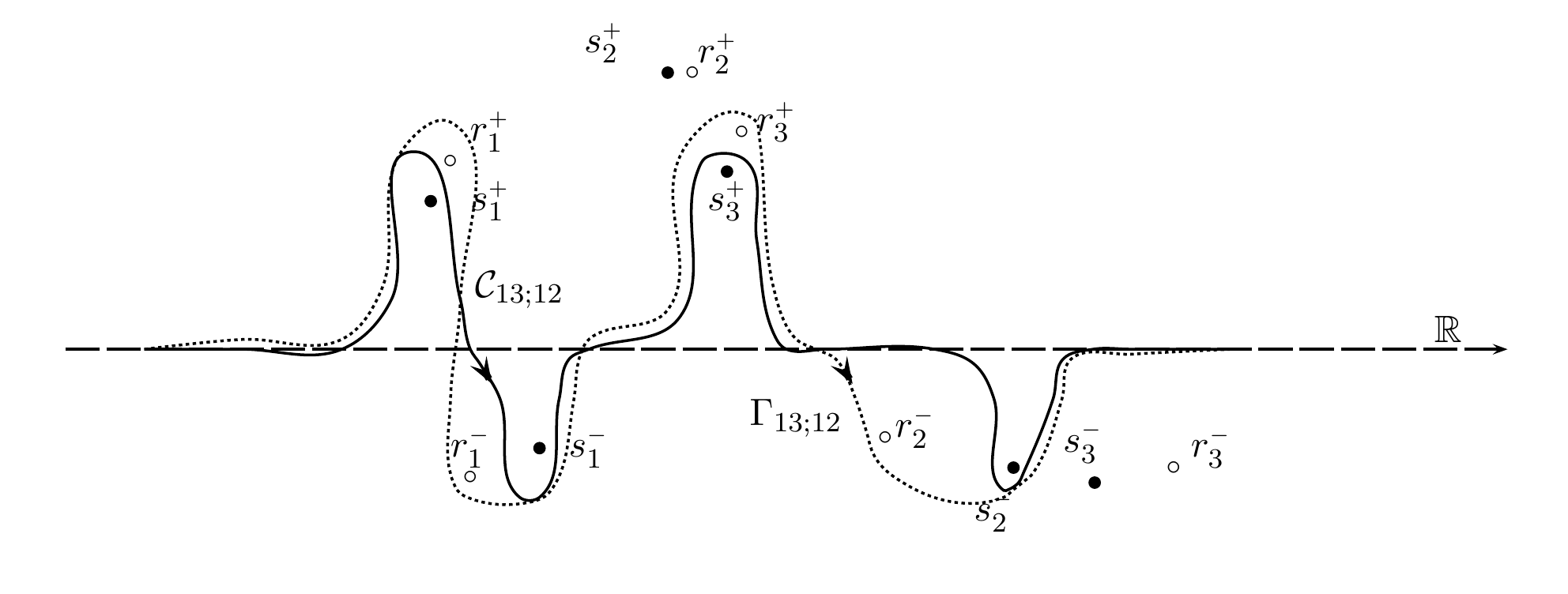}
\caption{\label{G1212}{\small The poles of the Fermi weight $\Rpm_j$
are depicted by $\scriptstyle\circ$, the zeros $\qpm_j$ of $1+\gamma
\vartheta(\lambda) F(\lambda)=0$ by $\scriptstyle\bullet$.   The
integration contour $\Gamma_{13;12}$ (dotted line) bypasses the
points $\qp_1$, $\qp_3$, $\Rp_1$, $\Rp_3$ from above and $\qm_1$,
$\qm_2$, $\Rm_1$, $\Rm_2$ from below.  The contour ${\cal
C}_{13;12}$ (solid line) separates  the points $\qp_1$, $\qp_3$ from
$\Rp_1$, $\Rp_3$  and $\qm_1$, $\qm_2$ from $\Rm_1$, $\Rm_2$.} }
\end{center}
\end{figure}

\vspace{2mm}
In addition to the points $\Rpm_j$, we define the  points
$\qpm_j$ such that $1+\gamma \vartheta(\qpm_j)F(\qpm_j)=0$. For
$\gamma$ small enough, these points are slightly shifted in respect
to the poles $\Rpm_j$ (see Fig.~\ref{G1212}), but their overall
number (equal to $2M$) is the same in virtue of Rouch\'e theorem.

%%%%%%%%%%%%%%%%%%%%%%%%%%%%%%%%%%%%%%%%%%%%%%%%%%%%%%%%%%%%%%%
% HERE IS RISUNOK-1
%%%%%%%%%%%%%%%%%%%%%%%%%%%%%%%%%%%%%%%%%%%%%%%%%%%%%%%%%%%%%%
%
%

We also introduce the sets of contours
$\Gamma_{\boldsymbol{j};\boldsymbol{k}}$ and ${\cal
C}_{\boldsymbol{j};\boldsymbol{k}}$. Here $\boldsymbol{j}$ and
$\boldsymbol{k}$ are multi-indices:
$\boldsymbol{j}=\{j_1,\dots,j_p\}$  and
$\boldsymbol{k}=\{k_1,\dots,k_p\}$ with $1\le j_s,\;k_s\le M$. Also
$\# \boldsymbol{j}=\# \boldsymbol{k}=p$, with $p=0,1,\dots,M$. In the
following, we often denote the cardinality of the sets
$\boldsymbol{j}$ and $ \boldsymbol{k}$ by $|\boldsymbol{j}|$, $|\boldsymbol{k}|$, here with $|\boldsymbol{j}| = |\boldsymbol{k}|$.

The contour $\Gamma_{\boldsymbol{j};\boldsymbol{k}}$ is a
deformation of the real axis such that, when moving from  $\mathbb{R}$ to
$\Gamma_{\boldsymbol{j};\boldsymbol{k}}$, one only crosses the roots
$\qp_{\boldsymbol{j}}$, $\qm_{\boldsymbol{k}}$ and the associated poles
$\Rp_{\boldsymbol{j}}$, $\Rm_{\boldsymbol{k}}$, while all the other
roots $\qpm_\ell$ and poles $\Rpm_\ell$, $\ell \not= j_a$ or $k_a$, are not crossed (see
Fig.~\ref{G1212}). Note that here and below we agree upon the notation
\begin{equation}
\qp_{\boldsymbol{j}}\equiv \{\qp_{j_a} \}   \quad , \quad \qm_{\boldsymbol{k}}\equiv \{\qp_{k_a}\}  \quad , \quad
\Rp_{\boldsymbol{j}}\equiv \{\Rp_{j_a}\} \quad , \quad  \Rm_{\boldsymbol{k}}\equiv  \{\Rm_{k_a}\} ,
\qquad \text{with} \quad a=1, \dots, |\boldsymbol{j}|  \;.
\end{equation}
Similarly, the contours ${\cal C}_{\boldsymbol{j};\boldsymbol{k}}$ are such that when moving
from $\mathbb{R}$ to ${\cal C}_{\boldsymbol{j};\boldsymbol{k}}$, one only crosses the roots
$\qp_{\boldsymbol{j}}, \qm_{\boldsymbol{k}}$; \textit{all} poles and other roots $\qpm_{\ell}$ are not crossed.
In particular,  ${\cal C}_{\boldsymbol{j};\boldsymbol{k}}$ separates the roots
$\qp_{\boldsymbol{j}},\qm_{\boldsymbol{k}}$ from  their associated poles
$\Rp_{\boldsymbol{j}},\Rm_{\boldsymbol{k}}$ (see Fig.~\ref{G1212}).
We stress that ${\cal
C}_{\emptyset,\emptyset}=\Gamma_{\emptyset,\emptyset}=\mathbb{R}$.

Finally we define an  auxiliary function $\nu(\lambda)$ as
 \begin{equation}\label{nu}
 \nu(\lambda)=\frac{-1}{2\pi i}\log \bigl(1+\gamma \vartheta(\lambda)F(\lambda)\bigr),
 \end{equation}
and a contour dependent functional

 \begin{equation}\label{A-rep}
 {\cal A}_{\cal L}([g],[\nu])=-\int\limits_{\cal L}
 \bigl(ix+g'(\lambda)\bigr)\nu(\lambda)\,d\lambda +\iint\limits_{\cal L}
 \frac{\nu(\lambda)\nu(\mu)}{(\lambda-\mu_+)^2}\,d\lambda\,d\mu \; ,
 \end{equation}
where the oriented contour ${\cal L}$ is either equal to ${\cal
C}_{\boldsymbol{j};\boldsymbol{k}}$ or $\Gamma_{\boldsymbol{j};\boldsymbol{k}}$, for some $\boldsymbol{j}$, $\boldsymbol{k}$. In \eqref{A-rep},
we have stressed that the second integral is double. The symbol $\mu_+$ means that $\mu$ is
slightly shifted to the left from the oriented integration contour ${\cal L}$.

\begin{thm}\label{Main-thm-FD1}\cite{Sla10a} Let $|\gamma|$ be small enough. Then in the  $x\to\infty$ limit
and under the above assumptions, the Fredholm determinant of the
operator $I+\gamma V$, with V given by \eqref{Vjk-kern}, admits the
asymptotic expansion
 \begin{equation}\label{logdet0}
 \det[I+\gamma V]=\sum_{ \{\boldsymbol{j}; \boldsymbol{k} \}}
 \exp\Bigl({\cal A}_{{\cal C}_{\boldsymbol{j};\boldsymbol{k}}}([g],[\nu])\Bigr) \cdot \left[1+ O\left(e^{-ax}\right)
 \right]
 \;,
 \end{equation}
where the  sum is taken with respect to all the possible choices of
multi-indices $\boldsymbol{j};\boldsymbol{k}$, including
$\boldsymbol{j}=\emptyset$ and $\boldsymbol{k}=\emptyset$.
\end{thm}

{\sl Remark 1.}\label{Rem-AsyExp0} In the following, we will use the above formula \eqref{logdet0} to derive the asymptotic behavior of  ${\cal I}_{n}[{\cal F}_n]$ from the relation   \eqref{5-Is-Fredholm}. To do this we need to compute the $n^{th}$ $\gamma$-derivative of the above  Fredholm determinant's asymptotic behavior \eqref{logdet0}. In this process one can produce polynomial (in the distance $x$) contributions to the remainders of each term in  \eqref{logdet0}. As a consequence the remainders will no longer be $O\left(e^{-ax}\right)$. However, they will still be a $o(e^{-a'x})$ for arbitrary $a' < a$. Hence in the following the remainders in all asymptotic formulae will be given up to $o(e^{-a'x})$ terms for arbitrary $a' < a$.

{\sl Remark 2.}\label{Rem-AsyExp1} Moving all the contours ${\cal
C}_{\boldsymbol{j};\boldsymbol{k}}$ in \eqref{logdet0} to the real
axis, one can recast the asymptotic expansion of the Fredholm determinant
in the following form
 \begin{equation}\label{logdet-al} 
 \det[I+\gamma V]=\exp\Bigl({\cal A}_{\mathbb{R}}([g],[\nu])\Bigr)
 \sum_{ \{\boldsymbol{j}; \boldsymbol{k} \}} A_{\boldsymbol{j}; \boldsymbol{k}}
 \prod_{b=1}^{|\boldsymbol{j}|}  \{  e^{ix(\qp_{j_b}-\qm_{k_b})}  \}
 \cdot \left[1+ o\left(e^{-a'x}\right) \right],
 \end{equation}
where $A_{\boldsymbol{j}; \boldsymbol{k}}$ are some constant
coefficients (see \cite{Sla10a} for their explicit form). For
$\gamma$ small enough, all the roots $\qp_{\boldsymbol{j}}$ and
$\qm_{\boldsymbol{k}}$ belong to the strip $|\Im(\lambda)|<a$.
However, certain sums
$\sum_{b=1}^{|\boldsymbol{j}|}(\qp_{j_b}-\qm_{k_b})$ may have their
imaginary part greater then $a$. In such a case, the corresponding
terms in the expansion \eqref{logdet-al} (and hence, in
\eqref{logdet0}) can be sub-leading in respect to the correction
$O(e^{-ax})$. Should such an effect occur, we will
simply drop the corresponding contributions.

{\sl Remark 3.}\label{Rem-AsyExp2} In the case of interest the
parameters $a$ (and $a'$) may depend on the coupling constant $c$ only. At the
same time, for $\gamma$ small enough, the roots $\qpm$ are close
to the poles of the Fermi weight $\Rpm$, whose positions depend not
only on the coupling constant, but also on the temperature and the
chemical potential. In particular, for $h>0$ and small $T$ all these
poles (and thus, the roots $\qpm$) collapse on the real axis. Hence,
for arbitrary choice of multi-indexes $\boldsymbol{j}$ and
$\boldsymbol{k}$ one can always make
$\Im(\sum_{b=1}^{|\boldsymbol{j}|}(\qp_{j_b}-\qm_{k_b}))<a$
provided the temperature is small enough.

The asymptotic expansion  \eqref{logdet0} is uniform in $\gamma$ small enough. However, it becomes inconvenient for the calculation of
the $\gamma$-derivatives of the Fredholm determinant at $\gamma=0$. Indeed, the
contours ${\cal C}_{\boldsymbol{j};\boldsymbol{k}}$ are squeezed between the
roots $\qpm$ and the poles $\Rpm$. As $\qpm_j\to\Rpm_j$ when
$\gamma\to 0$, strictly speaking, the contours ${\cal C}_{\boldsymbol{j};\boldsymbol{k}}$ do not exist at $\gamma=0$.

A possible way to overcome this problem is to move all the contours ${\cal
C}_{\boldsymbol{j};\boldsymbol{k}}$ in \eqref{logdet0} to the real axis (see
\cite{Sla10a}). For our goal, it is however more convenient to deform each contour
${\cal C}_{\boldsymbol{j};\boldsymbol{k}}$ in \eqref{logdet0} to its associated
contour $\Gamma_{\boldsymbol{j};\boldsymbol{k}}$ that does exist in the above limit $\gamma\to 0$. In doing so, one crosses the poles
$\Rp_{\boldsymbol{j}}$ and $\Rm_{\boldsymbol{k}}$. This produces additional
contributions to the functional ${\cal A}_{{\cal
C}_{\boldsymbol{j};\boldsymbol{k}}}([g],[\nu])$ and makes the formula
\eqref{logdet0} more cumbersome.

To describe the asymptotic expansion of $\det[I+\gamma V]$ in terms
of a sum over the contours
$\Gamma_{\boldsymbol{j};\boldsymbol{k}}$ we introduce new notations.
Let $L_{\cal L}[f](\lambda)$ be the Cauchy transform over a contour
$\cal L$ of a function $f(\lambda)$:
\begin{equation}\label{def-CauT}
 L_{\cal L}[f](\lambda)
  = \int\limits_{\cal L} \frac{f(\mu) d\mu}{\mu-\lambda}.
\end{equation}
 Then the asymptotic expansion \eqref{logdet0}
can be written in the form
 \begin{equation}\label{logdet1}
 \det[I+\gamma V]=\sum_{ \{\boldsymbol{j}; \boldsymbol{k} \}}
 \mathcal{U}_{\boldsymbol{j};\boldsymbol{k}}([g],[\nu],[F])\left[1+
 o\big( e^{-a'x} \big)\right]
 \;,
 \end{equation}
Here, as before, the sum is taken over all the possible choices of
multi-indices $\boldsymbol{j};\boldsymbol{k}$. The functional
$\mathcal{U}_{\boldsymbol{j};\boldsymbol{k}}$ reads
\begin{equation}
 \mathcal{U}_{\boldsymbol{j};\boldsymbol{k}}([g],[\nu],[F])=
\exp\biggl\{-\hspace{-1mm}\int\limits_{\Gamma_{\boldsymbol{j};\boldsymbol{k}}}
\hspace{-2mm} g^{\prime}(\lambda) \nu(\lambda) d \lambda\biggr\}
\prod_{a=1}^{|\boldsymbol{j}|}%
 \left\{ \gamma^2  e^{g(\Rm_{k_a}) - g(\Rp_{j_a})  }F(\Rp_{j_a}) F(\Rm_{k_a})
\right\}
\cdot U_{\boldsymbol{j};\boldsymbol{k}}[\nu] \; ,
\label{def-calU}
\end{equation}
where
\begin{multline}
U_{\boldsymbol{j};\boldsymbol{k}}[\nu]=
\left\{\det_{n} \left[ \frac{1}{\Rp_{j_a} - \Rm_{k_b}} \right] \right\}^2
\exp\Bigl({\cal A}_{\Gamma_{\boldsymbol{j};\boldsymbol{k}}}([0],[\nu])\Bigr)\\
\prod_{a=1}^{|\boldsymbol{j}|}  \left\{
\vartheta_{reg}(\Rp_{j_a})\vartheta_{reg}(\Rm_{k_a})
 e^{ix(\Rm_{k_a}-\Rp_{j_a}) +2L_{\Gamma_{\boldsymbol{j};\boldsymbol{k}}}[\nu]
 (\Rm_{k_a})
 -2L_{\Gamma_{\boldsymbol{j};\boldsymbol{k}}}[\nu](\Rp_{j_a})}
 \right\}.
\label{def-dirU}
\end{multline}
Above, we have introduced the residues of the Fermi weight
$\vartheta(\lambda)$ at the poles $\Rpm_\ell$:
$\vartheta_{reg}(r_{\ell}^{\pm})=\Res\bigl(\vartheta(\lambda),
\lambda=r_{\ell}^{\pm} \bigr)$. The detailed derivation of
\eqref{def-calU}, \eqref{def-dirU} is presented in
appendix~\ref{PrF}.

\subsection{Asymptotic expansion of multiple integrals}

It follows from \eqref{5-Is-Fredholm} and the from of the expansion
\eqref{logdet0}, that $\mathcal{I}_n[\mathcal{F}_n]$ can be presented as
 \begin{equation}\label{sum-C}
 {\cal I}_n[{\cal F}_n]=\sum_{  \{ \boldsymbol{j} ;\boldsymbol{k} \} }{\cal J}_{ \boldsymbol{j} ;\boldsymbol{k}  }[{\cal F}_n]
 \left[1+o\big(e^{-a^{\prime} x}\big)\right],
 \end{equation}
where ${\cal J}_{ \boldsymbol{j} ;\boldsymbol{k} }$ are some
functionals that we will construct below. They can be associated
with the  contours $\mathcal{C}_{\boldsymbol{j};\boldsymbol{k}}$ (or
$\Gamma_{\boldsymbol{j};\boldsymbol{k}}$), therefore we label them
by multi-indices $\boldsymbol{j};\boldsymbol{k}$. In order to obtain
their explicit form  one has to perform the following steps:
\begin{itemize}
\item compute the $n^{\text{th}}$ $\gamma$-derivative of $\mathcal{U}_{\boldsymbol{j};\boldsymbol{k}}([g],[\nu],[F])$ at $\gamma=0$ .
\item Substitute in this final expression $F\hookrightarrow F_k$ and $g\hookrightarrow g_k$.
\item Take the sum over $k$ and use the decomposition \eqref{CI-factor} so as to express the result in terms of $\mathcal{F}_n$
or its partial derivatives.
\end{itemize}
We stress that the derivation of this action as given in \eqref{InC-G} and
\eqref{InC} is essentially a consequence of the representation \eqref{6-F-VW}
for ${\cal F}_n$ along with the reduction properties of the functions $W_n$
and  $\mathcal{V}_n$ as described in \eqref{8-recurs}.

\begin{prop}\label{prop-mulInt}
Let ${\cal F}_n$ be as  given in \eqref{6-F-VW} with ${\cal V}_n$ and $W_n$
satisfying to the reduction properties \eqref{8-recurs}. Then
%{
\begin{equation}\label{InC-G}
  {\cal J}_{\boldsymbol{j}; \boldsymbol{k}}[{\cal F}_n]= \Bigl.\partial^n_\gamma
  G_{\boldsymbol{j}; \boldsymbol{k}}(\gamma)\Bigr|_{\gamma=0} \; ,
  \end{equation}
where, for $\gamma$ small enough,
\begin{equation}\label{InC}
 %\begin{multline}\label
 G_{\boldsymbol{j}; \boldsymbol{k}}(\gamma)=\sum_{m=0}^\infty\frac{1}{m!}
 \prod_{i=1}^m\partial_{\epsilon_i} \int\limits_{\Gamma_{\boldsymbol{j}; \boldsymbol{k}} } \prod_{i=1}^m
 \hat\nu(\lambda_i)\cdot W_{m+|\boldsymbol{j}|}
 \cdot
 \left.  \mathcal{U}_{\boldsymbol{j};\boldsymbol{k}}\left([0],[\hat\nu],[\mathcal{V}_{m+|\boldsymbol{j}|}] \right) \,d^m\lambda
 \right|_{\epsilon_i=0}.
 \end{equation}%nd{multline}
Here
$\mathcal{U}_{\boldsymbol{j};\boldsymbol{k}}\left([0],[\hat\nu],[\mathcal{V}_{m+|\boldsymbol{j}|}]
\right)$ is given by \eqref{def-calU}, where one should set
$g(\lambda)=0$ and replace the original functions $\nu$ and $F$ by
$\hat\nu$ and $\mathcal{V}_{m+|\boldsymbol{j}|}$ respectively. These
new functions have the form
 \begin{equation}\label{new-V}
 \mathcal{V}_{m+|\boldsymbol{j}|}(\omega)\equiv
 {\cal V}_{m+|\boldsymbol{j}|}\biggl(\omega\mid\!\begin{array}{ l}
   r_{\boldsymbol{k}}^- ,\{\lambda_a\} \\
  r_{\boldsymbol{j}}^+ , \{\lambda_a+\epsilon_a\}
 \end{array}\! \biggr),
 \end{equation}
and
 \begin{equation}\label{nu-hat}
 \hat\nu(\omega)\equiv \hat\nu\biggl(\omega\mid\!\begin{array}{l }
   r_{\boldsymbol{k}}^-  , \{\lambda_a\} \\
   r_{\boldsymbol{j}}^+  , \{\lambda_a+\epsilon_a\}
 \end{array}\! \biggr)
 =\frac{-1}{2\pi i}\log\left[1+\gamma\vartheta(\omega)
 {\cal V}_{m+|\boldsymbol{j}|}(\omega)\right].
 \end{equation}
Similarly, the notation $W_{m+|\boldsymbol{j}|}$ means
 \begin{equation}\label{W-sh}
 W_{m+|\boldsymbol{j}|}\equiv W_{m+|\boldsymbol{j}|}
 \biggl(\begin{array}{l}  r_{\boldsymbol{k}}^- , \{\lambda_a\} \\
 r_{\boldsymbol{j}}^+ , \{\lambda_a+\epsilon_a\}\end{array}\!
 \biggr).
 \end{equation}

\end{prop}

We remind here that $ r_{\boldsymbol{j}} = \{ r_{j_a} \}_{a=1}^{\abs{j}}$, \textit{etc}.

{\sl Proof.} The explicit form of the  functional
$U_{\boldsymbol{j}; \boldsymbol{k}}[\nu]$ is given by
\eqref{def-dirU}. However, to prove Proposition~\ref{prop-mulInt},
it is enough to know that it can be written in the following, quite general,
form:
 \begin{equation}\label{Tnu}
 U_{\boldsymbol{j}; \boldsymbol{k}}[\nu]=\sum_{p=0}^\infty \;  \int\limits_{ \Gamma_{\boldsymbol{j};\boldsymbol{k}} }
 U_p(\xi_1,\dots,\xi_p)\prod_{j=1}^p\nu(\xi_j)\,d\xi_j \;,
 \end{equation}
where  $U_p$ are some functions or distributions.  Their explicit
form as well as the fact that they depend on the poles $\Rpm_{\boldsymbol{j}/\boldsymbol{k}}$ does
not play an essential role. Substituting the above expansion into
\eqref{def-calU} and expanding the exponent of $\int
g'(\lambda)\nu(\lambda)\,d\lambda$ into a series we obtain
 \begin{multline}
\mathcal{U}_{\boldsymbol{j};\boldsymbol{k}}([g],[\nu],[F] ) =
\sum_{m,p=0}^{\infty} \frac{ \gamma^{2|\boldsymbol{j}|} }{ m! }
\int\limits_{\Gamma_{\boldsymbol{j};\boldsymbol{k}}}\,d^m \lambda
\,d^p \xi\cdot
U_p(\{\xi\}) \prod_{a=1}^{m}\{-g^{\prime}(\lambda_a)
\nu(\lambda_a)\} \cdot
\prod_{a=1}^{p}\nu(\xi_a) \\
\times \prod_{a=1}^{|\boldsymbol{j}|} \left\{ e^{g(\Rm_{k_a}) -
g(\Rp_{j_a})} F(\Rp_{j_a}) F(\Rm_{k_a}) \right\}  .
\label{devel-calU}
\end{multline}
We should now differentiate \eqref{devel-calU} over $\gamma$. The $n^{\e{th}}$
$\gamma$-derivative at $\gamma=0$ of the above expression is non-vanishing only
when $n\geq 2|\boldsymbol{j}|$, and then
 \begin{multline}\label{derg-exp-A}
 \left.  \frac{\partial^n\mathcal{U}_{\boldsymbol{j};\boldsymbol{k}} }{\partial\gamma^n} %([g],[\nu],[F] )
 \right|_{\gamma=0}
 =  \sum_{m,p=0}^\infty  {\sum_{\{\ell\},\{q\}}} \hspace{-2mm} ^{\prime} C_n(\{\ell\},\{q\})
 \int\limits_{\Gamma_{\boldsymbol{j};\boldsymbol{k}}} \frac{ d^m \lambda \, d^p \xi }{m!}\; U_p(\{\xi\})
 \prod_{a=1}^{|\boldsymbol{j}|} \left( e^{g(\Rm_{k_a})- g(\Rp_{j_a})}
 F(\Rp_{j_a}) F(\Rm_{k_a}) \right)\\
 \times
 \prod_{a=1}^m \left(-g'(\lambda_a)\vartheta^{\ell_a}(\lambda_a)
  F^{\ell_a}(\lambda_a)
 \frac{\partial^{\ell_a}\nu_0}{\partial\gamma^{\ell_a}}\right)
 \left.\prod_{a=1}^p\left( \vartheta^{q_a}(\xi_a)F^{q_a}(\xi_a)
 \frac{\partial^{q_a}\nu_0}{\partial\gamma^{q_a}}\right)
  \right|_{\gamma=0},
 \end{multline}
where
 \begin{equation}\label{nu0}
 \nu_0=\frac{-1}{2\pi i}\log(1+\gamma),\qquad
 C_n(\{\ell\},\{q\})=(2|\boldsymbol{j}|)!(n-2|\boldsymbol{j}|)!
 \left(\prod_{a=1}^m \ell_a! \prod_{a=1}^p q_a!\right)^{-1},
 \end{equation}
and the symbol ${\sum}'$ means that the sum over $\ell_a \geq 0$ and $q_a \geq  0$ in
\eqref{derg-exp-A} is constrained by
$\sum_{a=1}^m\ell_a+\sum_{a=1}^p q_a=n-2|\boldsymbol{j}|$.

We have brought the result of the action of the $n^{\e{th}}$
$\gamma$-derivative into a form convenient for the application of
the density procedure. We first substitute
 \begin{equation}\label{subst-01}
 F(\omega) \, \hookrightarrow \, \varphi_k(\omega)e^{-g_k(\omega)},\qquad
 g'(\omega) \, \hookrightarrow \, g'_k(\omega)
 \end{equation}
in \eqref{derg-exp-A}. Then the sum over $k$ is computed due to the identity
%
%\vartheta^{\ell_a}(\lambda_a) \vartheta^{q_a}(\xi_a)    \prod_{a=1}^p \vartheta^{q_a}(\xi_a)\prod_{a=1}^m \vartheta^{\ell_a}(\lambda_a)
%
 \begin{multline}\label{den-proc-osc}
 \sum_{k=1}^\infty\prod_{a=1}^m
 \left\{\varphi_k^{\ell_a}(\lambda_a)
  e^{-\ell_a g_k(\lambda_a)} [-g'_k(\lambda_a)] \right\}
 \prod_{a=1}^p
 \left\{\varphi_k^{q_a}(\xi_a) e^{-q_a g_k(\xi_a)} \right\}
  \prod_{a=1}^{|\boldsymbol{j}|} \left\{\varphi_k(\Rp_{j_a})
\varphi_k(\Rm_{k_a})e^{-2g_k(\Rp_{j_a})} \right\} \num
 = \prod_{a=1}^m \partial_{\epsilon_a} \cdot
  \biggl.{\cal F}_n\!\biggl(\!\!\begin{array}{llc}
   r_{\boldsymbol{j}}^+ , r_{\boldsymbol{k}}^- ,&
  \{\overbrace{\lambda_a,\dots,\lambda_a}^{\ell_a~\mbox{\scriptsize times}}\}_{a=1,\dots,m},
  &\{\overbrace{\xi_b,\dots,\xi_b}^{q_b~\mbox{\scriptsize times}}\}_{b=1,\dots,p}\\
    r_{\boldsymbol{j}}^+ , r_{\boldsymbol{j}}^+ ,& \{\lambda_a+\epsilon_a,
    \underbrace{\lambda_a,\dots,\lambda_a}_{\ell_a-1~\mbox{\scriptsize times}}\}_{a=1,\dots,m},&
   \{\underbrace{\xi_b,\dots,\xi_b}_{q_b~\mbox{\scriptsize times}}\}_{b=1,\dots,p}
  \end{array}\!\!\biggr)\biggr|_{\epsilon_a=0}.
 \end{multline}
Now, we use the representation \eqref{6-F-VW} for ${\cal F}_n$ together with the
reduction properties \eqref{8-recurs} of the functions ${\cal V}_n$ and $W_n$.
This leads to
 \begin{multline}\label{reduct-F}
%
%{\cal F}_n\!\biggl(\!\!\begin{array}{lllc}
%
%   \{r_{\boldsymbol{j}}^+\},\{r_{\boldsymbol{k}}^-\},&
%
%  \lambda_1,\qquad\{\lambda_1\}^{\ell_1-1},&
%
%  \ldots,\lambda_m,\qquad\;\{\lambda_m\}^{\ell_m-1},&
%                     \{\xi_1\}^{q_1},\ldots,\{\xi_p\}^{q_p}\\
%
%
%    \{r_{\boldsymbol{k}}^-\},\{r_{\boldsymbol{k}}^-\} ,& \lambda_1+\epsilon_1,\{\lambda_1\}^{\ell_1-1},&
%  \ldots,\lambda_m+\epsilon_m,\{\lambda_m\}^{\ell_m-1},
%                    & \{\xi_1\}^{q_1},\ldots,\{\xi_p\}^{q_p}
%  \end{array}\!\!\biggr)\\
{\cal F}_n\!\biggl(\!\!\begin{array}{llc}
   r_{\boldsymbol{j}}^+ , r_{\boldsymbol{k}}^- ,&
  \{\overbrace{\lambda_a,\dots,\lambda_a}^{\ell_a~\mbox{\scriptsize times}}\}_{a=1,\dots,m},
  &\{\overbrace{\xi_b,\dots,\xi_b}^{q_b~\mbox{\scriptsize times}}\}_{b=1,\dots,p}\\
    r_{\boldsymbol{j}}^+ , r_{\boldsymbol{j}}^+ ,& \{\lambda_a+\epsilon_a,
    \underbrace{\lambda_a,\dots,\lambda_a}_{\ell_a-1~\mbox{\scriptsize times}}\}_{a=1,\dots,m},&
   \{\underbrace{\xi_b,\dots,\xi_b}_{q_b~\mbox{\scriptsize times}}\}_{b=1,\dots,p}
  \end{array}\!\!\biggr)\\
  =W_{m+|\boldsymbol{j}|}%\biggl(\begin{array}{c} \{r^+_{\boldsymbol{j}}\},  \{\lambda_a\} \\
%
%
% \{r^-_{\boldsymbol{k}}\},  \{\lambda_a+\epsilon_a\}\end{array}\! \biggr)
%
 \cdot\prod_{b=1}^{m}
 \mathcal{V}_{m+|\boldsymbol{j}|}^{\ell_b}(\lambda_b) \cdot
\prod_{b=1}^{p} \mathcal{V}_{m+|\boldsymbol{j}|}^{q_b}(\xi_b) \cdot
\prod_{b=1}^{|\boldsymbol{j}|}
\{ \mathcal{V}_{m+|\boldsymbol{j}|}(\Rp_{j_b})\mathcal{V}_{m+|\boldsymbol{j}|}(\Rm_{k_b}) \}
%
 %\prod_{j=1}^m{\cal V}_m^{\ell_j}\biggl(\lambda_j\mid\!\begin{array}{c} \{\lambda_a\} \\
% \{\lambda_a+\epsilon_a\}
% \end{array}\! \biggr)\prod_{j=1}^p{\cal V}_m^{r_j}\biggl(\xi_j\mid\!\begin{array}{c} \{\lambda_a\} \\ \{\lambda_a+\epsilon_a\}
% \end{array}\! \biggr),
%
 \end{multline}
where the functions $\mathcal{V}_{m+|\boldsymbol{j}|}$ and
$W_{m+|\boldsymbol{j}|}$ should be understood as in \eqref{new-V},
\eqref{W-sh}. Substituting this identity into \eqref{derg-exp-A} we obtain
 \begin{multline}\label{asy-In}
 {\cal J}_{\boldsymbol{j}; \boldsymbol{k}}[{\cal F}_n]=\sum_{m,p=0}^\infty\frac{ \partial^n_\gamma}{m!}
 \prod_{a=1}^m\partial_{\epsilon_a}% \Biggl\{\;
 \int\limits_{\Gamma_{\boldsymbol{j};\boldsymbol{k}}} d^m\lambda \, d^p\xi \; U_p(\{\xi_j\})
   W_{m+|\boldsymbol{j}|}%\biggl(\begin{array}{c} \{r^+_{\boldsymbol{j}}\},  \{\lambda_a\} \\
%
%
%             \{r^-_{\boldsymbol{k}}\},  \{\lambda_a+\epsilon_a\}\end{array}\! \biggr)  \Biggr.
%
%
\\
\biggl. \times
 \prod_{b=1}^{|\boldsymbol{j}|}
\{  \mathcal{V}_{m+|\boldsymbol{j}|}(\Rp_{j_b})\mathcal{V}_{m+|\boldsymbol{j}|}(\Rm_{k_b}) \}
%
%\Biggl.
  \prod_{a=1}^m \hat\nu(\lambda_a)
 \prod_{a=1}^p\hat\nu(\xi_a)
        %\Biggr\}
        \biggr|_{\gamma=0\atop{\epsilon_a=0}},
 \end{multline}
where $\hat\nu(\omega)$ is given by \eqref{nu-hat}. The sum over $p$ in
\eqref{asy-In} produces
$\mathcal{U}_{\boldsymbol{j};\boldsymbol{k}}\left([0],[\hat\nu],
[\mathcal{V}_{m+|\boldsymbol{j}|}] \right)$
%
% \begin{equation}\label{asy-In1}
% {\cal J}_{\boldsymbol{j}; \boldsymbol{k}}[{\cal F}_n]=\partial^n_\gamma\sum_{m=0}^\infty\frac{1 }{m!}
% \prod_{a=1}^m\partial_{\epsilon_a} \int\limits_{\cal C}
% e^{{\cal A}_{\boldsymbol{j}; \boldsymbol{k}}([0],[\hat\nu])}W_m\biggl(\begin{array}{c} \{\lambda_a\} \\
% \{\lambda_a+\epsilon_a\}\end{array}\! \biggr)
% \left.
%  \prod_{a=1}^m \pox{\hat\nu^{\pa{m}}(\lambda_a)\,d\lambda_a} \;
% \right|_{\gamma=0\atop{\epsilon_a=0}},
% \end{equation}
%
and we arrive at the statement of Proposition.\qed

%%%%%%%%%%%%%%%%%%%%%%%%%%%%%%%%%%%%%%%%%%%%%%%%%%%%%%%%%%%%%%%%%%%%%%%%%%%%%%%%%%%%%%%%%%%%%%%%%%%%%%%%%%%%%%%%%%%%%%%%%%%%%%%%%%%%%%%%%%%%%%%%%%%%%%
%%%%%%%%%%%%%%%%%%%%%%%%%%%%%%%%%%%%%%%%%%%%%%%%%%%%%%%%%%%%%%%%%%%%%%%%%%%%%%%%%%%%%%%%%%%%%%%%%%%%%%%%%%%%%%%%%%%%%%%%%%%%%%%%%%%%%%%%%%%%%%%%%%%%%%
%%%%%%%%%%%%%%%%%%%%%%%%%%%%%%%%%%%%%%%%%%%%%%%%%%%%%%%%%%%%%%%%%%%%%%%%%%%%%%%%%%%%%%%%%%%%%%%%%%%%%%%%%%%%%%%%%%%%%%%%%%%%%%%%%%%%%%%%%%%%%%%%%%%%%%
%%%%%%%%%%%%%%%%%%%%%%%%%%%%%%%%%%%%%%%%%%%%%%%%%%%%%%%%%%%%%%%%%%%%%%%%%%%%%%%%%%%%%%%%%%%%%%%%%%%%%%%%%%%%%%%%%%%%%%%%%%%%%%%%%%%%%%%%%%%%%%%%%%%%%%

\section{Lagrange series}
\label{section Lagrange series}

We have presented the action of the functional ${\cal J}_{\boldsymbol{j};
\boldsymbol{k}}$ on the function ${\cal F}_n$ as the $n^{\e{th}}$
$\gamma$-derivative at $\gamma=0$ of the function $G_{\boldsymbol{j};
\boldsymbol{k}}(\gamma)$  defined in terms of an infinite series \eqref{InC}. In this
section, we show that this series is nothing else but the continuous generalization
of a Lagrange series, in complete analogy with the $T=0$ case studied in
\cite{KitKMST09a}. It can be shown that, for $\gamma$ small enough, this series
is absolutely convergent and that its sum can be expressed in
terms of a solution of a non-linear integral equation. This is a difference in respect to the $T=0$ case where the integral equation was, eventually, linear.

The series \eqref{InC} only contains the first order derivatives
in respect to each $\epsilon_a$. Once that these derivatives are computed,
one should set all $\epsilon_a=0$. Therefore, starting from \eqref{InC}, one
can restrict to linear in $\epsilon_a$, $a=1,\dots,m$ contributions to the
functions $\hat\nu$, $\mathcal{V}_{m+|\boldsymbol{j}|}$, and
$W_{m+|\boldsymbol{j}|}$. It means that all $\epsilon_a^n$ terms with $n>1$ can
be dropped, while the terms $\epsilon_a\epsilon_b$ for $a\not=b$ \textit{etc} should
be kept.  After such a linearization, the functions
$\mathcal{V}_{m+|\boldsymbol{j}|}$ and $\hat\nu(\omega)$ go to
\begin{equation}\label{calV-lin}
\mathcal{V}_{m+|\boldsymbol{j}|} \biggl(\omega\mid\!\begin{array}{c
c}
  r_{\boldsymbol{k}}^- & \{\lambda_a\} \\
 r_{\boldsymbol{j}}^+ & \{\lambda_a+\epsilon_a\} \end{array} \biggr)
 \hookrightarrow
 {\Phi}_{\boldsymbol{j};\boldsymbol{k}}\Bigl(\omega\;\Bigl|\; 2\pi i\alpha
-i\sum_{a=1}^m\epsilon_a K(\omega-\lambda_a)\Bigr.\Bigr)
\end{equation}
and
 \begin{equation}\label{nu-hat-lin}
\hat\nu\biggl(\omega\mid\!\begin{array}{c c}
  r_{\boldsymbol{k}}^- & \{\lambda_a\} \\
   r_{\boldsymbol{j}}^+ &  \{\lambda_a+\epsilon_a\}
 \end{array}\! \biggr)
 \hookrightarrow {\phi}_{\boldsymbol{j};\boldsymbol{k}}\Bigl(\omega\;\Bigl|\; 2\pi i\alpha
 -i\sum_{a=1}^m\epsilon_a K(\omega-\lambda_a)\Bigr.\Bigr) \;,
 \end{equation}
where
\begin{equation}\label{Phi-phijk}
{\Phi}_{\boldsymbol{j};\boldsymbol{k}}(\omega\mid X)= e^{X}
\prod_{a=1}^{|\boldsymbol{j}|} \frac{ e^{i\theta(\omega-\Rp_{j_a})}
}{ e^{i\theta(\omega-\Rm_{k_a})} } -1 \; ,\qquad
%
%\quad \e{and} \quad
%
{\phi}_{\boldsymbol{j};\boldsymbol{k}}(\omega\mid X)= \frac{-1
}{2\pi i} \log[ 1+\gamma \vartheta(\omega)
{\Phi}_{\boldsymbol{j};\boldsymbol{k}}(\omega\mid X) ] \; .
\end{equation}
We recall that $K(\lambda)$ and $\theta(\lambda)$ are given resp. by
\eqref{Lieb-kern}, \eqref{def-theta}.

The linearization of the function $W_{m+|\boldsymbol{j}|}$ is
carried out in appendix~\ref{AEebQ}. Even though its explicit form
is involved, this matters little at this stage of our calculations.
Therefore we present it in a rather symbolic form
\begin{equation}\label{W-lin}
 W_{m+|\boldsymbol{j}|} \biggl(\begin{array}{l} r_{\boldsymbol{k}}^- , \{\lambda_a\} \\
 r_{\boldsymbol{j}}^+ ,\{\lambda_a+\epsilon_a\}\end{array}\! \biggr)
 \hookrightarrow
 \widetilde{\cal W}_{\boldsymbol{j};\boldsymbol{k}} \left( \sum_{a=1}^m\epsilon_a\, g^{(1)}_\sigma(\lambda_a) ;
\sum_{a=1}^m \hspace{-.2mm}\sum_{b=1}^m \epsilon_{a}\,
\epsilon_{b}\, g^{(2)}(\lambda_{a},\lambda_{b}) \right)
\end{equation}
where $g^{(1)}_\sigma$ (with $\sigma=0,\pm$) and $g^{(2)}$   are
some functions. Their explicit expressions  can be found in
appendix~\ref{AEebQ}. Here, we point out that $ \widetilde{\cal W}_{\boldsymbol{j};\boldsymbol{k}} $
depend on the three functions $g_{\sigma}^{(1)}(\lambda_a)$, $\sigma =0, \pm $
 (\textit{cf} appendix~\ref{AEebQ}).

In all of the above notations, the subscripts
${\boldsymbol{j},\boldsymbol{k}}$ indicate that the functions
$\phi_{\boldsymbol{j};\boldsymbol{k}}$,
$\Phi_{\boldsymbol{j};\boldsymbol{k}}$, and
$\widetilde{\mathcal{W}}_{\boldsymbol{j};\boldsymbol{k}}$  depend on
the sets  of poles $r_{\boldsymbol{j}}$ and $r_{\boldsymbol{k}}$. If these sets are
empty, then we will omit the subscripts and write simply $\phi$,
$\Phi$, and $\widetilde{\mathcal{W}}$.

Substituting \eqref{calV-lin}, \eqref{nu-hat-lin}, and \eqref{W-lin} into
\eqref{InC} we obtain
 \begin{multline}\label{Lagr-ser}
 G_{\boldsymbol{j};\boldsymbol{k}}(\gamma)=\sum_{m=0}^\infty\frac{1}{m!}
 \prod_{a=1}^m\partial_{\epsilon_a} \int\limits_{\Gamma_{\boldsymbol{j};\boldsymbol{k}}} \prod_{a=1}^m
 {\phi}_{\boldsymbol{j};\boldsymbol{k}}\Bigl(\lambda_a\mid 2\pi i\alpha -i\sum_{b=1}^m\epsilon_b K(\lambda_a-\lambda_b)\Bigr)\\
 \times \widetilde{\cal W}_{\boldsymbol{j};\boldsymbol{k}} \left( \sum_{a=1}^m\epsilon_a\, g^{(1)}_\sigma(\lambda_a);
  \sum_{a=1}^m \hspace{-.2mm}\sum_{b=1}^m \epsilon_{a}\,
 \epsilon_{b}\, g^{(2)}(\lambda_{a},\lambda_{b}) \right)
 \biggl.  \mathcal{U}_{\boldsymbol{j};\boldsymbol{k}}([0],[{\phi}_{\boldsymbol{j};\boldsymbol{k}}], [{\Phi}_{\boldsymbol{j};\boldsymbol{k}}] )
    \,d^m\lambda
 \biggr|_{\epsilon_a=0}.
 \end{multline}

The above series is the continuous generalization of the Lagrange series
(\textit{cf} \cite{WhiW} for the scalar case, \cite{AizYL78} for the
multi-dimensional case, \cite{KitKMST09a} and Appendix~\ref{S-CgMLS} for the
continuous generalization). This series is absolutely convergent if there
exists a $R_0>0$ such that for all $\lambda \in \Gamma_{\boldsymbol{j};\boldsymbol{k}}$
 \begin{equation}\label{Conv-cond}
   \left|
 \frac{1}{2\pi} \log\left( 1+ |\gamma \vartheta(\lambda)|  \cdot \biggl|
\prod_{j=1}^{n} \frac{ e^{i\theta(\lambda-\Rp_{j_a})} }{
e^{i\theta(\lambda-\Rm_{k_a})} } \cdot  e^{2i\pi \alpha}
e^{R_0 \int_{\Gamma_{\boldsymbol{j};\boldsymbol{k}}}^{} |K(\lambda-\omega)| d \omega } -1 \biggr|   \right)    \right|
 <R_0  \;.
 \end{equation}
This clearly holds for $\gamma$ small enough and hence the series \eqref{Lagr-ser} is then absolutely convergent. In this case, the
result of summation reads (see appendix~\ref{S-CgMLS})
 \begin{multline}\label{Sum-Lagr-ser1}
 G_{\boldsymbol{j};\boldsymbol{k}}(\gamma)=
   \frac{\mathcal{U}_{\boldsymbol{j};\boldsymbol{k}}([0],[z_{\boldsymbol{j};\boldsymbol{k}}],
   \left[ (e^{-2i\pi z_{\boldsymbol{j};\boldsymbol{k}}}-1)/(\gamma\vartheta)
  \right])}
    { \det_{\Gamma_{\boldsymbol{j};\boldsymbol{k}} }[I+{\cal K}_{\boldsymbol{j};\boldsymbol{k}}] } \\
 \times \widetilde{\cal W}_{\boldsymbol{j};\boldsymbol{k}} \biggl(\;\;
    \int\limits_{ \Gamma_{\boldsymbol{j};\boldsymbol{k}} }g^{(1)}_\sigma(\lambda)z_{\boldsymbol{j};\boldsymbol{k}}(\lambda)\,
 d\lambda  ;
 \int\limits_{ \Gamma_{\boldsymbol{j};\boldsymbol{k}} } g^{(2)}(\lambda,\mu)z_{\boldsymbol{j};\boldsymbol{k}}(\lambda)
 z_{\boldsymbol{j};\boldsymbol{k}}(\mu)\,d\lambda\,d\mu  \biggr) \;.
 \end{multline}
The function $z_{\boldsymbol{j};\boldsymbol{k}}(\lambda)$ appearing above is the unique solution to the
non-linear integral equation
 \begin{equation}\label{8-Int-eq0}
 z_{\boldsymbol{j};\boldsymbol{k}}(\lambda)-{\phi}_{\boldsymbol{j};\boldsymbol{k}}\biggl(\lambda\mid 2\pi i\alpha
 -i\int\limits_{\Gamma_{\boldsymbol{j};\boldsymbol{k}}}
 K(\lambda-\omega)z_{\boldsymbol{j};\boldsymbol{k}}(\omega)\,d\omega\biggr)=0.
 \end{equation}
Note that, the uniqueness and existence of this solution is provided by the convergence of
the Lagrange series. The integral operator $I+{\cal K}_{\boldsymbol{j};\boldsymbol{k}}$ acts on the contour
$\Gamma_{\boldsymbol{j};\boldsymbol{k}}$ with the kernel
\begin{equation}
 {\cal K}_{\boldsymbol{j};\boldsymbol{k}}(\lambda,\mu)=iK(\lambda-\mu){\phi}'_{\boldsymbol{j};\boldsymbol{k}}
 \biggl(\lambda\mid 2i\pi \alpha  -i\int\limits_{\Gamma_{\boldsymbol{j};\boldsymbol{k}}} K(\lambda-\tau)
 z_{\boldsymbol{j};\boldsymbol{k}}(\tau)\,d\tau\biggr),
\label{Om}
\end{equation}
where
 \begin{equation}\label{pi-prime}
 {\phi}'_{\boldsymbol{j};\boldsymbol{k}}(\omega\mid X)=
 \partial_{X}{\phi}_{\boldsymbol{j};\boldsymbol{k}}(\omega\mid X).
 \end{equation}
It is easy to see that  its Fredholm determinant coincides with  the functional Jacobian
of equation \eqref{8-Int-eq0}.

Observe that apart from the appearance of the Fredholm determinant described above,
the result of the Lagrange series summation  reduces to the replacement of the
discrete sums involving the $\epsilon_a$'s by integrals  over the contour
$\Gamma_{\boldsymbol{j};\boldsymbol{k}}$ with a weight function
$z_{\boldsymbol{j};\boldsymbol{k}}(\lambda)$. Due to this fact,  and the form of
equation \eqref{8-Int-eq0}, one has that the original arguments
$\phi_{\boldsymbol{j};\boldsymbol{k}}$ and
$\Phi_{\boldsymbol{j};\boldsymbol{k}}$ of the functional
$\mathcal{U}_{\boldsymbol{j};\boldsymbol{k}}$ are replaced by
$z_{\boldsymbol{j};\boldsymbol{k}}$ and $(e^{-2i\pi
z_{\boldsymbol{j};\boldsymbol{k}}}-1)/(\gamma\vartheta))$ respectively (taking
into account that $\Phi_{\boldsymbol{j};\boldsymbol{k}}=(e^{-2i\pi
\phi_{\boldsymbol{j};\boldsymbol{k}}}-1)/(\gamma\vartheta))$.

The obtained result can be slightly simplified by deforming the integration
contours $\Gamma_{\boldsymbol{j};\boldsymbol{k}}$. Indeed, it follows
straightforwardly from \eqref{8-Int-eq0}, \eqref{Phi-phijk} that $e^{-2\pi i
z_{\boldsymbol{j};\boldsymbol{k}}(\lambda)}$ has simple poles at
$\Rp_{\boldsymbol{j}}$ and $\Rm_{\boldsymbol{k}}$.
%
%
%
%\begin{equation}\label{poles-z}
%
% (\lambda-\Rp_{j_a}) e^{-2\pi iz_{\boldsymbol{j};\boldsymbol{k}}(\lambda)} \; = \;
% \underset{\lambda\rightarrow \Rp_{j_a}}{\e{O}}(1) \; , \qquad
%
% (\lambda-\Rm_{k_a}) e^{-2\pi iz_{\boldsymbol{j};\boldsymbol{k}}(\lambda)} \; = \;   \underset{\lambda\rightarrow \Rm_{k_a}}{\e{O}}(1) \;.
%
%\end{equation}
%
Define also the sets of zeros $\hqp_{\boldsymbol{j}}$ and
$\hqm_{\boldsymbol{k}}$ of $e^{-2\pi i
z_{\boldsymbol{j};\boldsymbol{k}}(\lambda)}$ such that:
\begin{equation}\label{zeros-z}
 e^{-2\pi iz_{\boldsymbol{j};\boldsymbol{k}}(\hqp_{j_a})}=0  \;, \qquad
 e^{-2\pi iz_{\boldsymbol{j};\boldsymbol{k}}(\hqm_{k_a})}=0  \;,\qquad
 \hqpm_{\boldsymbol{j}/\boldsymbol{k}}\to\Rpm_{\boldsymbol{j}/\boldsymbol{k}}\quad \mbox{as}\quad\gamma\to 0.
\end{equation}
We now introduce the contour $\hat{\cal C}_{\boldsymbol{j};\boldsymbol{k}}$. It
is a deformation of the real axis such that moving from $\mathbb{R}$ to
$\hat{\cal C}_{\boldsymbol{j};\boldsymbol{k}}$ one only crosses the roots
$\hqp_{j_1},\dots, \hqp_{j_n}$ and $\hqm_{k_1},\dots, \hqm_{k_n}$, while the
other roots $\hqpm_\ell$ and all the poles $\Rpm_\ell$ are not crossed (see
Fig.~\ref{hG1212}). In particular,  this contour separates the poles
$\Rpm_{\boldsymbol{j}/\boldsymbol{k}}$ from the roots
$\hqpm_{\boldsymbol{j}/\boldsymbol{k}}$.

Let us shift the integration contour
$\Gamma_{\boldsymbol{j};\boldsymbol{k}}$ to the contour $\hat{\cal C}_{\boldsymbol{j};\boldsymbol{k}}$  everywhere in \eqref{Sum-Lagr-ser1}. Observe that, for $\gamma$ small enough, the
pole and zero structure of the function $e^{-2\pi
iz_{\boldsymbol{j};\boldsymbol{k}}(\lambda)}$ in a neighborhood of the contours
$\Gamma_{\boldsymbol{j};\boldsymbol{k}}$ and $\hat{\cal
C}_{\boldsymbol{j};\boldsymbol{k}}$ is completely analogous to the one of the
function $e^{-2i\pi \nu(\lambda)}$ \eqref{nu}. This allows us to use the method described
in appendix~\ref{PrF}. However, now, all the calculations should be done in a reverse
order.

In particular, the equation \eqref{GtoC-f} applied to the contours
$\Gamma_{\boldsymbol{j};\boldsymbol{k}}$ and $\hat{\cal
C}_{\boldsymbol{j};\boldsymbol{k}}$ and the function
$z_{\boldsymbol{j};\boldsymbol{k}}$ gives
\begin{equation}\label{GtoC-K}
\int\limits_{\Gamma_{\boldsymbol{j};\boldsymbol{k}}}
K(\lambda-\omega)z_{\boldsymbol{j};\boldsymbol{k}}(\omega)\,d\omega =
\int\limits_{\hat{\cal C}_{\boldsymbol{j};\boldsymbol{k}}}
K(\lambda-\omega)z_{\boldsymbol{j};\boldsymbol{k}}(\omega)\,d\omega -
\sum_{a=1}^{|\boldsymbol{j}|}\Bigl[
\theta(\lambda-\Rm_{k_a})-\theta(\lambda-\Rp_{j_a}) \Bigr] \;.
\end{equation}
Substituting this into \eqref{8-Int-eq0}, we find that
$z_{\boldsymbol{j};\boldsymbol{k}}(\lambda)$ solves the integral equation
 \begin{equation}\label{8-Int-eq0-mod}
 z_{\boldsymbol{j};\boldsymbol{k}}(\lambda)-
 {\phi}\biggl(\lambda\mid 2\pi i\alpha -i\int\limits_{\hat{\cal C}_{\boldsymbol{j};\boldsymbol{k}}}
 K(\lambda-\omega)z_{\boldsymbol{j};\boldsymbol{k}}(\omega)\,d\omega\biggr)=0,
 \end{equation}
(recall that ${\phi}(\lambda|X)$ is given by \eqref{Phi-phijk} with
$|\boldsymbol{j}|=0$).  The Jacobian of equation \eqref{8-Int-eq0}
naturally turns into the Jacobian of the equation \eqref{8-Int-eq0-mod}, and
hence,
 \begin{equation}\label{jac-jac}
 \det_{\Gamma_{\boldsymbol{j};\boldsymbol{k}}
 }[I+{\cal K}_{\boldsymbol{j};\boldsymbol{k}}]=
 \det_{\hat{\cal C}_{\boldsymbol{j};\boldsymbol{k}} }[I+{\cal K}],
 \end{equation}
where
\begin{equation}
 {\cal K}(\lambda,\mu)=iK(\lambda-\mu){\phi}'
 \biggl(\lambda\mid 2i\pi \alpha
 -i\int\limits_{{\cal C}_{\boldsymbol{j};\boldsymbol{k}}} K(\lambda-\tau)
 z_{\boldsymbol{j};\boldsymbol{k}}(\tau)\,d\tau \biggr),
\label{Om-new}
\end{equation}
and $\phi'$ is given by \eqref{pi-prime}.

Formula \eqref{GtoC-f} also yields
 \begin{equation}\label{WG-WC}
 \widetilde{\cal W}_{\boldsymbol{j};\boldsymbol{k}} \biggl(\;\;
 \int\limits_{ \Gamma_{\boldsymbol{j};\boldsymbol{k}}}g^{(1)}_\sigma(\lambda)
 z_{\boldsymbol{j};\boldsymbol{k}}(\lambda)\, d\lambda  ;\dots \biggr)
 =\widetilde{\cal W}\biggl(\;\;
 \int\limits_{ \hat{\cal C}_{\boldsymbol{j};\boldsymbol{k}} }g^{(1)}_\sigma(\lambda)
 z_{\boldsymbol{j};\boldsymbol{k}}(\lambda)\,d\lambda  ;\dots
 \biggr),
 \end{equation}
(see appendix~\ref{AEebQ} for more details). Finally, making the calculations of
appendix~\ref{PrF} in the reverse order we find that
 \begin{equation}\label{Ujk-A}
 \mathcal{U}_{\boldsymbol{j};\boldsymbol{k}}([0],[z_{\boldsymbol{j};\boldsymbol{k}}],
   \left[ (e^{-2i\pi z_{\boldsymbol{j};\boldsymbol{k}}}-1)/(\gamma\vartheta))
  \right])=\exp\bigl\{ {\cal A}_{\hat{\cal
  C}_{\boldsymbol{j};\boldsymbol{k}}}([0],[z_{\boldsymbol{j};\boldsymbol{k}}])\bigr\},
  \end{equation}
where the functional $\mathcal{A}_{\hat{\cal
C}_{\boldsymbol{j};\boldsymbol{k}}}([0],[z])$ is given by \eqref{A-rep} with
${\cal L}=\hat{\cal C}_{\boldsymbol{j};\boldsymbol{k}}$ and $g(\lambda)=0$.

Thus, the sum of the series \eqref{InC} for
$G_{\boldsymbol{j};\boldsymbol{k}}(\gamma)$ takes the form
 \begin{equation}\label{Sum-Lagr-ser}
 G_{\boldsymbol{j};\boldsymbol{k}}(\gamma)=
 %\exp\left\
  \frac{ \exp\bigl\{ {\cal A}_{\hat{\cal C}_{\boldsymbol{j};\boldsymbol{k}}}([0],[z_{\boldsymbol{j};\boldsymbol{k}}])\bigr\} }
  { \det_{\hat{\cal C}_{\boldsymbol{j};\boldsymbol{k}} }[I+{\cal K}] } \cdot %\right\}
 \widetilde{\cal W} \biggl(\;\;
    \int\limits_{ \hat{\cal C}_{\boldsymbol{j};\boldsymbol{k}} }g^{(1)}_\sigma(\lambda)z_{\boldsymbol{j};\boldsymbol{k}}(\lambda)\,
 d\lambda  ;
 \int\limits_{ \hat{\cal C}_{\boldsymbol{j};\boldsymbol{k}} } g^{(2)}(\lambda,\mu)z_{\boldsymbol{j};\boldsymbol{k}}(\lambda)
 z_{\boldsymbol{j};\boldsymbol{k}}(\mu)\,d\lambda\,d\mu  \biggr).
 \end{equation}

We have expressed $G_{\boldsymbol{j};\boldsymbol{k}}(\gamma)$ as a
functional of the function
$z_{\boldsymbol{j};\boldsymbol{k}}(\lambda)$ satisfying the
non-linear integral equation \eqref{8-Int-eq0-mod}. We now show that
this last integral equation can be transformed into a TBA-like
equation. For this, we introduce a new function
$u_{\boldsymbol{j};\boldsymbol{k}}(\lambda)$:
 \begin{equation}\label{def-u}
  \frac{u_{\boldsymbol{j};\boldsymbol{k}}(\lambda)}T=i\int\limits_{\hat{\cal C}_{\boldsymbol{j};\boldsymbol{k}}}
 K(\lambda-\omega)z_{\boldsymbol{j};\boldsymbol{k}}(\omega)\,d\omega -2\pi i\alpha+\frac{\varepsilon(\lambda)}T,
 \end{equation}
where $\varepsilon(\lambda)$ solves the Yang--Yang equation \eqref{YY-eq}. Then,
equation \eqref{8-Int-eq0-mod} yields
 \begin{equation}\label{sol-z}
  z_{\boldsymbol{j};\boldsymbol{k}}(\lambda)=-\frac1{2\pi i}\log\left[1+\gamma\vartheta(\lambda)\left(
 e^{\frac{\varepsilon(\lambda)-u_{\boldsymbol{j};\boldsymbol{k}}(\lambda)}T}-1\right) \right].
 \end{equation}
Multiplying both parts of \eqref{sol-z} by $iK(\mu-\lambda)$ and integrating
with respect to $\lambda$ over the contour $\hat{\cal
C}_{\boldsymbol{j};\boldsymbol{k}}$, after some simple algebra, we obtain
 \begin{equation}\label{inteq-u1}
 u_{\boldsymbol{j};\boldsymbol{k}}(\mu)-\varepsilon(\mu)+2\pi i\alpha T=
 -\frac{T}{2\pi}\int\limits_{\hat{\cal C}_{\boldsymbol{j};\boldsymbol{k}}}
 K(\mu-\lambda)\log\left(\frac{1+\gamma e^{-\frac{u_{\boldsymbol{j};\boldsymbol{k}}(\lambda)}T}+(1-\gamma)e^{-\frac{\varepsilon(\lambda)}T}}
 {1+e^{-\frac{\varepsilon(\lambda)}T}}\right)  \; d \, \lambda.%\\
 %
 %\left.- \log\left(1+e^{-\frac{\varepsilon(\lambda)}T}\right)\right]\,d\lambda.
%
 \end{equation}
It is easy to see that the roots $\hqp_{\boldsymbol{j}}$ and
$\hqm_{\boldsymbol{k}}$ satisfy the equations
 \begin{equation}\label{demand}
 1+\gamma\exp\Bigl(-{u_{\boldsymbol{j};\boldsymbol{k}}(\hqpm_{\boldsymbol{j}/\boldsymbol{k}})}/T\Bigr)
 +(1-\gamma)\exp\Bigl(-{\varepsilon(\hqpm_{\boldsymbol{j}/\boldsymbol{k}})/T}\Bigr)=0.
 %\qquad  \hqpm_{j_a/k_a} \to\Rpm_{j_a/k_a}\quad\mbox{if}\quad\gamma\to0.
 \end{equation}
When deforming the contour $\hat{\cal C}_{\boldsymbol{j};\boldsymbol{k}}$ to the
real axis in \eqref{inteq-u1} we cross these roots, but do not cross the poles of the Fermi
weight $\Rpm_\ell$, i.e. the points where $1+\exp\bigl(-\varepsilon(\Rpm_\ell)/T\bigr)=0$. Therefore, we obtain
 \begin{multline}\label{inteq-u2}
 u_{\boldsymbol{j};\boldsymbol{k}}(\mu)-\varepsilon(\mu)+2\pi i\alpha T=-\frac{T}{2\pi}\int\limits_{\mathbb{R}}
 K(\mu-\lambda)\log\left[1+\gamma e^{-\frac{u_{\boldsymbol{j};\boldsymbol{k}}(\lambda)}T}
 +(1-\gamma)e^{-\frac{\varepsilon(\lambda)}T}\right]\,d\lambda\\
 +\frac{T}{2\pi}\int\limits_{\mathbb{R}}K(\mu-\lambda)\log\left[1+
 e^{-\frac{\varepsilon(\lambda)}T}\right]\,d\lambda
 -iT\sum_{a=1}^n \Bigl[ \theta(\hqp_{j_a}-\mu)-\theta(\hqm_{k_a}-\mu) \Bigr].
 \end{multline}
Finally, using the Yang--Yang equation we arrive at
 \begin{multline}\label{inteq-u3}
 u_{\boldsymbol{j};\boldsymbol{k}}(\mu)=\mu^2-(h+2\pi i\alpha T)-\frac{T}{2\pi}\int\limits_{\mathbb{R}}
 K(\mu-\lambda)\log\left[1+\gamma e^{-\frac{u_{\boldsymbol{j};\boldsymbol{k}}(\lambda)}T}
 +(1-\gamma)e^{-\frac{\varepsilon(\lambda)}T}\right]\,d\lambda\\
 -iT\sum_{a=1}^n \Bigl[ \theta(\hqp_{j_a}-\mu)-\theta(\hqm_{k_a}-\mu) \Bigr].
 \end{multline}
Thus, we have reduced equation \eqref{8-Int-eq0-mod} for the function
$z_{\boldsymbol{j};\boldsymbol{k}}(\lambda)$ to equation \eqref{inteq-u3}
for the function $u_{\boldsymbol{j};\boldsymbol{k}}(\lambda)$. In the following,  we shall
consider the function $u_{\boldsymbol{j};\boldsymbol{k}}(\lambda)$ as the
primary object, while the function $z_{\boldsymbol{j};\boldsymbol{k}}(\lambda)$
will be defined by \eqref{sol-z}.

\vspace{2mm}

Let us summarize the obtained results. We have expressed the large $x$
asymptotic behavior of the multiple integrals ${\cal I}_n[{\cal F}_n]$
\eqref{def-CI} in terms of the function
$G_{\boldsymbol{j};\boldsymbol{k}}(\gamma)$ \textit{via} \eqref{sum-C}, \eqref{InC-G}.
The function $G_{\boldsymbol{j};\boldsymbol{k}}(\gamma)$ is given by
\eqref{Sum-Lagr-ser}. It is a functional of $u_{\boldsymbol{j};\boldsymbol{k}}(\lambda)$, the solution
to the non-linear integral equation \eqref{inteq-u3}. In its turn,
$u_{\boldsymbol{j};\boldsymbol{k}}(\lambda)$ depends on the sets of parameters
$\hqp_{\boldsymbol{j}}$ and $\hqm_{\boldsymbol{k}}$ satisfying to the condition
\eqref{demand}.

Observe that the roots $\hqpm_{\boldsymbol{j}/\boldsymbol{k}}$ depend on
$\gamma$:
$\hqpm_{\boldsymbol{j}/\boldsymbol{k}}=\hqpm_{\boldsymbol{j}/\boldsymbol{k}}(\gamma)$.
One can treat them as $\gamma$-deformations of the poles
$\Rpm_{\boldsymbol{j}/\boldsymbol{k}}$ such that
$\hqpm_{\boldsymbol{j}/\boldsymbol{k}}(0)=
\Rpm_{\boldsymbol{j}/\boldsymbol{k}}$. Similarly, the function
$u_{\boldsymbol{j};\boldsymbol{k}}(\lambda)=u_{\boldsymbol{j};\boldsymbol{k}}(\lambda,\gamma)$
can be considered as a $\gamma$-deformation of the function
$u_{\boldsymbol{j};\boldsymbol{k}}(\lambda,0)$ which satisfies equation
\eqref{inteq-u3} at $\gamma=0$:
 \begin{equation}\label{u0}
 u_{\boldsymbol{j};\boldsymbol{k}}(\lambda,0)=\varepsilon(\lambda)-2\pi i\alpha T
 -iT\sum_{a=1}^n \Bigl[
 \theta(\Rp_{j_a}-\lambda)-\theta(\Rm_{k_a}-\lambda)\Bigr].
 \end{equation}
The above $\gamma$-deformations are analytic in $\gamma$, at least for $\gamma$ small enough.

\vspace{2mm}
To conclude this section we simplify the notations used above. Recall that we
have originally  used the multi-indices ${\boldsymbol{j}}$ and ${\boldsymbol{k}}$
in order to denote certain subsets of the Fermi weight's poles
$\Rp_{\boldsymbol{j}}$ and $\Rm_{\boldsymbol{k}}$. Let us enumerate all
subsets of multi-indices $\{\boldsymbol{j};\boldsymbol{k}\}$ by one number, say
$i$. Then every subset
$\Rpm_{\boldsymbol{j}/\boldsymbol{k}}$ can be enumerated as $\{\Rpm\}_i$.
In particular, we agree upon $\{\Rpm\}_0=\emptyset$. However, in all other respects, the
order of enumeration is not essential. Given a subset $\{\Rpm\}_i$ we can
uniquely determine the roots $\{\hqpm\}_i$ and the corresponding function
$u_i(\lambda)$ as analytical $\gamma$-deformations of $\{\Rpm\}_i$ and
$u_i(\lambda,0)$ \eqref{u0}. In its turn, given $\{\hqpm\}_i$, we can define the
integration contours $\hat{\cal C}_i$, and hence, find the function
$G_i(\gamma)$. Thus, the asymptotic behavior of the multiple integrals
\eqref{def-CI}  can be written in the form
 \begin{equation}\label{sum-i}
 {\cal I}_n[{\cal F}_n]=\sum_i \Bigl.\partial^n_\gamma
 G_i(\gamma)\left[1+o\big(e^{-a^{\prime} x}\big)\right]\Bigr|_{\gamma=0},
 \end{equation}
where $i$ enumerates the subsets of multi-indices ${\boldsymbol{j}}$
and ${\boldsymbol{k}}$.

We do not write explicitly the upper limit of summation over $i$ in
the formula \eqref{sum-i}. It depends on how many poles of the Fermi
weight belong to the strip $|\Im(\lambda)|<a'$. In particular, it
follows from the Remark 3 given on page \pageref{Rem-AsyExp2} that for
$h>0$ and $T$ small enough and arbitrary $i_0$ there exists
$T(i_0)$ such that
 \begin{equation}\label{Est}
 \lim_{x\to\infty}e^{a^{\prime} x}\Bigl.\partial^n_\gamma
 G_{i_0}(\gamma)\Bigr|_{\gamma=0}=0, \quad\mbox{for}\quad T<T(i_0).
 \end{equation}
Therefore in this case the sum in \eqref{sum-i} may contain an
arbitrary number of terms.

%%%%%%%%%%%%%%%%%%%%%%%%%%%%%%%%%%%%%%%%%%%%%%%%%%%%%%%%%%%%%%%%%%%%%%%%%%%%%%%%%%%%%%%%%%%%%%%%%%%%%%%%%%%%%%%%%%%%%%%%%%%%%%%%%%%%%%%%%%%%%%%%%%%%
%%%%%%%%%%%%%%%%%%%%%%%%%%%%%%%%%%%%%%%%%%%%%%%%%%%%%%%%%%%%%%%%%%%%%%%%%%%%%%%%%%%%%%%%%%%%%%%%%%%%%%%%%%%%%%%%%%%%%%%%%%%%%%%%%%%%%%%%%%%%%%%%%%%%

\section{Asymptotic behavior of the correlation function\label{ABoCF}}
%\label{Section Global Asymptotic Behavior}

We have presented the large $x$ asymptotic behavior of the integrals
${\cal I}_n[{\cal F}_n]$ \eqref{def-CI} in the form \eqref{sum-i}.
We should now substitute this result  into the series
\eqref{3-fin-answ-TD}, what leads us to
 \begin{equation}\label{3-prefin-res-as}
 \Mmoy{e^{2\pi i\alpha {\cal Q}_x}}=\sum_{n=0}^\infty\sum_i \Bigl.\frac1{n!}\partial^n_\gamma
 G_i(\gamma)\Bigr|_{\gamma=0}+
 \sum_{n=0}^\infty\sum_i \Bigl.\frac1{n!}\partial^n_\gamma
 G_i(\gamma)\cdot o\Bigl(e^{-a^{\prime}x}\Bigr)\Bigr|_{\gamma=0} \; .
 \end{equation}
Since $G_i(\gamma)$ does not depend on $n$, the first series gives
the Taylor expansion for $G_i(\gamma)$ at $\gamma=1$. Hence, this
series should result in $G_i(1)$ provided it is convergent at
$\gamma=1$.  One can easily convince himself that it is true at
least for the case $c=\infty$. Indeed, in this case the sum of
$G_i(1)$ coincides with the asymptotic expansion of the Fredholm
determinant \eqref{4-kernel} obtained in \cite{Sla10a}. Therefore
considering the QNLS model with $c<\infty$ as a smooth deformation
of the free fermion case we assume that  this series is convergent
at $\gamma=1$, and thus we obtain
 \begin{equation}\label{first-ser}
 \sum_{n=0}^\infty\sum_i \Bigl.\frac1{n!}\partial^n_\gamma
 G_i(\gamma)\Bigr|_{\gamma=0}=\sum_i G_i(1).
 \end{equation}
Note that the convergence of the series \eqref{first-ser} is
related to the convergence of the Lagrange series
\eqref{Lagr-ser} at $\gamma=1$. This means that the analytic
$\gamma$-deformations described in the end of the previous section
can be continued from a vicinity of $\gamma=0$ to the point
$\gamma=1$. Hence, for the computation $G_i(1)$ it is enough to set
$\gamma=1$ in the equations  for the functions $u_i(\lambda)$ and
$z_i(\lambda)$. The non-linear integral equation \eqref{inteq-u3}
for $u_i(\lambda)$ takes the form \eqref{inteq-u-main}, while the
equation \eqref{sol-z} for $z_i(\lambda)$ turns into
 \begin{equation}\label{sol-Z}
 z_i(\lambda)=-\frac1{2\pi i}\log\left(\frac{1+
 e^{-\frac{u_i(\lambda)}T} }{
 1+ e^{-\frac{\varepsilon(\lambda)}T}} \right) \; .
 \end{equation}
Substituting this into the functional ${\cal A}_{\hat{\cal
C}_i}([0],[z_i])$  we find
 \begin{equation}\label{A-zi}
 {\cal A}_{\hat{\cal C}_i}([0],[z_i])=-ix\int\limits_{\hat{\cal C}_i}
 z_i(\lambda)\,d\lambda +\iint\limits_{\hat{\cal C}_i}
 \frac{z_i(\lambda)z_i(\mu)}{(\lambda-\mu_+)^2}\,d\lambda\,d\mu \; ,
 \end{equation}
and hence, every $G_i(1)$ is proportional to $e^{-xp_i}$ with $p_i$
given by \eqref{p-decay}. Note that the generating function
$\moy{e^{2\pi i\alpha {\cal Q}_x}}$ is not necessary a decreasing
function of $x$ at arbitrary complex value of $\alpha$. Therefore
the real parts of the obtained $p_i$ are not necessary non-negative.
However we should have $\Re(p_i)\ge 0$ at $\alpha=0$, since this
case describes the physical two-point correlation function
$\moy{j(x)j(0)}$. This property is confirmed by  numerical
computations (\textit{cf} appendix~\ref{Appendix Numerics}), and the
analysis of the low-temperature limit \cite{KozMS10c}.

Concerning the  series  corresponding to the reminder in
\eqref{3-prefin-res-as}, we assume that it remains exponentially
small with respect to the first term, provided the sum over $i$ in
\eqref{first-ser} is restricted by some $i_0$. Similarly to
\eqref{sum-i} we do not write this upper limit of summation
explicitly, because we can not find its exact value for arbitrary
temperature. One can show however (\textit{cf}
\cite{KozMS10c}) that  $i_0$ grows as the temperature decreases. In
particular, it goes to infinity when $T\to 0$.

Thus, we reproduce the expansion \eqref{fin-answ}
 \begin{equation}\label{fin-answ1}
 \moy{e^{2\pi i\alpha {{\cal Q}_x}}}=    \sum_{i}e^{-xp_i}B[u_i]
 +o\Bigl(e^{-xp_{max}}\Bigr),\qquad x\to\infty.
 \end{equation}
where $p_{max}=\max_i(\Re(p_i))$.

It remains to describe the constant coefficients $B[u_i]$ in \eqref{fin-answ}.
One part of these coefficients comes from the constant term in the functional
${\cal A}_{\hat{\cal C}_i}([0],[z_i])$ \eqref{A-zi}. Another part is equal to
the Fredholm determinant $\det_{\hat{\cal C}_i }[I+{\cal K}]$ appearing in equation
\eqref{Sum-Lagr-ser}. Observe that, at $\gamma=1$, the derivative ${\phi}'$
\eqref{pi-prime} reads
 \begin{equation}\label{deriv-f}
 {\phi}'\left(\lambda\mid  \frac{\varepsilon(\lambda)-u_i(\lambda)}T\right)=
  -\frac1{2\pi i}\frac1{1+
 e^{-\frac{u_i(\lambda)}T} } \; .
 \end{equation}
Thus,
 \begin{equation}\label{detO-detK}
 \det_{\hat{\cal C}_i }[I+{\cal K}]=\det_{\hat{\cal C}_i
 }\Bigl[I-\frac1{2\pi}K^{(u_i)}\Bigr],
 \end{equation}
where the kernel $K^{(u_i)}$ is the analog of $K^{(\varepsilon)}$
where $\varepsilon$ has been replaced by $u_i$:
 \begin{equation}\label{Ku}
 K^{(u_i)}(\lambda,\mu)=\frac{K(\lambda-\mu)}{1+e^{\frac{u_i(\mu)}T}}\; .
 \end{equation}
This operator acts on the contour $\hat{\cal C}_i$ and not on
$\mathbb{R}$ as it was the case for $I+K^{(\varepsilon)}$.

Finally, the most complicated part of the coefficients $B[u_i]$
comes from the function $\widetilde{\cal W}$ (this part of the computations is given in
appendix~\ref{AEebQ}). Combining all these results we get
 \begin{multline}\label{C2-W}
 B[u_i]=\frac{({e^{2\pi i\alpha}}-1)^2\exp\biggl(\dis\int_{\hat{\cal C}_i}
 \frac{z_i(\lambda)z_i(\mu)}{(\lambda-\mu_+)^2}\,d\lambda\,d\mu-C_0[z_i,{\hat{\cal C}_i}]
 \biggr) }
 {[e^{ L_{\hat{\cal C}_i}[z_i](\theta_1+i c)}  -e^{2\pi i\alpha+L_{\hat{\cal C}_i}[z_i](\theta_1-i c)}]
 [e^{ -L_{\hat{\cal C}_i}[z_i](\theta_2-i c)} -e^{2\pi i\alpha-L_{\hat{\cal C}_i}[z_i](\theta_2+i c)}]} \numa{35}
 \times\frac{\det\left[I+\frac1{2\pi i}\hat U^{(1)}[z_i]\right]
 \det\left[I+\frac1{2\pi i} \hat U^{(2)}[z_i]\right]}
 {\det\left[I-{\textstyle\frac1{2\pi}}K^{(\varepsilon)}\right]
 \det\left[I-{\textstyle\frac1{2\pi}}K^{(u_i)}\right]} \; .
 \end{multline}
Recall that  $L_{\hat{\cal C}_i}[z_i]$ is  the Cauchy transform
\eqref{def-CauT} of the function $z_i(\lambda)$ on the contour $\hat{\cal
C}_i$. The functional $C_0$ as well as  the kernels of the integral operators
$\hat U^{(1)}(w,w',[z_i])$ and $\hat U^{(2)}(w,w',[z_i])$ are given in
appendix~\ref{AEebQ} (see resp.\eqref{7-C0}, \eqref{C2-U1}, \eqref{C2-U2}).
The latter operators act on an counterclockwise oriented contour surrounding $\hat{\cal
C}_i$.

\section{The leading term and corrections\label{LTaC}}

Taking the second $\alpha$-derivative of the expansion \eqref{fin-answ}
for $\moy{e^{2\pi i\alpha {{\cal Q}_x}}}$
at $\alpha=0$ we obtain the asymptotic behavior of the density-density
temperature correlation function
$\langle j(x)j(0)\rangle_T$. We show below that the leading term of this
asymptotic expansion is produced  by $e^{-xp_0}$ corresponding to the
contour $\hat{\cal C}_{0}=
\mathbb{R}$ while the other $p_i$'s lead to sub-leading corrections.

\subsection{The leading term\label{LT}}
As expected on general grounds, we obtain the following  leading term of the asymptotic behavior.
\begin{prop}
The leading term of the asymptotic behavior of the two-point function
\begin{equation}
 \langle j(x)j(0)\rangle_T =  \langle j(x)j(0)\rangle_T^{(0)} + \text{O}(x^{-\infty}) \;.
\end{equation}
is given by the square of the density of particles in the state of
thermal equilibrium:
 \begin{equation}\label{Asy-cor-fun2}
 \langle j(x)j(0)\rangle_T^{(0)} =
 \biggl(\;\int\limits_{\mathbb{R}}\rho_p(\lambda)\,d\lambda\biggr)^2 = \langle j(0)\rangle_T^{2} .
 \end{equation}

This contribution stems from the choice of the contour $\hat{\cal
C}_0=\mathbb{R}$.
\end{prop}

\proof Setting $\hat{\cal C}_0=\mathbb{R}$ in \eqref{inteq-u-main1} we obtain
 \begin{equation}\label{inteq-uR}
 u_0(\mu)=\mu^2-h_\alpha -\frac{T}{2\pi}\int\limits_{\mathbb{R}}
 K(\mu-\lambda)\log\left(1+e^{-\frac{u_0(\lambda)}T}\right)\,d\lambda.
 \end{equation}
The form of this equation coincides with
the one of the Yang--Yang equation up to the shift of the chemical
potential. Thus, we have
 \begin{equation}\label{solution}
 u_0(\lambda)=\varepsilon(\lambda|h_\alpha),
  \end{equation}
where we have insisted explicitly on the dependence of $\varepsilon(\lambda)$
on $h$. Substituting this into \eqref{p-decay}  we find
 \begin{equation}\label{fin-answ-LT}
  p_0=-\frac 1{2\pi}\int\limits_{\mathbb{R}}
 \log\left(\frac{1+e^{-\varepsilon(\lambda|h_\alpha)/T}}
 {1+e^{-\varepsilon(\lambda|h)/T}}\right)\,d\lambda=\frac 1T({\cal P}-{\cal P}_\alpha),
 \end{equation}
where we have defined
 \begin{equation}\label{press}
 {\cal P}=\frac T{2\pi}\int\limits_{\mathbb{R}}
 \log\left(1+e^{-\varepsilon(\lambda|h)/T}\right)\,d\lambda \; .
 \end{equation}
The constant ${\cal P}_\alpha$ in \eqref{fin-answ-LT} corresponds to the
shifted chemical potential, i.e. it is defined  by \eqref{press} with
$\varepsilon(\lambda|h)$ replaced by $\varepsilon(\lambda|h_\alpha)$. This
result reproduces the prediction of \cite{BogIK93L}. The quantity
$\mathcal{P}$ can be interpreted as the pressure in the gas, whereas
$\mathcal{P}_{\alpha}$ would correspond to the pressure in the presence of a
shifted complex valued chemical potential.

Thus, we see that $p_0\to0$ as $\alpha\to0$ meaning that the contribution
stemming from the choice $i=0$ (that is
$(\boldsymbol{j};\boldsymbol{k})=(\emptyset,\emptyset)$) to the
density-density correlation function does not have an exponential decay.
Applying the second order derivatives over $\alpha$ and $x$ to the presumed leading
order term of $\moy{e^{2\pi i\alpha {{\cal Q}_x}}}$ we obtain
 \begin{equation}\label{Asy-cor-fun}
 \langle j(x)j(0)\rangle_T^{(0)} =\left.-\frac{1}{8\pi^2}\frac{\partial^2}{\partial x^2}
 \frac{\partial^2}{\partial \alpha^2} e^{-xp_0}B[u_0]\right|_{\alpha=0}
 =\left.-\frac{B[u_0]}{4\pi^2}
 \left(\frac{\partial p_0}{\partial\alpha}\right)^2\right|_{\alpha=0},
 \end{equation}
and thus, to proceed further we should find $u_0(\lambda)$ and $z_0(\lambda)$
up to linear in $\alpha$ terms. Differentiating \eqref{inteq-uR} over
$\alpha$ at $\alpha=0$ we obtain
 \begin{equation}\label{diff-inteq-uR}
 \left.\frac{\partial u_0(\mu)}{\partial\alpha}\right|_{\alpha=0}=
 -2\pi iT+ \frac1{2\pi}\int\limits_{\mathbb{R}}K^{(\varepsilon)}(\mu,\lambda)
 \left.\frac{\partial u_0(\lambda)}{\partial\alpha}\right|_{\alpha=0}\,
 d\lambda,
 \end{equation}
and hence,
 \begin{equation}\label{u-expans}
 u_0(\lambda)=\varepsilon(\lambda)-4\pi^2i\alpha T\rho_t(\lambda)+O(\alpha^2),
 \end{equation}
where $\rho_t(\lambda)$ is the total density  satisfying the integral equation
\eqref{inteq-rho}. Equation \eqref{inteq-rho} can be solved in terms of the
resolvent $R(\lambda,\mu)$ defined by
$\left(I-\frac1{2\pi}K^{(\varepsilon)}\right)(I+R)=I$:
 \begin{equation}\label{inteq-Z-sol}
 2\pi\rho_t(\mu) =1+\int\limits_{\mathbb{R}}R(\mu,\lambda)\, d\lambda.
 \end{equation}
Substituting \eqref{u-expans} into \eqref{sol-Z} we obtain
 \begin{equation}\label{sol-Zbeta}
 z_0(\lambda)=-2\pi\alpha\rho_t(\lambda)\vartheta(\lambda)+O(\alpha^2),
 \end{equation}
and thus, due to \eqref{p-decay},
 \begin{equation}\label{deriv-pR}
 \left.\frac{\partial p_0}{\partial\alpha}\right|_{\alpha=0}=-
 2\pi i\int\limits_{\mathbb{R}}\rho_t(\lambda)\vartheta(\lambda)\,d\lambda=-
 2\pi i\int\limits_{\mathbb{R}}\rho_p(\lambda)\,d\lambda.
 \end{equation}
Hence,
 \begin{equation}\label{Asy-cor-fun1}
 \langle j(x)j(0)\rangle_T^{(0)} =\left(\int\limits_{\mathbb{R}}\rho_p(\lambda)\,d\lambda
 \right)^2 \cdot \left. \;  B[u_0]\right|_{\alpha=0},
 \end{equation}
and it remains to apply Lemma \ref{lemma constante etat contour R} in order to
conclude that $B[u_0]=1$ at $\alpha=0$. \qed

 \begin{lemma}
\label{lemma constante etat contour R}
 $B[u_0]=1$ at $\alpha=0$.\end{lemma}
{\sl Proof.} Since $z_0(\lambda)\to 0$ at $\alpha\to 0$, we conclude
that
 \begin{equation}\label{exp-to0}
 \exp\biggl(\dis\int_{\mathbb{R}}
 \frac{z_0(\lambda)z_0(\mu)}{(\lambda-\mu_+)^2}\,d\lambda\,d\mu-C_0[z_0,{\mathbb{R} }]
 \biggr) =1,\quad\mbox{at}\quad\alpha=0.
 \end{equation}
For the calculation of the limit of the remaining part of equation \eqref{C2-W},
we use the obvious properties of the Cauchy transform
$L_{\mathbb{R}}[z_0](\omega)$:
 \begin{equation}\label{Cauchy-Tr-Pr}
 \begin{array}{l}
 {\dis L_{\mathbb{R}}[z_0](\omega+i0)-L_{\mathbb{R}}[z_0](\omega-i0)=2\pi iz_0(\omega),}\num
  {\dis L_{\mathbb{R}}[z_0](\omega+ic)-L_{\mathbb{R}}[z_0](\omega-ic)=
 \frac1{T}(u_0(\lambda)-\varepsilon(\lambda)+2\pi i\alpha T)=
 2\pi i\alpha(1-2\pi\rho_t(\lambda))+O(\alpha^2).}
 \end{array}
 \end{equation}
The last property follows from \eqref{def-u}. Using \eqref{Cauchy-Tr-Pr} we
find at $\alpha\to 0$
 \begin{equation}\label{pref-1}
 \frac{{e^{2\pi i\alpha}}-1}{e^{ L_{\mathbb{R}}[z_0](\theta_1+i c)}  -e^{2\pi i\alpha+L_{\mathbb{R}}[z_0](\theta_1-i c)}}=
  \frac{\left({e^{2\pi i\alpha}}-1\right)e^{-2\pi i\alpha-L_{\mathbb{R}}[z_0](\theta_1-i c)}}
 {e^{ \bigl(u_0(\theta_1)-\varepsilon(\theta_1)\bigr)/T}  -1}\to -\frac1{2\pi\rho_t(\theta_1)},
  \end{equation}
and
 \begin{equation}\label{pref-2}
 \frac{{e^{2\pi i\alpha}}-1}{e^{ -L_{\mathbb{R}}[z_0](\theta_2-i c)}  -e^{2\pi i\alpha-L_{\mathbb{R}}[z_0](\theta_2+i c)}}=
  \frac{\left({e^{2\pi i\alpha}}-1\right)e^{-2\pi i\alpha+L_{\mathbb{R}}[z_0](\theta_2-i c)}}
 {e^{ \bigl(u_0(\theta_2)-\varepsilon(\theta_2)\bigr)/T}-1  }\to -\frac1{2\pi\rho_t(\theta_2)}.
  \end{equation}

Finally we should compute the ratio of determinants in the second
line of \eqref{C2-W}. Consider, for example the determinant of the
operator $I+\frac1{2\pi i}U^{(1)}$.  Due to the factor $e^{L_{\mathbb{R}}[z_0](w)}$, the
kernel $\hat U^{(1)}(w,w')$ has a cut on the real axis. Hence the
action of the integral operator $\hat U^{(1)}$ can be reduced to an
action on the real axis with the kernel
 \begin{equation}\label{7-G-qq}
 \left.\hat U^{(1)}\right|_{\Gamma(\mathbb{R})} \to \left.\tilde U^{(1)}\right|_{\mathbb{R}}=
  \left.(\hat U^{(1)}_--\hat U^{(1)}_+)\right|_{\mathbb{R}},
  \end{equation}
where $\hat U^{(1)}_\pm$ are the limiting values of $\hat U^{(1)}$
from the upper (lower) half-planes. Using equations
\eqref{Cauchy-Tr-Pr} we obtain
 \begin{equation}\label{7-UU}
 \det_{\Gamma(\mathbb{R})} \Bigl[I+\frac1{2\pi i}\hat U^{(1)}[z_0]\Bigr]=
  \det_{\mathbb{R}}\Bigl[I+\frac1{2\pi i}\tilde U^{(1)}[z_0] \Bigr]  \; ,
 \end{equation}
where the operator in the r.h.s. acts on $\mathbb{R}$ and its kernel reads
 \begin{equation}\label{7-tUl}
 \tilde  U^{(1)}(w,w')=
 -e^{L_{\mathbb{R}}[z_0](w-i 0)-L_{\mathbb{R}}[z_0](w-i c)-2\pi i\alpha}
  \frac{\left(1-e^{2\pi iz_0(w)}\right)(K_\alpha(w-w')-K_\alpha(\theta_1-w'))}
 {e^{ \bigl(u_0(w)-\varepsilon(w)\bigr)/T}-1  }\;.
 \end{equation}
Setting now  $\alpha=0$ and using \eqref{u-expans}, \eqref{sol-Zbeta} we obtain
 \begin{equation}\label{7-b0}
 \left. \det_{\Gamma(\mathbb{R})} \Bigl[ I+\frac1{2\pi i}\hat U^{(1)}[z_0]\Bigr]\right|_{\alpha=0}=
 \det_{\mathbb{R}}\Bigl[I-\frac1{2\pi }(K^{(\varepsilon)}(w,w')-K^{(\varepsilon)}(\theta_1,w')) \Bigr].
 \end{equation}
Thus, in terms of the resolvent  $R(\lambda,\mu)$ the ratio of
determinants can be presented as
 \begin{multline}\label{7-rat-det1}
 \frac{  \det_{\mathbb{R}}[ I-\frac1{2\pi }(K^{(\varepsilon)}(w,w')-K^{(\varepsilon)}(\theta_1,w')) ]  }
 { \det_{\mathbb{R}}[I-\frac1{2\pi }K^{(\varepsilon)}(w,w') ]  }
 =\det_{\mathbb{R}}[I+R(\theta_1,w') ]\num
 =1+\int\limits_{\mathbb{R}}R(\theta_1,w)\,dw=2\pi\rho_t(\theta_1),
 \end{multline}
where we have used \eqref{inteq-Z-sol}. Similarly one has
 \begin{equation}\label{7-rat-det2}
 \left.\frac{ \det_{\Gamma(\mathbb{R}) }[I+\frac1{2\pi i}\hat U^{(2)}[z_0] ] }
 {\det_{\mathbb{R}}[ I-\frac1{2\pi }K^{(u)}(w,w') ]  }\right|_{\alpha=0}
 =2\pi\rho_t(\theta_2),
 \end{equation}
and we arrive at the statement of Lemma \ref{lemma constante etat contour R}.\qed

%%%%%%%%%%%%%%%%%%%%%%%%%%%%%%%%%%%%%%%%%%%%%%%%%%%%%%%%%%%%%%%%
\subsection{The corrections\label{aC}}

It is clear that for other terms in the expansion \eqref{fin-answ}
$u_i(\lambda)\ne\varepsilon(\lambda)$ even at $\alpha=0$. This yields $p_i\ne
0$ at $\alpha=0$ and $\hat{\cal C}\ne\mathbb{R}$. Therefore after taking the
second $\alpha$-derivative and setting $\alpha=0$ the corresponding terms of
the expansion  contain exponential factors $e^{-xp_i}$. Our numerical computations together with the low-temperature limit
support that the  real part of the $p_i$'s are indeed positive and, hence, the corresponding terms in the asymptotic expansion are sub-leading. In
particular, in the low-temperature limit, equation \eqref{inteq-u-main1} can be solved for
all possible choices of subsets $\boldsymbol{j},\boldsymbol{k}$. Respectively,
one can find all $p_i$. It follows from this analysis, that the main sub-leading
contribution to the asymptotic behavior of $\moy{e^{2\pi i\alpha {{\cal
Q}_x}}}$ originates from the contours $\hat{\cal C}_i$, where $\{\hqpm\}_i$
are $\gamma$-deformations of the poles $\{\Rpm\}_i$ lying nearest to the real
axis. We conjecture that this property holds for the finite temperature as
well.

Finally, we would like to give an interpretation of the results obtained from
the viewpoint of the quantum transfer matrix approach \cite{Klu04}. As we
have mentioned already this method was developed for the description of quantum
spin chains, but it was shown in \cite{SeeBGK07} that in a special continuous
limit it can be applied to the description of thermodynamics in the
one-dimensional Bose gas. The central object of this method is a function
$\frak{a}_0(\lambda)$ constructed via certain solution of Bethe equations and
satisfying a non-linear integral equation. Knowing this function allows one to
calculate the maximal eigenvalue $\Lambda_0$ of the quantum transfer matrix
$T_q$. In the continuous limit mentioned above, one can establish the following
correspondence
 \begin{equation}\label{Correspond}
 \begin{array}{l}
 {\dis \frak{a}_0(\lambda)\mapsto e^{-\varepsilon(\lambda)/T},}\num
 {\dis \Lambda_0 \mapsto \frac 1{2\pi}\int\limits_{\mathbb{R}}
 \log\left(1+e^{-\varepsilon(\lambda)/T}\right)\,d\lambda.}
 \end{array}
 \end{equation}
The sub-leading eigenvalues $\Lambda_i$ of $T_q$ can be expressed in terms of
functions $\frak{a}_i(\lambda)$ satisfying the same type of the integral
equation, but on deformed integration contours surrounding some zeros of
$1+\frak{a}_i (\lambda)$. It is easy to show that, in the
continuous limit, these contours  go into the contours  $\hat{\cal C}_i$. Then
one should have the following correspondence
 \begin{equation}\label{Correspond-i}
 \begin{array}{l}
 {\dis \frak{a}_i(\lambda)\mapsto e^{-u_i(\lambda)/T},}\num
 {\dis \Lambda_i \mapsto \frac 1{2\pi}\int\limits_{\hat{\cal C}_i}
 \log\left(1+e^{-u_i(\lambda)/T}\right)\,d\lambda.}
 \end{array}
 \end{equation}
Taking into account \eqref{p-decay} we obtain the following mapping
 \begin{equation}\label{map}
 \left(\frac{\Lambda_i(h_\alpha)}{\Lambda_0(h)}\right)^m\mapsto
 e^{-xp_i},
 \end{equation}
where $m$ is the lattice distance that scales to the distance $x$ in the
continuous limit. In \eqref{map}, we have also stressed that the eigenvalue
$\Lambda_0(h)$ corresponds to the quantum transfer matrix $T_q(h)$ with the
chemical potential $h$, while the eigenvalues $\Lambda_i(h_\alpha)$ correspond
to the quantum transfer matrix $T_q(h_\alpha)$ with the shifted chemical
potential $h_\alpha$. Thus, the rates of the exponential decay in the
asymptotic behavior of $\moy{e^{\alpha {{\cal Q}_x}}}$ appear to be nothing
else but the ratios of the eigenvalues of the quantum transfer matrix (in the
continuous limit).

\section*{Conclusion}

%In the present paper we have considered the finite temperature correlation
%function of the densities in the QNLS model. We have expressed its long-distance asymptotic expansion in terms of solutions
%to non-linear integral equations of Yang-Yang type. These equations are closely
%related to the ones appearing in the quantum transfer matrix
%approach. In the framework of our method, the non-linear integral
%equations arise as the result of summation of the generalized
%Lagrange series. Thus, our results can be considered as a link
%between two methods.{\bf what are these two methods????}

The main goal of this article was to derive the long distance asymptotic
behavior of temperature correlation functions of one-dimensional bosons in
the framework of the algebraic Bethe ansatz approach. Considering
the example of the two-point function of densities we showed that this
asymptotic behavior can be expressed in terms of solutions of  the TBA
equations. This result is quite expected from the viewpoint of the QTM
method.

Comparing to the QTM approach,  we should say that our derivation is certainly
more involved technically.
However in our opinion these two methods are complementary, in particular speaking about their range of applications. Our starting point is the master
equation, which is a multiple integral representation for various
correlation functions, including the dynamical ones. Recall that such type
of representations is known now for an arbitrary integrable model
possessing the six-vertex $R$-matrix \cite{KitKMST07}. The expansion of the
master equation into the series allows us to take the thermodynamic limit.
Then in the asymptotic regime this series turns into the generalized
Lagrange series, which can be summed up explicitly in terms of solutions
to some integral equations. The last ones appear to be just the equations
providing the dressing of bare variables originally entering the
representation for the correlation function. We have demonstrated
such a mechanism already in the case of zero temperature correlations
\cite{KitKMST09a}. We have now shown how it works in the case of thermal
correlation functions.

Our derivation however is not free of several assumptions of technical kind. For
example, such questions as the convergence of the series
\eqref{3-fin-answ-TD} and \eqref{first-ser} formally remain unsolved (although these series are absolutely convergent at the free fermion point). We
think nevertheless that the results obtained under these assumptions are themselves  strong arguments in favor of  their validity. As an additional
test for our results
we will consider the low-temperature limit of our asymptotic expansion in
a forthcoming publication.
In this limit, the non-linear integral equations determining the
correlation functions in the asymptotic regime can be solved explicitly in
terms of  dressed physical quantities such as the energy, momentum, and charge.  In particular,
we will show \cite{KozMS10c} that the low-temperature limit of
our results reproduces the conformal field theory predictions for the
exponential decay of the correlation functions. We are however able to go well
beyond these predictions as our results not only hold for all ranges
of temperatures but also provide explicit formulae for the
corresponding amplitudes. Moreover, it will be shown that the dependence of these amplitudes in terms of powers of the temperature in the low-temperature limit can be computed explicitly, in complete agreement with the CFT predictions \cite{KozMS10c}.

%%%%%%%%%%%%%%%%%%%%%%%%%%%%%%%%%%%%

\section*{Acknowledgements}

We are very grateful to N. Kitanine and V. Terras for useful and numerous discussions.
J. M. M. and  N. S.  are supported by CNRS.  We also acknowledge the
support from the GDRI-471 of CNRS "French-Russian network in
Theoretical and Mathematical  Physics" and RFBR-CNRS-09-01-93106L-a.
N. S. is also supported by the Program of RAS Mathematical Methods
of the Nonlinear Dynamics, RFBR-08-01-00501a, RFBR-09-01-12150ofi-m.
K. K. K. is supported by the EU Marie-Curie Excellence Grant
MEXT-CT-2006-042695. N. S. and K. K. K. would like to thank the
Theoretical Physics group of the Laboratory of Physics at ENS Lyon
for hospitality, which makes this collaboration possible.

%%%%%%%%%%%%%%%%%%%%%%%%%%%%%%%%%%%%%%%%%%%%%%%%%%%%%%%%%%%%%%%%%%%%%%%%%%%%%%%%%%%%%%%%
%%%%%%%%%%%%%%%%%%%%%%%%%%%%%%%%%%%%%%%%%%%%%%%%%%%%%%%%%%%%%%%%%%%%%%%%%%%%%%%%%%%%%%%%
%%%%%%%%%%%%%%%%%%%%%%%%%%%%%%%%%%%%%%%%%%%%%%%%%%%%%%%%%%%%%%%%%%%%%%%%%%%%%%%%%%%%%%%%
%%%%%%%%%%%%%%%%%%%%%%%%%%%%%%%%%%%%%%%%%%%%%%%%%%%%%%%%%%%%%%%%%%%%%%%%%%%%%%%%%%%%%%%%
%%%%%%%%%%%%%%%%%%%%%%%%%%%%%%%%%%%%%%%%%%%%%%%%%%%%%%%%%%%%%%%%%%%%%%%%%%%%%%%%%%%%%%%%
%%%%%%%%%%%%%%%%%%%%%%%%%%%%%%%%%%%%%%%%%%%%%%%%%%%%%%%%%%%%%%%%%%%%%%%%%%%%%%%%%%%%%%%%

\appendix

\section{Numerical analysis of the correlation lengths}
\label{Appendix Numerics}

The inverse correlation lengths $\Re(p_i)$ are defined in terms of the integral
\eqref{p-decay} involving the solution $u_i$ to the TBA non-linear integral equation  \eqref{inteq-u-main1}.
Hence, the $p_i$ depend on the coupling constant $c$, the temperature $T$, and the chemical
potential $h$. Apart from these parameters, they also depend on the choice of the system of roots
$\{\hat s^\pm\}_i$ and it is to this dependence that the subscript $i$ refers to. In this appendix, we gather four plots resulting from our numerical
computations of several correlation lengths (the real part of the $p_i$) as functions of their
parameters.

\begin{figure}[!h]
\begin{tabular}{ll}
\includegraphics[width=7cm]{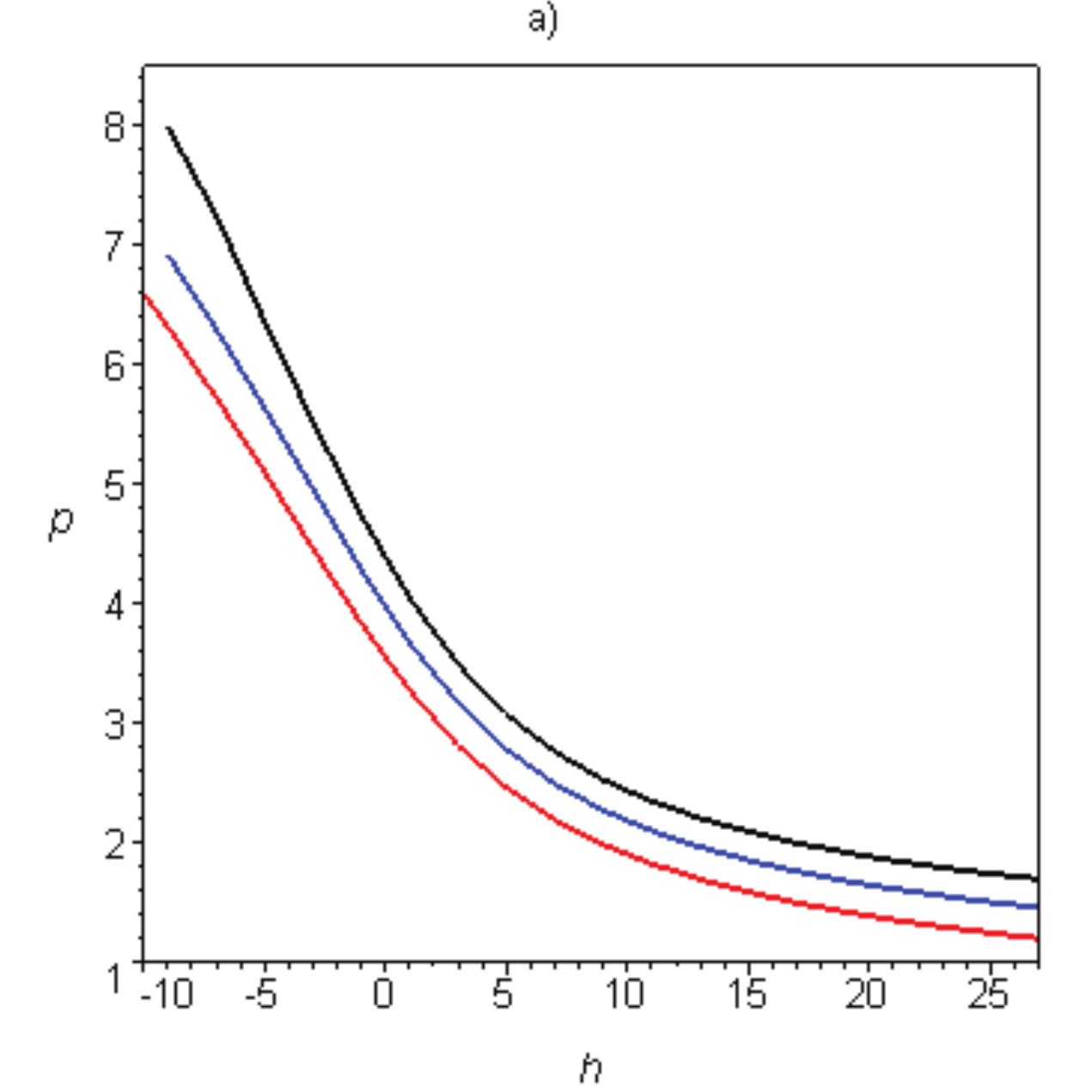} & 
\includegraphics[width=7cm]{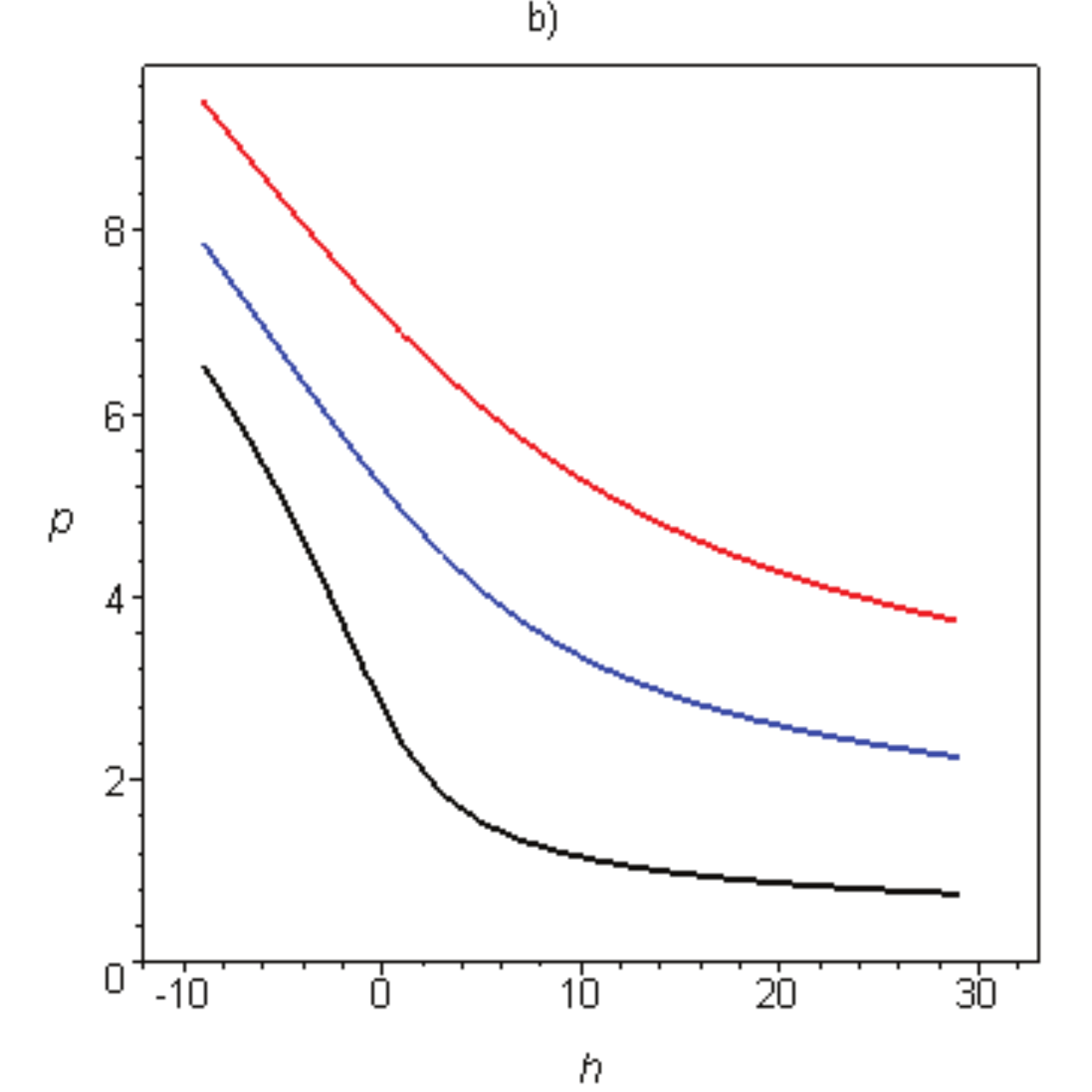}
\end{tabular} \caption{\small
The dependence $p=p(h)$ in a) for a fixed temperature $T=2$ and different  coupling constants $c=7$ (black),  $c=10$ (blue) and
$c=\infty$ (red). In b)
the coupling constant is fixed at $c=10$ and the temperature is  set to
 $T=1$ (black), $T=3$ (blue) and $T=5$ (red).}
 \label{fi:n1}
\end{figure}
%
%\vspace{5mm}
%
%\begin{figure}[h] % FIG 2

\begin{figure}[!ht]
\begin{tabular}{ll}
\includegraphics[width=7cm]{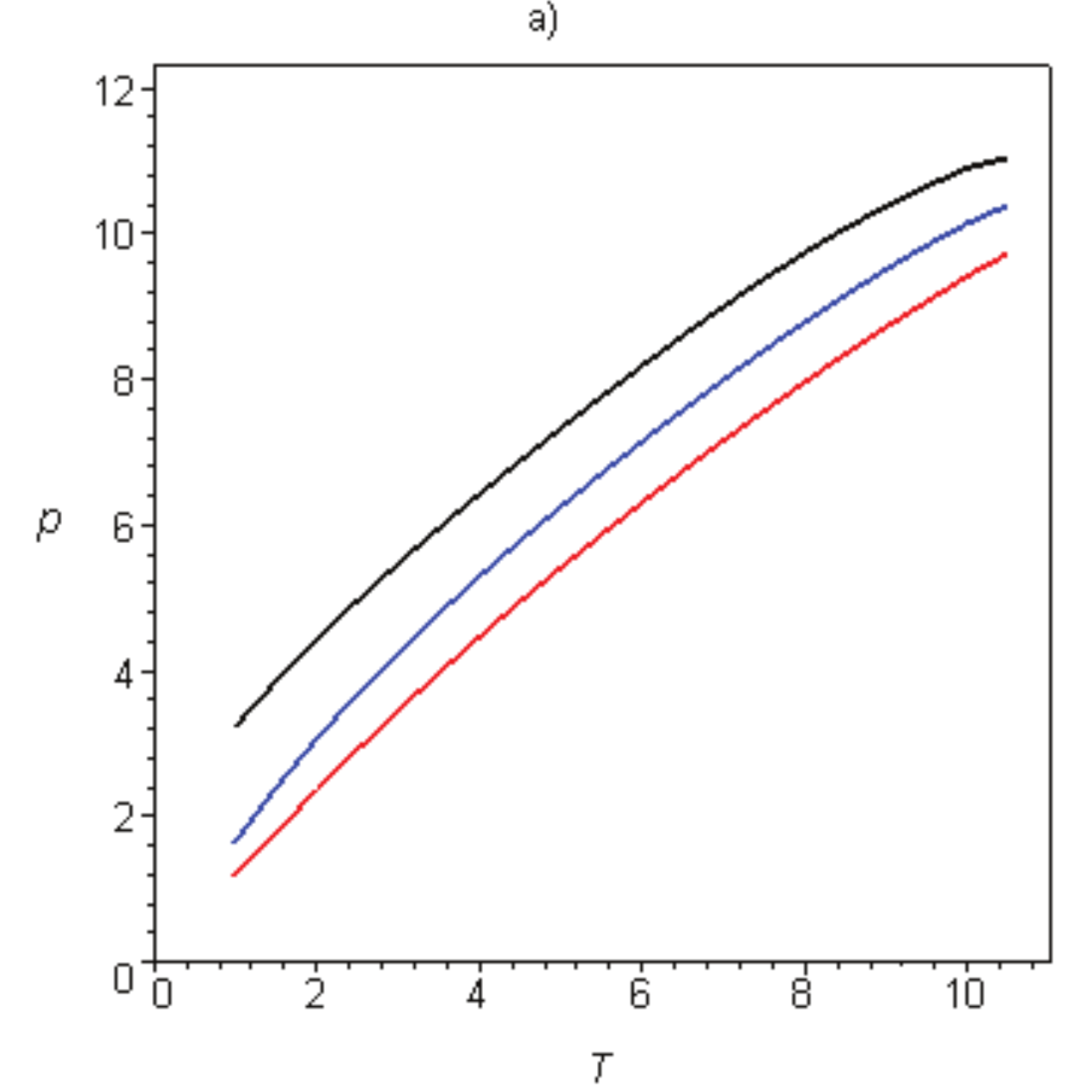} &  
\includegraphics[width=7cm]{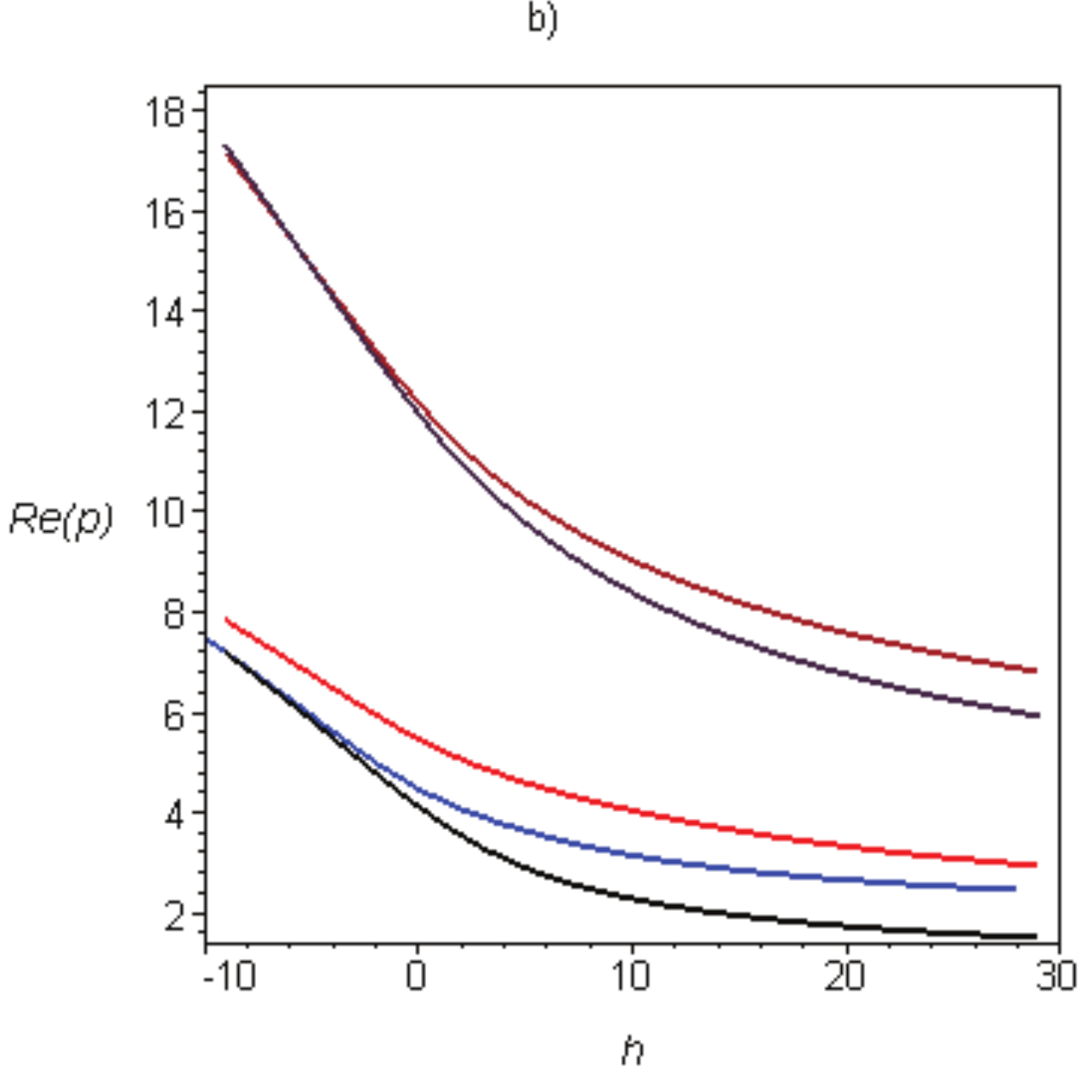}
\end{tabular} \caption{\small
In a), the dependence $p=p(T)$ is presented for the coupling constant $c=10$ and at different values of the  chemical potential
$h=-1$ (black), $h=4$ (blue) and  $h=9$ (red). In b) we have represented $\Re(p)$ as a function of the chemical potential $h$ for the coupling constant  fixed to
$c=10$ and  the temperature to $T=2$. The three cases correspond to  one pair of roots
$\hat s^\pm$ in various positions : (i) $\hat s^+$ is the nearest root in the first
quadrant, $\hat s^-$ is the nearest root in the fourth quadrant
(black); (ii) $\hat s^+$ is the nearest root in the first quadrant,
$\hat s^-$ is the nearest root in the third quadrant (blue); (iii)
$\hat  s^+$ is the next after the nearest root in the first
quadrant, $\hat s^-$ is the nearest root in the fourth quadrant
(red). Two cases with two pairs of roots $\hat s^\pm$: (iv) $\hat
s^\pm_1$ are the nearest roots in the first and the fourth
quadrants, $\hat s^\pm_2$ are the next after the nearest roots in
the first and the fourth quadrants (violet); (v) $\hat s^\pm_1$ are
the nearest roots in the first and the fourth quadrants, $\hat
s^\pm_2$ are the next after the nearest roots in the first and the
third quadrants (brown).}
\label{fi:n2}
\end{figure}

The situation considered in the first
three figures (Fig.~\ref{fi:n1} a) and b),  Fig.~\ref{fi:n2} a)) corresponds to
the correlation length determined in terms of
the solution to the non-linear integral equation with one root $\hat s^+$ in the first
quadrant of $\mathbb{C}$ and one root $\hat s^-$ in the fourth
quadrant. These roots are built as $\gamma$ deformations of the
poles of the Fermi weight lying closest to the real axis and located
in these quadrants. For this case, the corresponding $p$ is real.

On Fig.~\ref{fi:n1} a) we have represented  $p$ as a
function of chemical potential for a fixed temperature and
several values of the coupling constant $c$. Similarly, in Fig.~\ref{fi:n1} b)
we have also represented  $p$  as a function of
chemical potential but for a fixed value of the coupling
constant $c$ and different values of the temperature.
On Fig.~\ref{fi:n2} a) we have represented   $p$  as a function of
temperature for a fixed value of the coupling constant and several
values of the chemical potential.
For all these quantities, the numerical values are given in dimensionless units after proper
rescaling of the TBA non-linear integral equation \eqref{inteq-u-main1}.

The last plot Fig.~\ref{fi:n2} b)  represents different correlation lengths $\Re(p)$ as a function of the chemical potential and for a fixed value of the coupling constant and the temperature. There, we have considered several possible configurations of the system of roots $\{\hat{s}^{\pm}\}$ for which $p$ are complex numbers.

%%%%%%%%%%%%%%%%%%%%%%%%%%%%%%%%%%%%%%%%%%%%%%%%%%%%%%%%%%%%%%%%%%%%%%%%%%%%%%%%%%%%%%%%%%%%%%%%%%%%%%%%%%%%%%%%%%%%%%%%%%%%%%%%%%%%%%%%%%%%%%%%%%%%
%%%%%%%%%%%%%%%%%%%%%%%%%%%%%%%%%%%%%%%%%%%%%%%%%%%%%%%%%%%%%%%%%%%%%%%%%%%%%%%%%%%%%%%%%%%%%%%%%%%%%%%%%%%%%%%%%%%%%%%%%%%%%%%%%%%%%%%%%%%%%%%%%%%%

%%%%%%%%%%%%%%%%%%%%%%%%%%%%%%%%%%%%%%%%%%%%%%%%%%%%%%%%%%%%%%%%%%%%%%%%%%%%%%%%%%%%%%%%%%%%%%%%%%%%%%%%%%%%%%%%%%%%%%%%%%%%%%%%%%%%%%%%%%%%%%%%%%%%
%%%%%%%%%%%%%%%%%%%%%%%%%%%%%%%%%%%%%%%%%%%%%%%%%%%%%%%%%%%%%%%%%%%%%%%%%%%%%%%%%%%%%%%%%%%%%%%%%%%%%%%%%%%%%%%%%%%%%%%%%%%%%%%%%%%%%%%%%%%%%%%%%%%%

\section{Linearized form of $W_{m}$}
\label{AEebQ}

The explicit expression for the function $W_{m}$ is given by
\eqref{2-W}--\eqref{2-U2}. It contains Fredholm determinants whose
kernels depend on the products \eqref{1-Vpm}. At
$\{z_{k}\}=\{\lambda_{k}+\epsilon_{k}\}$ we have, to the linear
order in each of the $\epsilon_a$'s
 \begin{equation}\label{9-linPa1}
 V^{(m+|\boldsymbol{j}|)}_\sigma(w)
 \hookrightarrow
   \prod_{a=1}^{|\boldsymbol{j}|} \frac{ w-\Rm_{k_a} + ic\sigma }{   w-\Rp_{j_a} + ic\sigma   }
     \; \cdot \;  \exp\left\{\sum_{a=1}^m\ \epsilon_{a}  g^{(1)}_\sigma(\lambda_a)\right\},
 \qquad \sigma=0,\pm,
 \end{equation}
where
 \begin{equation}\label{g1-fun}
 g^{(1)}_\sigma(\lambda)\equiv g^{(1)}_\sigma(w,\lambda)=\frac1{w-\lambda+i\sigma
 c},\qquad \sigma=0,\pm.
 \end{equation}

Apart form the products of the type \eqref{9-linPa1}, the function
$W_{m}$ also contains the  product of the functions
$V^{(m+|\boldsymbol{j}|)}_-$ (see \eqref{2-W}). Setting
$\{z_{k}\}=\{\lambda_{k}+\epsilon_{k}\}$ we obtain
 \begin{multline}\label{9-W0}
\prod_{a=1}^{m}\frac{V^{(m+|\boldsymbol{j}|)}_-(
z_a)}{V^{(m+|\boldsymbol{j}|)}_-( \lambda_a)}  \cdot
\prod_{a=1}^{|\boldsymbol{j}|} \frac{ V^{(m+|\boldsymbol{j}|)}_-(
\Rp_{j_a} )}{V^{(m+|\boldsymbol{j}|)}_-(  \Rm_{k_a}  )}
  \hookrightarrow
 \\
 \times \exp\left\{ \sum_{a=1}^{m} \sum_{b=1}^{ |\boldsymbol{j}| }
 \epsilon_a\Bigl[  g^{(1)}_+(  \Rp_{j_b},\lambda_a) + g^{(1)}_-(\Rp_{j_b},\lambda_a)
 -g^{(1)}_-(\Rm_{k_b},\lambda_a) - g^{(1)}_+( \Rm_{k_b},\lambda_a)    \Bigr]   \right\} \\
 \times  \prod_{a,b=1}^{|\boldsymbol{j}|} \frac{ (\Rm_{k_a}
 -\Rp_{j_b} -ic)(\Rp_{j_a} -\Rm_{k_b} -ic)   }{ (\Rm_{k_a} -\Rm_{k_b}-ic)
 (\Rp_{j_a} -\Rp_{j_b} -ic) } \cdot \exp\left\{\sum_{a=1}^m\sum_{b=1}^m
 \epsilon_{a}\,\epsilon_{b}\,
 g^{(2)}(\lambda_{a},\lambda_{b})\right\} ,
 \end{multline}
where
 \begin{equation}\label{g2-fun}
 g^{(2)}(\lambda,\mu)=- \frac1{(\lambda-\mu-i c)^2}.
 \end{equation}

The  expressions \eqref{9-linPa1}, \eqref{9-W0} should be
substituted into \eqref{2-W}, what gives us the linearized form
$\widetilde{\mathcal{W}}_{\boldsymbol{j};\boldsymbol{k}}$
\eqref{W-lin}.

The summation of the Lagrange series  leads to the replacement of
the sums by integrals over the original contour
$\Gamma_{\boldsymbol{j};\boldsymbol{k}}$
 \begin{equation}\label{C-sum-int}
 \begin{array}{l}
 {\dis \sum_{k=1}^m\epsilon_{k}\,g^{(1)}_\sigma(w,\lambda_{k}) \hookrightarrow
 \int\limits_{ \Gamma_{\boldsymbol{j};\boldsymbol{k}} }g^{(1)}_\sigma(w,\lambda)z_{\boldsymbol{j};\boldsymbol{k}}(\lambda)\,d\lambda
 =-L_{ \Gamma_{\boldsymbol{j};\boldsymbol{k}} }[z_{\boldsymbol{j};\boldsymbol{k}}](w+i\sigma c),}\num
 {\dis \sum_{k=1}^m\sum_{k'=1}^m \epsilon_{k}\,\epsilon_{k'}\,
 g^{(2)}(\lambda_{k},\lambda_{k'})\hookrightarrow \int\limits_{ \Gamma_{\boldsymbol{j};\boldsymbol{k}} }g^{(2)}(\lambda,\mu)
 z_{\boldsymbol{j};\boldsymbol{k}}(\lambda)z_{\boldsymbol{j};\boldsymbol{k}}(\mu)\,d\lambda\,d\mu
 =-C_0[z_{\boldsymbol{j};\boldsymbol{k}}, \Gamma_{\boldsymbol{j};\boldsymbol{k}} ],}
 \end{array}
 \end{equation}
where $z_{\boldsymbol{j};\boldsymbol{k}}(\lambda)$ solves the
integral equation \eqref{8-Int-eq0}, $L_{
\Gamma_{\boldsymbol{j};\boldsymbol{k}} }$ is the operator of Cauchy
transform \eqref{def-CauT}, and for any contour $\mathcal{L}$
 \begin{equation}\label{7-C0}
 C_0[z_{\boldsymbol{j};\boldsymbol{k}}, \mathcal{L}  ]=\int\limits_{ \mathcal{L}  }
 \frac{z_{\boldsymbol{j};\boldsymbol{k}}(\lambda)z_{\boldsymbol{j};\boldsymbol{k}}(\mu)}{(\lambda-\mu-i c)^2}\,d\lambda\,
 d\mu\, .
 \end{equation}

Substituting these formulae into \eqref{2-W} and  moving the contours from
$\Gamma_{\boldsymbol{j};\boldsymbol{k}} $ to $ \hat{\cal
C}_{\boldsymbol{j};\boldsymbol{k}} $ we get that all the explicit dependence on
the poles $\Rpm_{\boldsymbol{j}/\boldsymbol{k}}$ cancels out and
\begin{multline}
 \widetilde{\cal W}_{\boldsymbol{j};\boldsymbol{k}}\left( \int_{ \Gamma_{\boldsymbol{j};\boldsymbol{k}} }
 g^{(1)}_\sigma(\lambda)\, z_{\boldsymbol{j};\boldsymbol{k}}(\lambda) \,d \lambda ;
 \int_{ \Gamma_{\boldsymbol{j};\boldsymbol{k}} } g^{(2)}(\lambda,\mu)\, z_{\boldsymbol{j};\boldsymbol{k}}(\lambda)
 z_{\boldsymbol{j};\boldsymbol{k}}(\mu) \,d \lambda\,d\mu\right) \\
 = \widetilde{\cal W}\left( \int_{ \hat{\cal C}_{\boldsymbol{j};\boldsymbol{k}} } g^{(1)}_\sigma(\lambda)
 z_{\boldsymbol{j};\boldsymbol{k}}(\lambda) \,d \lambda ;
 \int_{ \hat{\cal C}_{\boldsymbol{j};\boldsymbol{k}} } g^{(2)}(\lambda,\mu)
  z_{\boldsymbol{j};\boldsymbol{k}}(\lambda)z_{\boldsymbol{j};\boldsymbol{k}}(\mu) \,d
  \lambda\,d\mu\right),
\end{multline}
where
 \begin{multline}\label{C2-WW}
 \widetilde{\cal W}\left( \left\{ \int_{ \hat{\cal C}_{\boldsymbol{j};\boldsymbol{k}} }
 g^{(1)}(\lambda)\, z_{\boldsymbol{j};\boldsymbol{k}}(\lambda) \,d \omega \right\};
 \int_{ \hat{\cal C}_{\boldsymbol{j};\boldsymbol{k}} } g^{(2)}(\lambda,\mu)
 z_{\boldsymbol{j};\boldsymbol{k}}(\lambda)z_{\boldsymbol{j};\boldsymbol{k}}(\mu) \,d \lambda\,d\mu\right)
 \numa{35}
 =\frac{e^{-C_0[z_{\boldsymbol{j};\boldsymbol{k}}, \hat{\cal C}_{\boldsymbol{j};\boldsymbol{k}}  ]}({e^{2\pi i\alpha}}-1)^2
 \det\left(I+\frac1{2\pi i}\hat U^{(1)}[z_{\boldsymbol{j};\boldsymbol{k}}]\right)
 \det\left(I+\frac1{2\pi i} \hat U^{(2)}[z_{\boldsymbol{j};\boldsymbol{k}}]\right)}
 {[e^{ L_{ \hat{\cal C}_{\boldsymbol{j};\boldsymbol{k}} }[z_{\boldsymbol{j};\boldsymbol{k}}](\theta_1+i c)}
 -e^{2\pi i\alpha+L_{ \hat{\cal C}_{\boldsymbol{j};\boldsymbol{k}} }[z_{\boldsymbol{j};\boldsymbol{k}}](\theta_1-i c)}]
 [e^{ -L_{ \hat{\cal C}_{\boldsymbol{j};\boldsymbol{k}}  }[z_{\boldsymbol{j};\boldsymbol{k}}](\theta_2-i c)}
 -e^{2\pi i\alpha-L_{ \hat{\cal C}_{\boldsymbol{j};\boldsymbol{k}} }[z_{\boldsymbol{j};\boldsymbol{k}}](\theta_2+i c)}]},
 \end{multline}
and the kernels of the integral operators $\hat
U^{(1)}(w,w',[z_{\boldsymbol{j};\boldsymbol{k}}])$ and $\hat
U^{(2)}(w,w',[z_{\boldsymbol{j};\boldsymbol{k}}])$ have the form
 \begin{equation}\label{C2-U1}
 \hat U^{(1)}(w,w',[z_{\boldsymbol{j};\boldsymbol{k}}])=-e^{ L_{ \hat{\cal C}_{\boldsymbol{j};\boldsymbol{k}} }[z_{\boldsymbol{j};\boldsymbol{k}}](w)}\cdot
 \frac{K_\alpha(w-w')-K_\alpha(\theta_1-w')}{
 e^{ L_{ \hat{\cal C}_{\boldsymbol{j};\boldsymbol{k}} }[z_{\boldsymbol{j};\boldsymbol{k}}](w+i c)}  -e^{2\pi i\alpha+L_{ \hat{\cal C}_{\boldsymbol{j};\boldsymbol{k}} }[z_{\boldsymbol{j};\boldsymbol{k}}](w-i c)}},
 \end{equation}
 \begin{equation}\label{C2-U2}
 \hat U^{(2)}(w,w',[z_{\boldsymbol{j};\boldsymbol{k}}])=e^{ -L_{ \hat{\cal C}_{\boldsymbol{j};\boldsymbol{k}} }[z_{\boldsymbol{j};\boldsymbol{k}}](w')}\cdot
 \frac{K_\alpha(w-w')-K_\alpha(w-\theta_2)}{
 e^{ -L_{ \hat{\cal C}_{\boldsymbol{j};\boldsymbol{k}} }[z_{\boldsymbol{j};\boldsymbol{k}}](w'-i c)}
 -e^{2\pi i\alpha-L_{ \hat{\cal C}_{\boldsymbol{j};\boldsymbol{k}} }[z_{\boldsymbol{j};\boldsymbol{k}}](w'+i c)}}.
 \end{equation}
Both integral operators act on a counterclockwise oriented closed
contour surrounding the contour $\hat{\cal
C}_{\boldsymbol{j};\boldsymbol{k}}$.

\section{Deformation of the contours ${\cal C}_{\boldsymbol{j};\boldsymbol{k}}$ \label{PrF}}
\label{appendix Deformation of integrals}

In this section we calculate the difference ${\cal A}_{{\cal
C}_{\boldsymbol{j};\boldsymbol{k}}}([g],[\nu])- {\cal A}_{\Gamma
_{\boldsymbol{j};\boldsymbol{k}}}([g],[\nu])$ (see \eqref{A-rep}). We shall
consider some  fixed contours ${\cal C}_{\boldsymbol{j};\boldsymbol{k}}$ and
$\Gamma_{\boldsymbol{j};\boldsymbol{k}}$. Hence, for brevity, we
omit the subscripts ${\boldsymbol{j};\boldsymbol{k}}$ in the following. Moreover, it is clear
that, without loss of any generality, we can set $\boldsymbol{j}=(1,\dots,n)$ and
$\boldsymbol{k}=(1,\dots,n)$. Also we note that we can carry out the intermediate
computations up to integer multiples of $2i\pi$. This is justified
in as much as we take the exponential at the end.

Observe that the contour ${\cal C}\cup -\Gamma$ surrounds the points
$\Rp_k$ in the counterclockwise direction and the points $\Rm_k$ in the
clockwise direction. Therefore if $f(\omega)$ is holomorphic in a
domain containing $\Gamma$ and $\cal C$, then
\begin{equation}\label{GtoC-f}
\int\limits_{\Gamma} f'(\omega)\nu(\omega)\,d\omega =
\int\limits_{\cal C} f'(\omega)\nu(\omega)\,d\omega +
\sum_{k=1}^{n}\Bigl( f(\Rm_{k})-f(\Rp_{k})\Bigr) \;.
\end{equation}
Using \eqref{GtoC-f} one can easily calculate  the difference of single
integrals entering \eqref{A-rep}:
 \begin{equation}\label{diff-singI}
 -\int\limits_{\cal C}
 \bigl(ix+g'(\lambda)\bigr)\nu(\lambda)\,d\lambda +
 \int\limits_{\Gamma}
 \bigl(ix+g'(\lambda)\bigr)\nu(\lambda)\,d\lambda=\sum_{k=1}^n\Bigl(
 ix(\Rm_k-\Rp_k)+g(\Rm_k)-g(\Rp_k)\Bigr).
 \end{equation}

The calculation of the difference of the double integrals in
\eqref{A-rep} is more involved. It is convenient to present
$\nu(\lambda)$ in the form
 \begin{equation}\label{nu-tnu}
 \nu(\lambda)=\tilde\nu(\lambda)-\frac1{2\pi
 i}\log \left\{ \prod_{k=1}^n\frac{(\lambda-\qp_k)(\lambda-\qm_k)}
 {(\lambda-\Rp_k)(\lambda-\Rm_k)} \right\},
 \end{equation}
where $\tilde\nu(\lambda)$ is holomorphic in a domain containing
both contours ${\cal C}$ and $\Gamma$. Let
 \begin{equation}\label{Doub-int}
 J({\cal C})=\iint\limits_{\cal
 C}\frac{\nu(\lambda)\nu(\mu)}{(\lambda-\mu_+)^2}\,d\lambda\,d\mu,
 \qquad
  J(\Gamma)=\iint\limits_{\Gamma}\frac{\nu(\lambda)\nu(\mu)}{(\lambda-\mu_+)^2}\,d\lambda\,d\mu.
  \end{equation}
Let us also introduce auxiliary functions
 \begin{equation}\label{alpha-C}
 \alpha_+(\lambda;{\cal C})=\log \left\{ \prod_{k=1}^n\frac{\lambda-\qp_k}
 {\lambda-\Rm_k} \right\} \; , \qquad
 \alpha_-(\lambda;{\cal C})=\log\left\{ \prod_{k=1}^n\frac{\lambda-\qm_k}
 {\lambda-\Rp_k} \right\} \; ,
 \end{equation}
and
 \begin{equation}\label{alpha-G}
 \alpha_+(\lambda;\Gamma)=\log\left\{ \prod_{k=1}^n\frac{(\lambda-\qp_k)}
 {(\lambda-\Rp_k)} \right\}  \; , \qquad
 \alpha_-(\lambda;\Gamma)=\log\left\{ \prod_{k=1}^n\frac{(\lambda-\qm_k)}
 {(\lambda-\Rm_k)} \right\} \; .
 \end{equation}
It is easy to see that the functions $\alpha_\pm(\lambda;{\cal C})$ are
holomorphic to the left (resp. to the right) from the contour ${\cal C}$.
Similarly the functions $\alpha_\pm(\lambda;\Gamma)$ are holomorphic to the
left (resp. to the right) from the contour $\Gamma$. All the functions
\eqref{alpha-C} and \eqref{alpha-G} behave as $O(\lambda^{-1})$ when
$\lambda\to\infty$, and
 \begin{equation}\label{rep-add-term}
  \alpha_+(\lambda;{\cal C})+ \alpha_-(\lambda;{\cal C})
 =\alpha_+(\lambda;\Gamma)+\alpha_-(\lambda;\Gamma)=
 \log\left\{\prod_{k=1}^n\frac{(\lambda-\qp_k)(\lambda-\qm_k)}{(\lambda-\Rp_k)(\lambda-\Rm_k)}\right\} \; .
 \end{equation}

Using \eqref{nu-tnu} and \eqref{alpha-C} we can present  $J({\cal C})$ as a sum
of four integrals $J({\cal C})=\sum_{j=1}^4 J_j({\cal C})$, where
 \begin{align}\label{J-JjC1}
 J_1({\cal C})&=\iint\limits_{\cal
 C}\frac{\tilde\nu(\lambda)\tilde\nu(\mu)}{(\lambda-\mu_+)^2}\,d\lambda\,d\mu,\\
 J_2({\cal C})&=-\frac1{2\pi i}\iint\limits_{\cal
 C}\frac{\tilde\nu(\lambda)\bigl(\alpha_+(\mu;{\cal C})+ \alpha_-(\mu;{\cal C})\bigr)}{(\lambda-\mu_+)^2}\,d\lambda\,d\mu,\\
 J_3({\cal C})&=-\frac1{2\pi i}\iint\limits_{\cal
 C}\frac{\bigl(\alpha_+(\lambda;{\cal C})+ \alpha_-(\lambda;{\cal C})\bigr)\tilde\nu(\mu)}{(\lambda-\mu_+)^2}\,d\lambda\,d\mu,\\
 J_4({\cal C})&=\frac1{(2\pi i)^2}\iint\limits_{\cal
 C}\frac{\bigl(\alpha_+(\lambda;{\cal C})+ \alpha_-(\lambda;{\cal C})\bigr)
 \bigl(\alpha_+(\mu;{\cal C})+ \alpha_-(\mu;{\cal
 C})\bigr)}{(\lambda-\mu_+)^2}\,d\lambda\,d\mu.\label{J-JjC4}
 \end{align}
Similarly $J(\Gamma)=\sum_{j=1}^4 J_j(\Gamma)$ where $J_j(\Gamma)$
is obtained from $J_j({ \cal C})$ by replacing everywhere $\cal C$ by
$\Gamma$ in \eqref{J-JjC1}--\eqref{J-JjC4}.

Since $\tilde\nu(\lambda)$ is holomorphic in a domain containing the
contours $\cal C$ and $\Gamma$, we conclude that $J_1({\cal
C})-J_1(\Gamma)=0$. The integrals $J_4({\cal C})$ and $J_4(\Gamma)$
can be taken explicitly. For example, using the analytic properties
of $\alpha_\pm(\lambda;{\cal C})$ we have
 \begin{multline}\label{J4-calc}
 J_4({\cal C})=\frac1{2\pi i}\int\limits_{\cal
 C}\alpha'_+(\mu;{\cal C})
 \bigl(\alpha_+(\mu;{\cal C})+ \alpha_-(\mu;{\cal
 C})\bigr)\,d\mu=\frac1{2\pi i}\int\limits_{\cal
 C}\alpha'_+(\mu;{\cal C})\alpha_-(\mu;{\cal
 C})\,d\mu\\
 =\frac1{2\pi i}\int\limits_{\cal
 C}\sum_{k=1}^n\left(\frac1{\mu-\qp_k}-\frac1{\mu-\Rm_k}\right)\alpha_-(\mu;{\cal
 C})\,d\mu=
 \log\left\{\prod_{j,k=1}^n\left(\frac{\Rm_k-\qm_j}{\Rm_k-\Rp_j}\cdot
 \frac{\qp_k-\Rp_j}{\qp_k-\qm_j}\right) \right\}.
 \end{multline}
The expression for $J_4(\Gamma)$ is obtained from \eqref{J4-calc} via the
replacement $\Rp\leftrightarrow\Rm$. Then we have
 \begin{equation}\label{J4-J4}
 J_4({\cal C})-J_4(\Gamma)=\log\left\{\prod_{j,k=1}^n\left(\frac{\Rm_k-\qm_j}{\Rm_k-\Rp_j}\cdot
 \frac{\qp_k-\Rp_j}{\qp_k-\Rm_j}\cdot\frac{\Rp_k-\Rm_j}{\Rp_k-\qm_j}\right)\right\} \;.
 \end{equation}

Calculating $J_2$ and $J_3$ we can take explicitly only one of two
integrals. We have
 \begin{equation}\label{J2-calc}
 J_2({\cal C})=\int\limits_{\cal
 C}\alpha'_-(\lambda;{\cal C})\tilde\nu(\lambda)\,d\lambda=
 \int\limits_{\cal
 C}\sum_{k=1}^n\left(\frac1{\lambda-\qm_k}-\frac1{\lambda-\Rp_k}\right)\tilde\nu(\lambda)\,d\lambda,
 \end{equation}
and
 \begin{equation}\label{J3-calc}
 J_3({\cal C})=-\int\limits_{\cal
 C}\alpha'_+(\mu;{\cal C})\tilde\nu(\mu)\,d\mu=
 \int\limits_{\cal
 C}\sum_{k=1}^n\left(\frac1{\mu-\Rm_k}-\frac1{\mu-\qp_k}\right)\tilde\nu(\mu)\,d\mu.
 \end{equation}
The expressions for $J_2(\Gamma)$ and $J_3(\Gamma)$ can be obtained from
\eqref{J2-calc} and \eqref{J3-calc} via the replacements ${\cal C}\to \Gamma$
and $\Rp\leftrightarrow\Rm$. Then we have
 \begin{multline}\label{diff2233}
 \sum_{s=2,3}\bigl(J_s({\cal C})-J_s(\Gamma)\bigr)=
 \int\limits_{{\cal C}\cup\,-\Gamma}\sum_{k=1}^n
 \left(\frac1{\lambda-\qm_k}-\frac1{\lambda-\qp_k}\right)\tilde\nu(\lambda)\,d\lambda\\
 + \int\limits_{{\cal C}\cup \Gamma}\sum_{k=1}^n\left(\frac1{\lambda-\Rm_k}-\frac1{\lambda-\Rp_k}\right)\tilde\nu(\lambda)\,d\lambda.
 \end{multline}
The integral in the first line of \eqref{diff2233} vanishes as the contour
${\cal C}\cup-\Gamma$ does not surround the points $\qpm_k$. The second
integral in \eqref{diff2233} gives
 \begin{equation}\label{second-int}
 \sum_{s=2,3}\bigl(J_s({\cal C})-J_s(\Gamma)\bigr)=2
 \int\limits_{\Gamma}\sum_{k=1}^n\left(\frac1{\lambda-\Rm_k}-\frac1{\lambda-\Rp_k}\right)\tilde\nu(\lambda)\,d\lambda-
 2\pi i\sum_{k=1}^n\bigl(\tilde\nu(\Rp_k)+\tilde\nu(\Rm_k)\bigr).
 \end{equation}
Here we have used that the contour ${\cal C}\cup -\Gamma$ surrounds the points
$\Rp_k$ in the counterclockwise direction and the points $\Rm_k$ in the clockwise
direction. Substituting into \eqref{second-int} the function
$\tilde\nu(\lambda)$ in terms of $\nu(\lambda)$ via \eqref{nu-tnu} we obtain
after simple algebra
 \begin{multline}\label{sec-int-Cau}
 \sum_{s=2,3}\bigl(J_s({\cal C})-J_s(\Gamma)\bigr)=
 \sum_{k=1}^n\bigl(2L_\Gamma[\nu](\Rm_k)-2L_\Gamma[\nu](\Rp_k)-2\pi i\tilde\nu(\Rp_k)-2\pi i\tilde\nu(\Rm_k)\bigr)\\
 +2 \log\prod_{j,k=1}^n\left(\frac{\Rm_k-\qp_j}{\Rm_k-\Rp_j}\cdot
 \frac{\Rp_k-\qm_j}{\Rp_k-\Rm_j}\right),
 \end{multline}
where we have used the Cauchy transform over the contour $\Gamma$ of the
function $\nu(\lambda)$ (see \eqref{def-CauT}).

It remains to find $\tilde\nu(\Rpm_k)$. These numbers can be
expressed in terms of the residues of the Fermi weight
$\vartheta_{reg}(\Rpm_k)$ in the points $\Rpm_k$.  We have
 \begin{equation}\label{thF-nu}
 1+\gamma\vartheta(\lambda)F(\lambda)=e^{-2\pi i\nu(\lambda)}=
 e^{-2\pi i\tilde\nu(\lambda)}\prod_{j=1}^n\frac{(\lambda-\qp_j)(\lambda-\qm_j)}
 {(\lambda-\Rp_j)(\lambda-\Rm_j)}.
 \end{equation}
Hence,
 \begin{equation}\label{thF-nu-rk}
 \gamma\vartheta_{reg}(\Rpm_k)F(\Rpm_k)=
 e^{-2\pi i\tilde\nu(\Rpm_k)}\prod_{j=1}^n\frac{(\Rpm_k-\qp_j)(\Rpm_k-\qm_j)}
 {(\Rpm_k-\Rmp_j)}\prod_{j=1\atop{j\ne k}}^n \frac1{\Rpm_k-\Rpm_j}.
 \end{equation}
Thus, combining \eqref{J4-J4}, \eqref{sec-int-Cau}, and \eqref{thF-nu-rk} we
obtain
 \begin{equation}\label{J-J}
 e^{J({\cal C})-J(\Gamma)}=\left(\det_{n}
 \frac{1}{\Rp_{j} - \Rm_{k}}  \right)^2\prod_{k=1}^n\left[
 \gamma^2e^{2L_\Gamma[\nu](\Rm_k)-2L_\Gamma[\nu](\Rp_k)}
 \vartheta_{reg}(\Rp_k)\vartheta_{reg}(\Rm_k)F(\Rp_k)F(\Rm_k)\right],
 \end{equation}
where we have used
 \begin{equation}\label{det-Cauchy}
 \left(\det_{n}\frac{1}{\Rp_{j} - \Rm_{k}}
 \right)^2=\prod_{j,k=1\atop{j\ne k}}^n (\Rp_{j} - \Rp_{k})(\Rm_{j} - \Rm_{k})
 \prod_{j,k=1}^n(\Rp_{j} - \Rm_{k})^{-2}.
 \end{equation}
Taking into account \eqref{diff-singI} we immediately arrive at the formulae
\eqref{def-calU}, \eqref{def-dirU}.

We insist that all the above computations have been done by using the sole pole/zero structure of $e^{-2i\pi \nu(\omega)}$.
They are thus  valid in the specific case where one considers $\nu= z_{\boldsymbol{j};\boldsymbol{k}}$.

%%%%%%%%%%%%%%%%%%%%%%%%%%%%%%%%%%%%%%%%%%%%%%%%%%%%%%%%%%%%%%%%%%%%%%%%%%%%%%%%%%%%%%%%%%%%%%%%%%%%%%%%%%%%%%%%%%%%%%%%%%%%%%%%%%%%%%%%%%%%%%%%%%%%
%%%%%%%%%%%%%%%%%%%%%%%%%%%%%%%%%%%%%%%%%%%%%%%%%%%%%%%%%%%%%%%%%%%%%%%%%%%%%%%%%%%%%%%%%%%%%%%%%%%%%%%%%%%%%%%%%%%%%%%%%%%%%%%%%%%%%%%%%%%%%%%%%%%%

%%%%%%%%%%%%%%%%%%%%%%%%%%%%%%%%%%%%%%%%%%%%%%%%%%%%%%%%%%%%%%%%%%%%%%%%%%%%%%%%%%%%%%%%%%%%%%%%%%%%%%%%%%%%%%%%%%%%%%%%%%%%%%%%%%%%%%%%%%%%%%%%%%%%
%%%%%%%%%%%%%%%%%%%%%%%%%%%%%%%%%%%%%%%%%%%%%%%%%%%%%%%%%%%%%%%%%%%%%%%%%%%%%%%%%%%%%%%%%%%%%%%%%%%%%%%%%%%%%%%%%%%%%%%%%%%%%%%%%%%%%%%%%%%%%%%%%%%%

\section{Continuous generalization of the multiple Lagrange series\label{S-CgMLS}}
\label{appendix Lagrange series}

The continuous generalization of the multiple Lagrange series has
the form
 \begin{equation}\label{A4-MI-ser}
 \hat G
 =\sum_{n=0}^\infty\frac{1}{n!}
  \int\limits_{{\cal L}}\, d^n\lambda \;
  \prod_{j=1}^n
  \frac{\partial}{\partial \epsilon_j}
  \left.\prod_{j=1}^n
  f\left(\sum_{a=1}^n\epsilon_a\, \xi(\lambda_a,\lambda_j)\right)
  \cdot
  F\left( \sum_{a=1}^n h_1(\lambda_a)\, \epsilon_a;\
          \dots,\right) \right|_{\epsilon_j=0}.
 \end{equation}
Here the integrals are taken over some contour ${\cal L}$, the
functions $f$, $F$, $\xi$, and $h_1$ are holomorphic the corresponding
neighborhoods. The argument of the function $F$ may
also contain double sums
$\sum_{a,b}h_2(\lambda_a,\lambda_b)\epsilon_a\epsilon_b $
\textit{etc}, like, for example, in \eqref{W-lin}. Apart from
obvious modifications, the presence of such multiple sums does not
affect the result. Hence, we have omitted these arguments of the
function $F$ for brevity.

This series was studied in \cite{KitKMST09a}. If there exists a
$R_0>0$ such that
\begin{equation}\label{A4-RE-conv}
\sup_{\phi\in [0; 2\pi]}  \;  \sup_{\mu \in {\cal L}}
 \biggl|f\!\biggl(R_0 e^{i\phi} \int\limits_{{\cal L}}  |\xi(\lambda,\mu)|\,d\lambda\biggr)\biggr|<R_0,
 \end{equation}
then it is absolutely convergent. In such a case, the result of the summation reads
 \begin{equation}\label{A4-RE-res}
 \hat G=\frac{
 F\bigg( \int_{{\cal L}}h_1(\mu)\,z(\mu)\,d\mu;\ \dots\bigg)
  }{ \det_{\cal L}\bigg[\delta(\lambda-\mu)- \xi(\mu,\lambda)\,
 f'\bigg(
 \int_{{\cal L}}  \xi(\nu,\lambda)\, z(\nu)\,d\nu\bigg)\bigg]}\, ,
 \end{equation}
where the function $z(\mu)$ is the unique solution to the integral equation
\begin{equation}\label{A4-RE-Int-eq}
 z(\mu)=f\!\biggl(
 \int\limits_{{\cal L}}  \xi(\lambda,\mu)\, z(\lambda)\,d\lambda\biggr).
 \end{equation}
The denominator of \eqref{A4-RE-res} contains the Fredholm determinant of a linear
integral operator acting on ${\cal L}$. If the function $F$ depends on multiple
sums, then these should be replaced by multiple integrals, like, for example,
 \begin{equation}\label{High-Sum}
 \sum_{a,b=1}^n h_2(\lambda_a,\lambda_b)\epsilon_a\epsilon_b \hookrightarrow
 \int\limits_{{\cal L}}h_2(\lambda,\mu)\,z(\lambda)z(\mu)\,d\lambda\,d\mu,
 \end{equation}
etc. It is important to note that the form of the integral equation
\eqref{A4-RE-Int-eq} does not depend on the function $F$.

The result \eqref{A4-RE-res} can be directly applied to the summation of the
series \eqref{Lagr-ser}, where the integration contour ${\cal L}$ coincides
with $\Gamma_{\boldsymbol{j};\boldsymbol{k}}$ and the function $f$ is
${\phi}_{\boldsymbol{j};\boldsymbol{k}}$ \eqref{Phi-phijk}.

%%%%%%%%%%%%%%%%%%%%%%%%%%%%%%%%%%%%%%%%%%%%%%%%%%%%%%%%%%%%%%%%%%%%%%%%%%%%%%%%%%%%%%%%%%%%%%%%%%%%%%%%%%%%%%%%%%%%%%%%%%%%%%%%%%%%%%%%%%%%%%%%%%%%
%%%%%%%%%%%%%%%%%%%%%%%%%%%%%%%%%%%%%%%%%%%%%%%%%%%%%%%%%%%%%%%%%%%%%%%%%%%%%%%%%%%%%%%%%%%%%%%%%%%%%%%%%%%%%%%%%%%%%%%%%%%%%%%%%%%%%%%%%%%%%%%%%%%%

%%%%%%%%%%%%%%%%%%%%%%%%%%%%%%%%%%%%%%%%%%%%%%%%%%%%%%%%%%%%%%%%%%%%%%%%%%%%%%%%%%%%%%%%%%%%%%%%%%%%%%%%%%%%%%%%%%%%%%%%%%%%%%%%%%%%%%%%%%%%%%%%%%%%
%%%%%%%%%%%%%%%%%%%%%%%%%%%%%%%%%%%%%%%%%%%%%%%%%%%%%%%%%%%%%%%%%%%%%%%%%%%%%%%%%%%%%%%%%%%%%%%%%%%%%%%%%%%%%%%%%%%%%%%%%%%%%%%%%%%%%%%%%%%%%%%%%%%%

\end{document}